\documentclass[preprint,prd,eqsecnum,aps,epsf]{revtex4}

\usepackage{axodraw}

\begin{document}

\def\aprge{\buildrel > \over {_{\sim}}}
\def\aprle{\buildrel < \over {_{\sim}}}

\def\etal{{\it et.~al.}}
\def\ie{{\it i.e.}}
\def\eg{{\it e.g.}}

\def\bwt{\begin{widetext}}
\def\ewt{\end{widetext}}
\def\be{\begin{equation}}
\def\ee{\end{equation}}
\def\bea{\begin{eqnarray}}
\def\eea{\end{eqnarray}}
\def\bean{\begin{eqnarray*}}
\def\eean{\end{eqnarray*}}
\def\bary{\begin{array}}
\def\eary{\end{array}}
\def\bi{\bibitem}
\def\bit{\begin{itemize}}
\def\eit{\end{itemize}}

\def\lan{\langle}
\def\ran{\rangle}
\def\lra{\leftrightarrow}
\def\la{\leftarrow}
\def\ra{\rightarrow}
\def\dash{\mbox{-}}
\def\ol{\overline}

\def\ub{\ol{u}}
\def\db{\ol{d}}
\def\sb{\ol{s}}
\def\cb{\ol{c}}

\def\re{\rm Re}
\def\im{\rm Im}

\def \b{{\cal B}}
\def \ca{{\cal A}}
\def \ko{K^0}
\def \ok{\overline{K}^0}
\def \s{\sqrt{2}}
\def \st{\sqrt{3}}
\def \sx{\sqrt{6}}
\title{\Large{\bf Anomaly and Anomaly-Free Treatment of QFTs \\ Based on Symmetry-preserving Loop Regularization}}
\author{Yong-Liang Ma and Yue-Liang Wu}
\address{Institute of Theoretical Physics, Chinese Academy of Sciences, Beijing
100080, China}

\date{\today}
\begin{abstract}
The triangle anomaly in massless and massive QED is investigated
by adopting the symmetry-preserving loop regularization method
proposed recently in \cite{LR}. The method is realized in the
initial dimension of theory without modifying the original
lagrangian, it preserves symmetries under non-Abelian gauge and
Poincare transformations in spite of the existence of two
intrinsic mass scales $M_c$ and $\mu_s$ which actually play the
roles of UV- and IR-cut off respectively. The
axialvector-vector-vector (AVV) triangle diagrams in massless and
massive QED are evaluated explicitly by using the loop
regularization. It is shown that when the momentum $k$ of external
state is soft with $k^2 \ll \mu_s^2, m^2 $ ($m$ is the mass of
loop fermions) and $ M_c \to \infty$ , both massless and massive
QED become anomaly free. The triangle anomaly is found to appear
as quantum corrections in the case that $ m^2, \mu_s^2 \ll k^2 $
and $M_c \to \infty$. Especially, it is justified that in the
massless QED with $\mu_s =0$ and $M_c\to \infty$, the triangle
anomaly naturally exists as quantum effects in the axial-vector
current when the ambiguity caused by the trace of gamma matrices
with $\gamma_5$ is eliminated by simply using the definition of
$\gamma_5$. It is explicitly demonstrated how the Ward identity
anomaly of currents depends on the treatment for the trace of
gamma matrices, which enables us to make a clarification whether
the ambiguity of triangle anomaly is caused by the regularization
scheme in the perturbation calculations or by the trace of gamma
matrices with $\gamma_5$. For comparison, an explicit calculation
based on the Pauli-Villars regularization and dimensional
regularization is carried out and the possible ambiguities of Ward
identity anomalies caused from these two regularization schemes
are carefully discussed, which include the ambiguities induced by
the treatment of the trace of gamma matrices with $\gamma_5$ and
the action of the external momentum on the amplitude before the
direct calculation of the AVV diagram.
\end{abstract}
\maketitle

\tableofcontents

\section{Introduction}

\label{sec intro}

Symmetry has played an important role and achieved great triumphs
in modern physics especially in elementary particle physics, such
as the construction of standard model. Theoretically it is known
from the Noether's theorem that if a system is invariant under a
continue global transformation, there is a conserved current
corresponding to the transformation. Although Noether's theorem is
a general conclusion which was verified mathematically, while it
is a classical conclusion. In quantum theory, the situation can
become different and the Noether's theorem may be violated by
quantum corrections, which is usually called quantum anomaly. The
importance of anomaly arises from the study of
$\pi^0\rightarrow2\gamma$, the decay of this process is forbidden
if the classical symmetry is preserved in quantum case, which is
in contrast with the experimental data for a large decay rate,
this is the so called Veltman-Sutherland paradox\cite{V-Spara}. It
implies that some of the conservation laws corresponding to the
symmetries in such a process are violated by the quantum
corrections. This is known as the triangle
anomaly\cite{anomaly1,anomaly2}. It has been investigated by many
groups\cite{anomaly3,anomaly4,anomaly5,anomaly6,anomaly7}.

Anomaly as quantum effects becomes significantly important in
quantum field theories. In perturbation theory, the anomaly has
been calculated by using different regularization schemes, such as
Pauli-Villars regularization\cite{PV}, dimensional
regularization\cite{DR} and point splitting method\cite{PS}. It
has also be evaluated in language of path integral\cite{PI} and
constructed from differential geometry and topological
method\cite{DG}. In perturbative calculations, one needs to
carefully deal with the divergences by choosing appropriate
regularization schemes. In the dimensional regularization, it is
known to be difficult to define $\gamma_5$. In the Pauli-Villars
regularization, the field contents of the theory are changed and
the non-Abelian gauge invariance cannot well be preserved due to
the introduction of massive gauge fields.

In this paper, we shall investigate the anomaly based on the
symmetry-preserving loop regularization method developed recently
in ref. \cite{LR}. Unlike the Pauli-Villars and dimensional
regularization schemes, the loop regularization does not change
either the contents of lagrangian or the dimension of system.
Though its prescription is quite similar to the Pauli-Villars
regularization, while the basic concept is quite different. This
is because the regularization prescription in the loop
regularization is applied to the so-called irreducible loop
integrals (ILIs) evaluated from the Feynman loop diagrams rather
than to the propagators of additional super-heavy fields made in
the Pauli-Villars regularization, which is the reason why loop
regularization can satisfy a set of consistency conditions for
preserving non-Abelian gauge invariance. In comparison with the
dimensional regularization, the loop regularization is carried out
in the four dimensional space-time, thus the so-called $\gamma_5$
problem in the dimensional regularization does not exist in the
loop regularization. In addition, two mass scales are safely
introduced in the loop regularization, one is corresponding to the
characterizing energy scale $M_c$ which can be taken to be
infinity $M_c\to \infty$ for underlying renormalizable quantum
field theories, another is the sliding energy scale $\mu_s$, they
actually play the roles of ultraviolet and infrared cut-off scales
respectively, so that the loop regularization maintains the
well-defined divergent behavior of original integrals when taking
$M_c \to \infty$ and $\mu_s \to 0$. In fact, it has been shown
that the two mass scales play important roles in understanding the
dynamically spontaneous symmetry breaking and the meson
spectrum\cite{DW} for low energy dynamics of QCD.

In the direct calculation of the triangle diagram, the first
problem one meets is how to deal with the trace of gamma matrices.
There are three possible ways to treat the trace of gamma
matrices: (i) directly calculate the trace with the definition of
$\gamma_5$; (ii) firstly by classifying the Lorentz indices of the
two vector currents with another gamma matrix into one group to
reduce the number of gamma matrices before the trace; (ii) firstly
by classifying the Lorentz indices of one vector current and the
axial-vector current with another gamma matrix into one group to
reduce the number of gamma matrices before the trace. We shall
show that different treatments on the trace of gamma matrices lead
to different forms of Ward identity anomaly of the currents. This
may be referred as the ambiguity caused by the trace of gamma
matrices with $\gamma_5$, which is independent of regularization.
It is shown that such an ambiguity can simply been eliminated by
directly calculating the trace with the definition of $\gamma_5$,
which enables us to clarify the ambiguity solely arising from the
regularization schemes. In all treatments of the gamma matrices,
it will be seen that the IR-cut off energy scale $\mu_s$ plays a
crucial role in understanding anomaly. As either the massless QED
or massive QED is a renormalizable quantum field theory, one can
always take UV-cut off energy scale $M_c$ to be infinity, which
does not affect the analysis of anomaly.

For massless QED, it will be demonstrated that the introduction of
IR-cut off scale $\mu_s$ makes both the vector and axial-vector
currents conserved. It is then implies that the loop
regularization with $\mu_s\neq 0$ ensures Ward identities for both
the vector and axial-vector currents. While in the absence of the
IR-cut off scale, i.e., $\mu_s = 0$, we will show that the
standard form of axial anomaly is obtained in the treatment where
the trace of gamma matrices was evaluated by simply using the
definition of $\gamma_5$ in four dimensional space-time. In such a
treatment, the trace of gamma matrices is unique without
ambiguities. For the other treatments of currents, we will see
that, in the treatment where the two gamma matrices which carry
the Lorentz indices of the vector currents and the gamma matrix
between them are classified into one group to reduce the number of
gamma matrices, the vector Ward identities are violated and the
axial-vector Ward identity is preserved. While in the treatment,
where the two gamma matrices with the Lorentz indices of one
vector and one axial-vector currents and the gamma between them
are classified into one group to reduce the number of gamma
matrices, it is found that the Ward identities for the two
currents which are grouped become violated while the remaining
vector Ward identity is kept. It is clear that such ambiguities
are caused by the trace of gamma matrices with $\gamma_5$.
Nevertheless, in both cases, by redefining the physical vector
current to be conserved, the anomaly appears in the axial-vector
Ward identity with a standard form.

Similarly, for massive QED, the calculation shows that in the
treatment where the definition of $\gamma_5$ are explicitly used,
two vector currents are conserved and the axial-vector Ward
identity is in general violated by quantum corrections in the case
that $ m^2, \mu_s^2 \ll k^2 $ and $M_c \to \infty$. But when the
mass $m$ or IR cut-off scale $\mu_s$ is sufficiently large in
comparison with the momentum of external states, no anomaly will
appear in massive QED. In the treatment where the two vector
indices are grouped, both the vector Ward identities have anomaly
terms and the axial-vector becomes conserved, but both vector and
axial-vector currents can be made to be conserved and consequently
no anomaly will appear in massive QED when the mass $m$ or IR
cut-off scale $\mu_s$ is sufficiently large in comparison with the
momentum of external states. In the treatment where one vector and
the axial-vector currents are grouped, there is anomaly in the
grouped vector and axial-vector currents while the other vector
current is conserved automatically. Similarly, by a redefinition,
two vector currents can be made conserved and anomaly arises in
the axial-vector Ward identity. But with the same condition that
the mass $m$ or IR cut-off scale $\mu_s$ is sufficiently large in
comparison with the momentum of external states, the massive QED
becomes anomaly free.

From the loop regularization, it is clearly seen that in general
the triangle anomaly appears when the external momentum scale
$k^2$ of axial-vector current becomes much larger than the mass
$m$ of loop fermions or the IR cut-off scale $\mu_s$. An explicit
calculation shows that anomaly terms arise from the convergent
integrals as well as the finite parts of the superficially
divergent integrals in loop regularization.

The paper is organized as follows: Section II is a brief outline
of the loop regularization. In section III, the AVV diagrams in
massless QED are calculated explicitly in the loop regularization
with $\mu_s\neq 0$ and $\mu_s = 0$. In section IV, we calculate
both the AVV and PVV diagrams in massive QED. In section V and VI,
we present the calculations based on Pauli-Villars and dimensional
regularization respectively with $m \neq 0$ and $m = 0$. In
section VII, we show how the Ward identity anomaly of currents
depends on the treatment for the trace of gamma matrices and thus
clarify a unique solution for Ward identity anomaly appearing in
the axial-vector current. The possible ambiguities of Ward
identity anomaly in the dimensional regularization and
Pauli-Villars scheme are carefully investigated. Our conclusions
and remarks will be given in section VIII. Some useful
formulations are listed in Appendix.

\section{A brief outline of loop regularization}

\label{sec revLR}

In the quantum field theory, it is inevitable to deal with the
infinity problem in the momentum integral. To carry out the
integral explicitly, one must first make the infinite integrals be
well-defined, which is the so called regularization. Several
regularization methods have been proposed, the typical ones
include dimensional regularization and Pauli-Villars
regularization as well as the most recently proposed
symmetry-preserving loop regularization\cite{LR}.

To propose a regularization, several elements should be
considered, such as the Lorentz invariance, gauge invariance,
chiral properties and the behavior of divergence. In the
dimensional regularization, the momentum integral of Feynman loops
is performed by an analytic continuation of dimensions, it does
preserve the Lorentz invariance and gauge invariance when
$\gamma_5$ is not involved. Once $\gamma_5$ is concerned, the
dimensional regularization faces problem since $\gamma_5$ is an
intrinsically four dimensional object. Although a redefinition of
$\gamma_5$ given in \cite{DR} can preserve gauge symmetry at one
loop level, while such a redefinition will destroy the gauge
symmetry at two loop level\cite{2loopgamma5}. In the Pauli-Villars
regularization, the momentum integral is carried out in four
dimensions, while the field content is changed by the introduction
of heavy massive fields in Pauli-Villars regularization in order
to make the integral finite. Although the Abelian gauge symmetry
can be preserved, while it is not applicable to the non-Abelian
gauge symmetries as the introduction of heavy massive nonabelian
gauge fields will destroy gauge symmetry.

It is then of interest to find out a regularization which does not
modify the Lagrangian of original theory, and meanwhile preserves
both the Lorentz and gauge symmetries, as well as maintains the
behavior of divergent integrals by the introduction of intrinsic
mass scales. The symmetry-preserving loop regularization\cite{LR}
has been found to satisfy the above requirements. In the loop
regularization, the key concept is the introduction of irreducible
loop integrals (ILIs) which are evaluated from Feynman integrals.
For instance, at the one-loop level, all Feynman integrals can be
evaluated into the following irreducible loop integrals (ILIs)
\begin{eqnarray}
I_{-2a}&=&\int\frac{d^4k}{(2\pi)^4}\frac{1}{(k^2-M^2)^{2+a}}\nonumber\\
I_{-2a~\mu\nu}&=&\int\frac{d^4k}{(2\pi)^4}\frac{k_\mu
k_\nu}{(k^2-M^2)^{3+a}},
~~~~~~~~a=-1,0,1,2,\cdots\label{ILIs}\\
I_{-2a~\mu\nu\alpha\beta}&=&\int\frac{d^4k}{(2\pi)^4}\frac{k_\mu
k_\nu k_\alpha k_\beta}{(k^2-M^2)^{2+a}}\nonumber
\end{eqnarray}

As shown in \cite{LR}, to maintain the Lorentz invariance and
gauge invariance, the regularized ILIs should satisfy a set of
consistent conditions\cite{LR}
\begin{eqnarray}
I^R_{2\mu\nu}&=&{1\over2}g_{\mu\nu}I^R_2, \quad
I^R_{2\mu\nu\rho\sigma}={1\over8}g_{\{\mu\nu} g_{{\rho\sigma}\}}I^R_2\nonumber\\
I^R_{0\mu\nu}&=&{1\over4}g_{\mu\nu}I^R_0, \quad
I^R_{0\mu\nu\rho\sigma}={1\over24}g_{\{\mu\nu} g_{{\rho\sigma}\}}I^R_0 \label{consistcondition} \\
I^R_{-2\mu\nu}&=&{1\over6}g_{\mu\nu} I^R_{-2}, \quad
I^R_{-2\mu\nu\rho\sigma}={1\over48}g_{\{\mu\nu}
g_{{\rho\sigma}\}}I^R_{-2}\nonumber
\end{eqnarray}
where $g_{\{\mu\nu} g_{{\rho\sigma}\}}\equiv
g_{\mu\nu}g_{\rho\sigma}+g_{\mu\rho}g_{\nu\sigma}+g_{\mu\sigma}g_{\nu\rho}$.

The prescription of loop regularization is simple: replacing the
integration variable $k^2$ and integration measure
$\int\frac{d^4k}{(2\pi)^4}$ by the regularized ones as\cite{LR}
\begin{eqnarray}
k^2\rightarrow[k^2]_l\equiv
k^2-M_l^2,~~~~\int\frac{d^4k}{(2\pi)^4}\rightarrow
\int[\frac{d^4k}{(2\pi)^4}]_l\equiv\lim_{N,M_i^2}
\sum_{l=0}^Nc_l^N\int\frac{d^4k}{(2\pi)^4}\label{precedure}
\end{eqnarray}
with conditions
\begin{eqnarray}
\lim_{N,M_i^2}\sum_{l=0}^Nc_l^N(M_l^2)^n=0, \quad c_0^N=1 ~~~~(
i=0,1,\cdots,N~~\mbox{and} ~~n=0,1,\cdots)\label{condition}
\end{eqnarray}
where $c_l^N$ is the coefficients determined by the above
conditions. The regularized ILIs are then given by\cite{LR}
\begin{eqnarray}
I^R_{-2a}&=&\lim_{N,M_i^2}\sum_{l=0}^Nc_l^N
\int\frac{d^4k}{(2\pi)^4}\frac{1}{(k^2-M_l^2)^{2+a}}\nonumber\\
I^R_{-2a~\mu\nu}&=&\lim_{N,M_i^2}\sum_{l=0}^Nc_l^N
\int\frac{d^4k}{(2\pi)^4}\frac{k_\mu k_\nu}{(k^2-M_l^2)^{3+a}},~~~~~~~~a=-1,0,1,2,\cdots\\
I^R_{-2a~\mu\nu\rho\sigma}&=&\lim_{N,M_i^2}\sum_{l=0}^Nc_l^N
\int\frac{d^4k}{(2\pi)^4}\frac{k_\mu k_\nu k_\rho
k_\sigma}{(k^2-M_l^2)^{2+a}}\nonumber
\end{eqnarray}
which can be shown to satisfy the consistency conditions.

An explicit and simple solution of the above equations has been
found in \cite{LR}
\begin{equation}
M_l^2 = \mu_s^2 + l M_R^2, \quad c_l^N = (-1)^l \frac{N!}{(N-l)!\
l!}
\end{equation}
Here $M_R$ is an arbitrary mass scale and $N$ represents the
regulator number. It should be noticed that this is different from
Pauli-Villars regularization scheme which introduces additional
propagators of new fields. Considering the consistent conditions
given in $(\ref{consistcondition})$, we only need to evaluate
$I_2^R, I_0^R$ and $I_{-2}^R$. Their explicit forms are given
by\cite{LR}
\begin{eqnarray}
I_2^R&=&-\frac{i}{16\pi^2}\{M_c^2-\mu^2[\ln\frac{M_c^2}{\mu^2}-\gamma_\omega+1+
y_2(\frac{\mu^2}{M_c^2})]\}\label{I2R}\\
I_0^R&=&\frac{i}{16\pi^2}[\ln\frac{M_c^2}{\mu^2}-\gamma_\omega
+y_0(\frac{\mu^2}{M_c^2})]\label{I0R}\\
I_{-2}^R&=&-\frac{i}{16\pi^2}\frac{1}{2\mu^2}[1-y_{-2}(\frac{\mu^2}{M_c^2})]\label{I-2R}
\end{eqnarray}
with $ \mu^2 = \mu_s^2 + M^2$, $\gamma_w = \gamma_E =
0.5772\cdots$, and
\begin{eqnarray}
& & y_{-2} (x) = 1 - e^{-x} \nonumber \\
& & y_0 (x) = \int_0^x d \sigma\ \frac{1 - e^{-\sigma} }{\sigma},
\quad  y_1 (x)  = \frac{e^{-x} - 1 + x}{x}\\
& & y_2(x) = y_0(x) - y_1(x),\quad  M_c^2 = \lim_{N,M_R} M_R^2/\ln
N \nonumber\label{y-Function}
\end{eqnarray}
Where $\mu_s $ sets an IR `cutoff' at $M^2 =0$ and $M_c$ provides
an UV `cutoff'. More generally speaking, $M_c$ and $\mu_s$ play
the role of characterizing and sliding energy scales respectively.
For renormalizable quantum field theories, $M_c$ can be taken to
be infinity $(M_c\rightarrow\infty)$. $\mu_s$ can safely runs to
$\mu_s=0$ in a theory without infrared divergence.  In fact,
taking $M_l\to \infty$ (or $M_R\to \infty$) and $\mu_s=0$ in the
regularization is to recover the initial integral. Also once $M_R$
and $N$ are taken to be infinity, the regularized theory becomes
independent of regulators. For a detailed description and proof on
loop regularization, it is referred to the references in
\cite{LR}.

As an illustration, we make an evaluation for a fermionic loop in
massless QED. In massless QED, the electronic loop is given by

\begin{picture}(400,60)(0,0)
\ArrowArcn(150,30)(15,180,0)\ArrowArcn(150,30)(15,0,180)
\Photon(115,30)(135,30){2}{3}\Photon(165,30)(185,30){2}{3}
\Text(125,25)[]{$\rightarrow$}\Text(125,15)[]{$q$}
\Text(150,4)[]{$k$}\Text(155,50)[]{$k+q$}\Text(250,30)[]{$\equiv
i\Pi^{\mu\nu}(q)$}
\end{picture}
\begin{center}{\sl Fig1. Self-energy diagram of photon in quantum electrodynamics}
\end{center}
One can write $i\Pi^{\mu\nu}(q)$ explicitly as
\begin{eqnarray}
i\Pi^{\mu\nu}(q)=(-ie)^2(-1)\int\frac{d^4k}{(2\pi)^4}{\rm
tr}[\gamma^\mu\frac{i}{k\hspace{-0.2cm}\slash}
\gamma^\nu\frac{i}{k\hspace{-0.2cm}\slash+q\hspace{-0.2cm}\slash}]
\end{eqnarray}
By using the Feynman parametrization, we can combine the
denominators and get the form
\begin{eqnarray}
i\Pi^{\mu\nu}(q)=-e^2\int_0^1dx\int\frac{d^4k}{(2\pi)^4}{\rm
tr}\{\frac{\gamma_\mu
k\hspace{-0.2cm}\slash\gamma_\nu(q\hspace{-0.2cm}
\slash+k\hspace{-0.2cm}\slash)}{[(k+xq)^2-M^2]^2}\}
\end{eqnarray}
where $M^2=-x(1-x)q^2$. After making a simple evaluation for Dirac
algebra and shifting the integration variable, we have
\begin{eqnarray}
i\Pi^{\mu\nu}(q)=-4e^2\int_0^1dx[(2I_{2,\mu\nu}-g_{\mu\nu}I_2)+2x(1-x)(q^2g_{\mu\nu}-q_\mu
q_\nu)I_0]
\end{eqnarray}
where $I_{2,\mu\nu}$, $I_2$ and $I_0$ are defined in (\ref{ILIs}).
Note that the shifting of the integration variable is allowed as
the loop regularization preserves the translational and Lorentz
invariance.

It is seen that the logarithmically divergent integral preserves
the gauge symmetry while the quadratically divergent part violates
the gauge symmetry. To preserve the gauge symmetry, the
regularized quadratically divergent integral should satisfy the
relation
\begin{eqnarray}
2I_{2,\mu\nu}^R-g_{\mu\nu}I_2^R=0
\end{eqnarray}

With the prescription of loop regularization, turning to the
Euclidean space, the regularized ILIs are given by
\begin{eqnarray}
I_{2,\mu\nu}^R&=&-i\sum_{l=0}^Nc_l^N\int\frac{d^4k}{(2\pi)^2}\frac{k_\mu
k_\nu}{(k^2+M_l^2)^2}\label{defI2munu}\\
I_2^R&=&-i\sum_{l=0}^Nc_l^N\int\frac{d^4k}{(2\pi)^2}\frac{1}{(k^2+M_l^2)}\label{defI2}\\
I_0^R&=&i\sum_{l=0}^Nc_l^N\int\frac{d^4k}{(2\pi)^2}\frac{1}{(k^2+M_l^2)^2}\label{defI0}
\end{eqnarray}
Their explicit forms are giving in eqs. (\ref{I2R}) and
(\ref{I0R}).

\section{ Anomaly and Anomaly-free treatment for massless QED in loop regularization}

\label{sec ano-and-fre-massless_LR}

We begin with the massless spinor electrodynamics with the
lagrangian
\begin{eqnarray}
{\cal L}=\bar{\psi}\gamma^\mu(i\partial_\mu-e{\cal A}_\mu)\psi
\end{eqnarray}
In the following calculation we will neglect the coupling constant
$e$. The vector current $V_\mu(x)$ and axial-vector current
$A_\mu(x)$ are defined as
\begin{eqnarray}
V_\mu(x)=\bar{\psi}(x)\gamma_\mu\psi(x),~~~~~~~A_\mu(x)=\bar{\psi}(x)\gamma_\mu\gamma_5\psi(x)
\end{eqnarray}
Classically, the above currents are conserved, that is
\begin{eqnarray}
\partial^\mu V_\mu(x)=0,~~~~~~~\partial^\mu
A_\mu(x)=0\label{divcurrents}
\end{eqnarray}


To investigate the quantum corrections, considering the following
Green function
\begin{eqnarray}
T^{AVV}_{\mu\nu\lambda}(p,q;(p+q))&=&\int d^4x_1d^4x_2e^{ipx_1+iq
x_2}\langle0|T[V_\mu(x_1) V_\nu(x_2) A_\lambda(0)]|0\rangle
\end{eqnarray}
The corresponding classical Ward identity (\ref{divcurrents})
leads to the following relations
\begin{eqnarray}
&&p^\mu T^{AVV}_{\mu\nu\lambda}(p,q;(p+q))=0\\
&&q^\nu T^{AVV}_{\mu\nu\lambda}(p,q;(p+q))=0\\
&&(p+q)^\lambda T^{AVV}_{\mu\nu\lambda}(p,q;(p+q))=0
\end{eqnarray}

We now calculate the quantum corrections at one loop level.
Diagrammatically, $T^{AVV}$ can be represented as follows with its
cross one

\begin{picture}(400,70)(0,0)
\ArrowLine(200,30)(250,50)\ArrowLine(250,10)(200,30)\ArrowLine(250,50)(250,10)
\Text(200,30)[]{$\bullet$}\Text(250,50)[]{$\bullet$}\Text(250,10)[]{$\bullet$}
\Text(180,35)[]{$\gamma_\lambda\gamma_5$}\Text(260,45)[]{$\gamma_\mu$}\Text(260,15)[]{$\gamma_\nu$}
\Text(220,50)[]{$k+k_3$}\Text(270,30)[]{$k+k_1$}\Text(220,10)[]{$k+k_2$}\DashArrowLine(150,30)(200,30){4}
\Photon(250,50)(290,50){2}{5}\Photon(250,10)(290,10){2}{5}\Text(275,60)[]
{$q_\mu=(k_3-k_1)_\mu$}\Text(275,0)[]{$p_\nu=(k_1-k_2)_\nu$}\Text(275,47)[]
{$\rightarrow$}\Text(275,13)[]{$\rightarrow$}
\end{picture}
\begin{center}{\sl Fig.2. One loop diagrammatical representation of correction to
$T^{AVV}$.}\end{center}

One can easily write down the corresponding loop contributions
$T^{(1),AVV}_{\lambda\mu\nu}$ from the above diagram
\begin{eqnarray}
T^{(1),AVV}_{\lambda\mu\nu}&=&(-1)\int\frac{d^4k}{(2\pi)^4}{\rm
tr}\{\gamma_\lambda\gamma_5\frac{i}{(k\hspace{-0.2cm}\slash+k\hspace{-0.2cm}\slash_2)}
\gamma_\nu\frac{i}{(k\hspace{-0.2cm}\slash+k\hspace{-0.2cm}\slash_1)}
\gamma_\mu\frac{i}{(k\hspace{-0.2cm}\slash+k\hspace{-0.2cm}\slash_3)}\}\nonumber\\
&=&-i\int\frac{d^4k}{(2\pi)^4}\frac{(k+k_2)_\alpha(k+k_1)_\beta(k+k_3)_\xi}{(k+k_2)^2(k+k_1)^2(k+k_3)^2}{\rm
tr}\{\gamma_5\gamma_\lambda\gamma_\alpha\gamma_\nu\gamma_\beta\gamma_\mu\gamma_\xi\}\label{TAVV}
\end{eqnarray}

For the trace of gamma matrix, there are several ways to deal with
it. Firstly, by using relation
\begin{eqnarray}
\gamma_\mu\gamma_\nu\gamma_\alpha=g_{\mu\nu}\gamma_\alpha-g_{\mu\alpha}\gamma_\nu+g_{\nu\alpha}\gamma_\mu-i\epsilon_{\mu\nu\alpha\beta}\gamma_5\gamma_\beta\label{gammaidentity}
\end{eqnarray}
we can reduce the number of gamma matrix in the trace to make the
trace simpler. Even in the case, there are several ways to
classify the gamma matrices. If we select
\begin{eqnarray}
\gamma_\nu\gamma_\beta\gamma_\mu=g_{\nu\beta}\gamma_\mu-g_{\mu\nu}\gamma_\beta+g_{\mu\beta}
\gamma_\nu-i\epsilon_{\nu\beta\mu\sigma}\gamma_5\gamma_\sigma\label{gammasymmetricvectors}
\end{eqnarray}
we have
\begin{eqnarray}
{\rm
tr}\{\gamma_5\gamma_\lambda\gamma_\alpha\gamma_\nu\gamma_\beta\gamma_\mu\gamma_\xi\}
&=&4i\{g_{\mu\nu}\epsilon_{\lambda\alpha\beta\xi}-g_{\nu\beta}\epsilon_{\lambda\alpha\mu\xi}
-g_{\mu\beta}\epsilon_{\lambda\alpha\nu\xi}\nonumber\\
&&-g_{\lambda\alpha}\epsilon_{\nu\beta\mu\xi}+g_{\alpha\xi}\epsilon_{\nu\beta\mu\lambda}
-g_{\lambda\xi}\epsilon_{\nu\beta\mu\alpha}\}\label{twovectorgamma}
\end{eqnarray}
While if we choose
\begin{eqnarray}
\gamma_\lambda\gamma_\alpha\gamma_\nu=g_{\lambda\alpha}\gamma_\nu-g_{\lambda\nu}\gamma_\alpha
+g_{\alpha\nu}\gamma_\lambda-i\epsilon_{\lambda\alpha\nu\sigma}\gamma_5\gamma_\sigma
\end{eqnarray}
we then arrive at the following result
\begin{eqnarray}
{\rm
tr}\{\gamma_5\gamma_\lambda\gamma_\alpha\gamma_\nu\gamma_\beta\gamma_\mu\gamma_\xi\}
&=&4i\{g_{\nu\lambda}\epsilon_{\alpha\beta\mu\xi}-g_{\alpha\lambda}\epsilon_{\nu\beta\mu\xi}
-g_{\alpha\nu}\epsilon_{\lambda\beta\mu\xi}\nonumber\\
&&-g_{\mu\xi}\epsilon_{\lambda\alpha\nu\beta}+g_{\beta\xi}\epsilon_{\lambda\alpha\nu\mu}
-g_{\beta\mu}\epsilon_{\lambda\alpha\nu\xi}\}\label{vector-axialvectorgamma}
\end{eqnarray}
where we have used the trace formula
\begin{eqnarray}
Tr\{\gamma_5\gamma_\mu\gamma_\nu\gamma_\alpha\gamma_\beta\}
&=&-4i\epsilon_{\mu\nu\alpha\beta},~~~~~~~~~\epsilon_{0123}=1
\end{eqnarray}
This means that the different classifications of gamma matrices
may yield different forms for the tensor structures even they are
from the same identity. Namely, although the different forms of
the trace should lead to the same result for the given Lorentz
indices, while they do not respect to the same symmetry properties
of the Lorentz indices for a general case. We will explicitly show
their influences on the forms of anomaly in the following
sections. This may be referred as the ambiguities caused by the
trace of gamma matrices with $\gamma_5$, which is independent of
any regularization.

 As the first step, we shall eliminate such kind of ambiguities
before applying for any regularization schemes. For that, we find
a unique solution by treating all the three currents symmetrically
with adopting the definition of $\gamma_5$
\begin{eqnarray}
\gamma_5={i\over
4!}\epsilon_{\mu\nu\alpha\beta}\gamma_\mu\gamma_\nu\gamma_\alpha\gamma_\beta
\end{eqnarray}

By repeatedly using the relation $\gamma_{\rho}\gamma_{\sigma} =
2g_{\rho\sigma} - \gamma_{\sigma}\gamma_{\rho}$ to reduce the
gamma matrices, eventually we have
\begin{eqnarray}
Tr\{\gamma_5\gamma_\lambda\gamma_\alpha\gamma_\nu\gamma_\beta\gamma_\mu\gamma_\xi\}
&=&{i\over 4!}\epsilon_{\mu\nu\alpha\beta}Tr\{\gamma_\mu\gamma_\nu\gamma_\alpha
\gamma_\beta\gamma_\lambda\gamma_\alpha\gamma_\nu\gamma_\beta\gamma_\mu\gamma_\xi\}\nonumber\\
&=&4i\{\epsilon_{\lambda\alpha\beta\xi}g_{\mu\nu}-\epsilon_{\lambda\alpha\nu\beta}
g_{\mu\xi}+\epsilon_{\lambda\alpha\nu\mu}g_{\beta\xi}-\epsilon_{\lambda\alpha\nu\xi}
g_{\mu\beta}-\epsilon_{\lambda\alpha\beta\mu}g_{\nu\xi}\nonumber\\
&&-\epsilon_{\lambda\alpha\mu\xi}g_{\nu\beta}+\epsilon_{\lambda\nu\beta\mu}
g_{\alpha\xi}-\epsilon_{\lambda\nu\beta\xi}g_{\alpha\mu}+\epsilon_{\lambda\nu\mu\xi}
g_{\alpha\beta}-\epsilon_{\lambda\beta\mu\xi}g_{\alpha\nu}\nonumber\\
&&-\epsilon_{\alpha\nu\beta\mu}g_{\lambda\xi}+\epsilon_{\alpha\nu\beta\xi}
g_{\lambda\mu}-\epsilon_{\alpha\nu\mu\xi}g_{\lambda\beta}+\epsilon_{\alpha\beta\mu\xi}
g_{\lambda\nu}+\epsilon_{\nu\mu\beta\xi}g_{\lambda\alpha}\}\label{trace10gamma}
\end{eqnarray}
which is the most general form respecting all the symmetries of
the Lorentz indices and eliminates the ambiguities caused by the
trace of gamma matrices with $\gamma_5$.

In this section, we shall concentrate on the most general case by
treating all the Lorentz indices symmetrically. With the above
relation (\ref{trace10gamma}), after performing the Dirac algebra,
the amplitude $T^{(1),AVV}_{\lambda\mu\nu}$ can be written as
follows
\begin{eqnarray}
T^{(1),\{AVV\}}_{\lambda\mu\nu}&=&T^{(1),\{AVV\}}_{L,\lambda\mu\nu}
+T^{(1),\{AVV\}}_{C,\lambda\mu\nu}\nonumber\\
T^{(1),\{AVV\}}_{L,\lambda\mu\nu}&=&4\int\frac{d^4k}{(2\pi)^4}
\bigg\{\nonumber\\
&&\times\{-\epsilon_{\lambda\alpha\nu\beta}(k+k_2)_\alpha(k+k_1)_\beta(k+k_3)_\mu
-\epsilon_{\lambda\alpha\nu\rho}(k+k_2)_\alpha(k+k_1)_\mu(k+k_3)_\rho\nonumber\\
&&-\epsilon_{\lambda\alpha\beta\mu}(k+k_2)_\alpha(k+k_1)_\beta(k+k_3)_\nu
+\epsilon_{\lambda\alpha\beta\rho}g_{\mu\nu}(k+k_2)_\alpha(k+k_1)_\beta(k+k_3)_\rho\nonumber\\
&&-\epsilon_{\lambda\alpha\mu\rho}(k+k_2)_\alpha(k+k_1)_\nu(k+k_3)_\rho
-\epsilon_{\lambda\nu\beta\rho}(k+k_2)_{\mu}(k+k_1)_\beta(k+k_3)_\rho\nonumber\\
&&-\epsilon_{\lambda\beta\mu\rho}(k+k_2)_\nu(k+k_1)_\beta(k+k_3)_\rho
-\epsilon_{\alpha\nu\beta\mu}(k+k_2)_\alpha(k+k_1)_\beta(k+k_3)_\lambda\nonumber\\
&&+\epsilon_{\alpha\nu\beta\rho}g_{\lambda\mu}(k+k_2)_\alpha(k+k_1)_\beta(k+k_3)_\rho
-\epsilon_{\alpha\nu\mu\rho}(k+k_2)_\alpha(k+k_1)_\lambda(k+k_3)_\rho\nonumber\\
&&+\epsilon_{\alpha\beta\mu\rho}g_{\lambda\nu}(k+k_2)_\alpha(k+k_1)_\beta(k+k_3)_\rho
+\epsilon_{\nu\mu\beta\rho}(k+k_2)_\lambda(k+k_1)_\beta(k+k_3)_\rho\}\nonumber\\
&&\times\bigg[\frac{1}{(k+k_1)^2(k+k_2)^2(k+k_3)^2}\bigg]\nonumber\\
&&+\frac{\epsilon_{\lambda\alpha\nu\mu}}{2}\bigg[\frac{(k+k_2)_\alpha}{(k+k_2)^2(k+k_3)^2}
+\frac{(k+k_2)_\alpha}{(k+k_2)^2(k+k_1)^2}\bigg]\nonumber\\
&&+\frac{\epsilon_{\lambda\nu\beta\mu}}{2}\bigg[\frac{(k+k_1)_\beta}{(k+k_1)^2(k+k_3)^2}
+\frac{(k+k_1)_\beta}{(k+k_2)^2(k+k_1)^2}\bigg]\nonumber\\
&&+\frac{\epsilon_{\lambda\nu\mu\rho}}{2}\bigg[\frac{(k+k_3)_\rho}{(k+k_1)^2(k+k_3)^2}
+\frac{(k+k_3)_\rho}{(k+k_2)^2(k+k_3)^2}\bigg]\bigg\}\\
T^{(1),\{AVV\}}_{C,\lambda\mu\nu}&=&-2\epsilon_{\lambda\alpha\nu\mu}\int\frac{d^4k}{(2\pi)^4}
\frac{(k_3-k_1)^2(k+k_2)_\alpha}{(k+k_1)^2(k+k_2)^2(k+k_3)^2}\nonumber\\
&&-2\epsilon_{\lambda\nu\beta\mu}\int\frac{d^4k}{(2\pi)^4}
\frac{(k_3-k_2)^2(k+k_1)_\beta}{(k+k_1)^2(k+k_2)^2(k+k_3)^2}\nonumber\\
&&-2\epsilon_{\lambda\nu\mu\rho}\int\frac{d^4k}{(2\pi)^4}
\frac{(k_1-k_2)^2(k+k_3)_\rho}{(k+k_1)^2(k+k_2)^2(k+k_3)^2}
\end{eqnarray}
where we have used the identity
\begin{eqnarray}
(k+k_i)\cdot
(k+k_j)={1\over2}(k+k_i)^2+{1\over2}(k+k_j)^2-{1\over2}(k_i-k_j)^2
\end{eqnarray}

Applying the loop regularization to the amplitude, as it maintains
the translational invariance and respects symmetric integration
rules, by shifting the integration variable and making some
algebra, the regularized amplitude is found to have the form
\begin{eqnarray}
T^{R,(1),\{AVV\}}_{\lambda\mu\nu}&=&T^{R,(1),\{AVV\}}_{0,\lambda\mu\nu}
+T^{R,(1),\{AVV\}}_{-2,\lambda\mu\nu}\nonumber\\
T^{R,(1),\{AVV\}}_{0,\lambda\mu\nu}&=&2\int_0^1dx[\epsilon_{\lambda\alpha\nu\mu}(k_2-k_1)_\alpha
I_0^R(x,\mu_1)+\epsilon_{\lambda\alpha\nu\mu}(2x-1)(k_3-k_2)_\alpha I_0^R(x,\mu_2)\nonumber\\
&&+\epsilon_{\lambda\mu\alpha\nu}(k_3-k_1)_\alpha I_0^R(x,\mu_3)]\nonumber\\
&&-2\epsilon_{\lambda\mu\nu\alpha}\int_0^1dx_1\int_0^{x_1}dx_2(-2k_3-2k_2+4k_1)_\alpha
I_0^R(x_i,\mu)\label{divermasslessTAVV}\\
T^{R,(1),\{AVV\}}_{-2,\lambda\mu\nu}&=&-8\int_0^1dx_1\int_0^{x_1}dx_2\nonumber\\
&&\times\bigg\{\epsilon_{\lambda\alpha\nu\beta}(-\Delta+k_2)_\alpha(-\Delta+k_1)_\beta(-\Delta+k_3)_\mu\nonumber\\
&&+\epsilon_{\lambda\alpha\nu\rho}(-\Delta+k_2)_\alpha(-\Delta+k_1)_\mu(-\Delta+k_3)_\rho\nonumber\\
&&+\epsilon_{\lambda\alpha\beta\mu}(-\Delta+k_2)_\alpha(-\Delta+k_1)_\beta(-\Delta+k_3)_\nu\nonumber\\
&&+\epsilon_{\lambda\alpha\mu\rho}(-\Delta+k_2)_\alpha(-\Delta+k_1)_\nu(-\Delta+k_3)_\rho\nonumber\\
&&+\epsilon_{\lambda\nu\beta\rho}(-\Delta+k_2)_{\mu}(-\Delta+k_1)_\beta(-\Delta+k_3)_\rho\nonumber\\
&&+\epsilon_{\lambda\beta\mu\rho}(-\Delta+k_2)_\nu(-\Delta+k_1)_\beta(-\Delta+k_3)_\rho\nonumber\\
&&+\epsilon_{\alpha\nu\beta\mu}(-\Delta+k_2)_\alpha(-\Delta+k_1)_\beta(-\Delta+k_3)_\lambda\nonumber\\
&&+\epsilon_{\alpha\nu\mu\rho}(-\Delta+k_2)_\alpha(-\Delta+k_1)_\lambda(-\Delta+k_3)_\rho\nonumber\\
&&-\epsilon_{\nu\mu\beta\rho}(-\Delta+k_2)_\lambda(-\Delta+k_1)_\beta(-\Delta+k_3)_\rho\nonumber\\
&&+\frac{\epsilon_{\lambda\alpha\nu\mu}}{2}
(k_3-k_1)^2(-\Delta+k_2)_\alpha+\frac{\epsilon_{\lambda\nu\beta\mu}}{2}(k_3-k_2)^2(-\Delta+k_1)_\beta\nonumber\\
&&+\frac{\epsilon_{\lambda\nu\mu\rho}}{2}(k_2-k_1)^2(-\Delta+k_3)_\rho\bigg\}I_{-2}^R(x_i,\mu)\label{convermasslessTAVV}
\end{eqnarray}
where $x$ and $x_i$ are the Feynman parameters and
$\mu^2=\mu_s^2+M^2$ and $\mu_i^2=\mu_s^2+M_i^2$. We have
introduced the following definitions which will also be used in
the calculations below
\begin{eqnarray}
\Delta&=&(1-x)k_1+(x_1-x_2)k_2+x_2k_3\\
\Delta_1&=&(1-x)k_1+xk_2\\
\Delta_2&=&(1-x)k_2+xk_3\\
\Delta_3&=&(1-x)k_1+xk_3\\
-M^2&=&(x_1-x_2)(1-x_1)(k_1-k_2)^2\nonumber \\
 & + & x_2(1-x_1)(k_3-k_1)^2+x_2(x_1-x_2)(k_3-k_2)^2 \\
-M_1^2&=&x(1-x)(k_1-k_2)^2\\
-M_2^2&=&x(1-x)(k_3-k_2)^2\\
-M_3^2&=&x(1-x)(k_3-k_1)^2
\end{eqnarray}
To this step, all the momentum integrals, both the divergent and
convergent ones, have been carried out explicitly.

Next, we will check all three ward identities. By using the
definitions of $\Delta$ and $\Delta_i$ and introducing the
integrals
\begin{eqnarray}
& & I_{-2,(ij)}(\mu_s^2)= {i\over32\pi^2}\int_0^1dx_1
\int_0^{x_1}dx_2{x_2^i(x_1-x_2)^j\over-M^2-\mu_s^2}[1-y_{-2}({\mu_s^2+M^2\over M_c^2})] \label{defineI(-2ij)}\\
& & I_{0,(00)}(\mu_s^2)\equiv
{i\over16\pi^2}\{\int^1_0dx_1\int_0^{x_1}dx_2
 [\ln(\frac{M_c^2}{M^2+\mu_s^2})-\gamma_\omega+y_{0}({\mu_s^2+M^2\over M_c^2})]\nonumber\\
&&\;\;\;\;\;\;\;\;\;\;\;-\int_0^1dxx[\ln(\frac{M_c^2}{M_3^2+\mu_s^2})-\gamma_\omega+y_{0}
({\mu_s^2+M_3^2\over M_c^2})]\}\label{defineI(00)}\\
& & I_{0,(00)}^\prime(\mu_s^2)\equiv
{i\over16\pi^2}\{\int^1_0dx_1\int_0^{x_1}dx_2
 [\ln(\frac{M_c^2}{M^2+\mu_s^2})-\gamma_\omega+y_{0}({\mu_s^2+M^2\over M_c^2})]\nonumber\\
&&\;\;\;\;\;\;\;\;\;\;\;-\int_0^1dxx[\ln(\frac{M_c^2}{M_1^2+\mu_s^2})-\gamma_\omega+y_{0}
({\mu_s^2+M_1^2\over M_c^2})]\}\label{defineI('00)}
\end{eqnarray}
 we obtain
\begin{eqnarray}
(k_1-k_2)_\nu
T^{R,(1),\{AVV\}}_{\lambda\mu\nu}&=&4\epsilon_{\lambda\mu\nu\alpha}(k_1-k_2)_\nu(k_3-k_1)_\alpha
I_{0,(00)}\nonumber\\
&&-8\epsilon_{\lambda\mu\nu\alpha}(k_1-k_2)_\nu(k_3-k_1)_\alpha\nonumber\\
&&\times\{(k_1-k_2)^2[2I_{-2,(01)}-2I_{-2,(02)}-{1\over2}I_{-2,(00)}]\nonumber\\
&&+(k_1-k_2)\cdot(k_3-k_1)[2I_{-2,(11)}-I_{-2,(10)}]\}\label{A-V-V-VWard1}\\
(k_3-k_1)_\mu
T^{R,(1),\{AVV\}}_{\lambda\mu\nu}&=&-4\epsilon_{\lambda\nu\alpha\mu}(k_1-k_2)_\alpha(k_3-k_1)_\mu
I_{0,(00)}^\prime\nonumber\\
&&+8\epsilon_{\lambda\nu\alpha\mu}(k_1-k_2)_\alpha(k_3-k_1)_\mu\nonumber\\
&&\times\{(k_1-k_2)\cdot(k_3-k_1)[2I_{-2,(11)}-I_{-2,(01)}]\nonumber\\
&&+(k_3-k_1)^2[2I_{-2,(10)}-2I_{-2,(20)}-{1\over2}I_{-2,(00)}]\}\label{A-V-V-VWard2}\\
(k_3-k_2)_\lambda
T^{R,(1),\{AVV\}}_{\lambda\mu\nu}&=&-4\epsilon_{\mu\nu\lambda\alpha}(k_3-k_1)_\lambda(k_1-k_2)_\alpha
I_{0,(00)}^\prime\nonumber\\
&&-4\epsilon_{\mu\nu\lambda\alpha}(k_3-k_1)_\lambda(k_1-k_2)_\alpha
I_{0,(00)}\nonumber\\
&&-4\epsilon_{\mu\nu\lambda\alpha}(k_3-k_1)_\lambda(k_1-k_2)_\alpha\nonumber\\
&&\times\{(k_3-k_1)^2I_{-2,(00)}+(k_1-k_2)^2I_{-2,(00)}\nonumber\\
&&+2(k_1-k_2)\cdot(k_3-k_1)[I_{-2,(10)}+I_{-2,(01)}]\}\label{A-V-V-AWard}
\end{eqnarray}
Note that $I_{0,(00)}$ is the difference between two
logarithemically divergent integrals, consequently the above
result becomes finite. The explicit expressions and relations for
$I_{-2,(ij)}$ and $I_{0,(00)}$ can be read off from App.~A. by
taking $m^2=0$.

In the following subsections, we consider two interesting
treatments corresponding to anomaly-free and anomaly in the loop
regularization.

\subsection{Anomaly-free treatment in loop regularization }

\label{sec ano-fr-LR}

By adopting the explicit relations given in App.~A, we obtain:
\begin{eqnarray}
(k_1-k_2)_\nu
T^{R,(1),\{AVV\}}_{\lambda\mu\nu}&=&4\epsilon_{\lambda\mu\nu\alpha}(k_1-k_2)_\nu(k_3-k_1)_\alpha\nonumber\\
&&\times\bigg\{{i\over16\pi^2}[e^{-\mu_s^2/M_c^2}\int_0^1dx_1\int_0^{x_1}dx_2e^{-M^2/M_c^2}
-{1\over2}Y((p+q)^2,q^2)]\nonumber\\
&&-q^2I_{-2,(10)}-p^2I_{-2,(01)}+2\mu_s^2I_{-2,(00)}\nonumber\\
&&-2\{p^2I_{-2,(01)}-{1\over2}p^2I_{-2,(00)}\nonumber\\
&&-2[{-i\over64\pi^2}e^{-\mu_s^2/M_c^2}\int_0^1dx_1\int_0^{x_1}dx_2e^{-M^2/M_c^2}-{\mu_s^2\over2}I_{-2,(00)}+{q^2\over4}I_{-2,(10)}\nonumber\\
&&+{3p^2\over4}I_{-2,(01)}]+[{-i\over64\pi^2}Y((p+q)^2,q^2)+{p^2\over2}I_{-2,(00)}]\}\bigg\}
\end{eqnarray}
where the first part in the bracket comes from the difference of
two logarithemically divergent integrals and the others come from
the convergent integrals. It is easy to check that all terms
cancel each other, namely
\begin{eqnarray}
(k_1-k_2)_\nu
T^{R,(1),\{AVV\}}_{\lambda\mu\nu}&=&0\label{A-V-Vvector1}
\end{eqnarray}
Similarly, we have
\begin{eqnarray}
(k_3-k_1)_\mu
T^{R,(1),\{AVV\}}_{\lambda\mu\nu}&=&0\label{A-V-Vvector2}
\end{eqnarray}
Both (\ref{A-V-Vvector1}) and (\ref{A-V-Vvector2}) show that, in
loop regularization, with the symmetric treatment of all the three
currents, the vector currents are conserved.

For the axial-vector current, by using the same method, we can
arrive at the following result
\begin{eqnarray}
(k_3-k_2)_\lambda
T^{R,(1),\{AVV\}}_{\lambda\mu\nu}&=&16\mu_s^2\epsilon_{\mu\nu\lambda\alpha}(k_1-k_2)_\lambda(k_3-k_1)_\alpha
I_{-2,(00)}(0,\mu_s^2)\label{SymmaxialWard}\\
&+&
{i\over2\pi^2}\epsilon_{\mu\nu\lambda\alpha}(k_1-k_2)_\lambda(k_3-k_1)_\alpha
\int_0^1dx_1\int_0^{x_1}dx_2e^{-(\mu_s^2 + M^2)/M_c^2}\nonumber
\end{eqnarray}

Once considering the conditions for massless QED that:
$p^2=(k_1-k_2)^2=0, q^2=(k_3-k_1)^2=0, (p+q)^2=2p\cdot q$ is soft
with $\mu_s^2 \gg (p+q)^2$, we then observe that, to leading order
electromagnetism\cite{Zuber}, in the loop regularization with
$\mu_s\neq0$, the axial-vector current is also conserved for soft
initial state, such as pion, with $\mu_s^2\gg(p+q)^2$. More
generally, as long as taking the conditions that $p^2$, $q^2$,
 $(p+q)^2$ $\ll \mu_s^2 \ll M_c^2\to \infty$, the axial-vector current also becomes
 conserved
 \begin{eqnarray}
(k_3-k_2)_\lambda T^{R,(1),\{AVV\}}_{\lambda\mu\nu}&=&0
\end{eqnarray}
This is because, under this condition, by considering the behavior
of $y_{-2}$ in (\ref{y-Function}), and the definition
(\ref{defineI(-2ij)}), the Ward identity (\ref{SymmaxialWard})
becomes
\begin{eqnarray}
(k_3-k_2)_\lambda T^{R,(1),\{AVV\}}_{\lambda\mu\nu}&
=&16\mu_s^2\epsilon_{\mu\nu\lambda\alpha}(k_1-k_2)_\lambda(k_3-k_1)_\alpha
[\  \int_0^1dx_1\int_0^{x_1}dx_2 \frac{-i}{32\pi^2(\mu_s^2+M^2)} \nonumber\\
& & + O\left((\mu_s^2 + M^2) /M_c^2 \right) \  ] \nonumber \\
&&+{i\over2\pi^2}\epsilon_{\mu\nu\lambda\alpha}(k_1-k_2)_\lambda(k_3-k_1)_\alpha
\int_0^1dx_1\int_0^{x_1}dx_2e^{-(\mu_s^2 + M^2)/M_c^2}\nonumber\\
&=&-{i\over4\pi^2}\epsilon_{\mu\nu\lambda\alpha}(k_1-k_2)_\lambda(k_3-k_1)_\alpha
[\ 1 + O\left( \frac{p\cdot q}{\mu_s^2}\right) + O\left(\frac{\mu_s^2}{M_c^2}\right) \  ]\nonumber\\
&&+{i\over4\pi^2}\epsilon_{\mu\nu\lambda\alpha}(k_1-k_2)_\lambda(k_3-k_1)_\alpha
[\ 1 +  O\left(\frac{\mu_s^2}{M_c^2} \right)
+ O\left( \frac{p\cdot q}{M_c^2} \right) \  ] \nonumber\\
&=&
{i\over4\pi^2}\epsilon_{\mu\nu\lambda\alpha}(k_1-k_2)_\lambda(k_3-k_1)_\alpha
O\left(\frac{p\cdot q}{\mu_s^2},\  \frac{\mu_s^2}{M_c^2},\  \frac{p\cdot q}{M_c^2}\right)  \nonumber \\
& = &  0  \qquad \qquad  \qquad \mbox{for} \quad  p\cdot q \ll
\mu_s^2 \ , \quad M_c^2 \to \infty
\end{eqnarray}
Here we have only considered, similar to the discussions in the
literature\cite{Zuber}, the leading order contribution and ignored
the higher order terms arising from $M^2/\mu_s^2$ in the soft
limit that $p^2=0$, $q^2=0$ and $(p+q)^2 = 2p\cdot q \ll \mu_s^2$.

With the above explicit calculations, we can now arrive at the
conclusion that in loop regularization with the introduction of
nonzero IR cut-off scale $\mu_s\neq 0$ for massless QED, the Ward
identities of both vector and axial-vector currents become
conserved in the conditions: $p^2$,$q^2$, $(p+q)^2 \ll \mu_s^2 \ll
M_c^2\to \infty$.

\subsection{Anomaly of massless QED in loop regularization}

\label{sec ano-massless-LR}

We now consider the case that $\mu_s =0$. Making a similar
evaluation, we obtain the following results
\begin{eqnarray}
(k_3-k_1)_\mu T^{R,(1),\{AVV\}}_{\lambda\mu\nu}&=&0\label{symmevector1}\\
(k_1-k_2)_\nu T^{R,(1),\{AVV\}}_{\lambda\mu\nu}&=&0\\
(k_3-k_2)_\lambda
T^{R,(1),\{AVV\}}_{\lambda\mu\nu}&=&{i\over2\pi^2}\epsilon_{\mu\nu\lambda\alpha}(k_1-k_2)_\lambda(k_3-k_1)_\alpha
\int_0^1dx_1\int_0^{x_1}dx_2e^{-M^2/M_c^2}\label{symmeaxialvector}
\end{eqnarray}
Taking $M_c\to \infty$, we yield
\begin{eqnarray}
(k_3-k_1)_\mu T^{R,(1),\{AVV\}}_{\lambda\mu\nu}&=&0\\
(k_1-k_2)_\nu T^{R,(1),\{AVV\}}_{\lambda\mu\nu}&=&0\\
(k_3-k_2)_\lambda
T^{R,(1),\{AVV\}}_{\lambda\mu\nu}&=&{i\over4\pi^2}\epsilon_{\mu\nu\lambda\alpha}(k_1-k_2)_\lambda(k_3-k_1)_\alpha
\end{eqnarray}

By including the cross diagrams, we finally obtain the Ward
identities with anomaly of axial-vector current
\begin{eqnarray}
(k_3-k_1)_\mu
T^{R,\{AVV\}}_{\lambda\mu\nu}&=&0\\
(k_1-k_2)_\nu
T^{R,\{AVV\}}_{\lambda\mu\nu}&=&0\\
(k_3-k_2)_\lambda T^{R,\{AVV\}}_{\lambda\mu\nu}&=&{i\over2\pi^2}
\epsilon_{\mu\nu\alpha\beta}(k_1-k_2)_\alpha(k_3-k_1)_\beta
\end{eqnarray}
Their operator forms are given by
\begin{eqnarray}
\partial_\mu V_\mu(x)&=&0,\\
\partial_\mu
A_\mu(x)&=&{e^2\over4\pi^2}\epsilon_{\mu\nu\alpha\beta}\partial^\alpha
\mathcal{A}^\mu(x)\partial^\beta \mathcal{A}^\nu(x)\nonumber\\
&=&{e^2\over8\pi^2}F^{\mu\nu}(x)\widetilde{F}_{\mu\nu}(x)\label{operatormasslessanomal}
\end{eqnarray}
where $\widetilde{F}_{\mu\nu}(x)={1\over2}
\epsilon_{\mu\nu\alpha\beta}F^{\alpha\beta}(x)$ and the coupling
constant $e$ was restored. It should be noticed that the
coefficient of $\partial^\alpha \mathcal{A}^\mu(x)\partial^\beta
\mathcal{A}^\nu(x)$ is $e^2/(4\pi^2)$ rather than $e^2/(2\pi^2)$
as the derivative operator can act on both the vector fields
$\mathcal{A}_\mu(x)$ and $\mathcal{A}_\nu(x)$.

In conclusion, it is seen from the above analysis that the IR
regulator $\mu_s$ plays an important role in understanding the
anomaly. Once introducing the IR cut-off scale $\mu_s$ in loop
regularization so that it satisfies the condition $p^2$, $q^2$,
$(p+q)^2 \ll \mu_s^2 \ll M_c^2 \to \infty$, then both the vector
current and axial-vector current are conserved and no anomaly
appears. Namely, loop regularization with sufficient large IR
cut-off scale $\mu_s$ becomes a completely symmetry-preserving
regularization. In the absence of the IR cut-off scale ($\mu_s =
0$) for massless QED, the loop regularization leads to the
well-known triangle anomaly for the axial-vector current. It is
also seen that, in loop regularization, when the trace of gamma
matrices are manipulated with the definition of $\gamma_5$
directly, the vector currents are automatically preserved, only
the axial-vector Ward identity is violated by quantum corrections.
That comes to the correct standard form of anomaly.

\section{ Anomaly-free condition in massive QED in loop regularization}

\label{sec massiveLR}

Considering the massive spinor electrodynamics with the lagrangian
\begin{eqnarray}
{\cal L}=\bar{\psi}\gamma^\mu(i\partial_\mu-e{\cal
A}_\mu)\psi-m\bar{\psi}\psi
\end{eqnarray}
 Classically, the ward identities of the vector and axial-vector currents are
\begin{eqnarray}
\partial_\mu V_\mu(x)=0,~~~~~~~\partial_\mu
A_\mu(x)=2imP(x)
\end{eqnarray}
where
\begin{eqnarray}
P(x)=\bar{\psi}\gamma_5\psi
\end{eqnarray}

To calculate the quantum corrections in perturbation theory,
besides the amplitude $(\ref{TAVV})$, one should consider the
amplitude
\begin{eqnarray}
T^{PVV}_{\mu\nu}(p,q;(p+q))&=&\int d^4x_1d^4x_2e^{ipx_1+iq
x_2}\langle0|T[V_\mu(x_1) V_\nu(x_2) P(0)]|0\rangle
\end{eqnarray}
The classical Ward identities $(\ref{divcurrents})$ for massive
case become
\begin{eqnarray}
&&p^\mu T^{AVV}_{\mu\nu\lambda}(p,q;(p+q))=0\\
&&q_\nu T^{AVV}_{\mu\nu\lambda}(p,q;(p+q))=0\\
&&(p+q)_\lambda
T^{AVV}_{\mu\nu\lambda}(p,q;(p+q))=2mT^{PVV}_{\mu\nu}
\end{eqnarray}
The corresponding diagrams and their cross ones are shown as
follows
\\
\begin{picture}(400,70)(0,0)
\ArrowLine(100,30)(150,50)\ArrowLine(150,10)(100,30)\ArrowLine(150,50)(150,10)
\Text(100,30)[]{$\bullet$}\Text(150,50)[]{$\bullet$}\Text(150,10)[]{$\bullet$}
\Text(80,35)[]{$\gamma_\lambda\gamma_5$}\Text(160,45)[]{$\gamma_\mu$}\Text(160,15)[]{$\gamma_\nu$}
\Text(120,50)[]{$k+k_3$}\Text(170,30)[]{$k+k_1$}\Text(120,10)[]{$k+k_2$}
\DashArrowLine(50,30)(100,30){4}\Photon(150,50)(190,50){2}{5}\Photon(150,10)(190,10){2}{5}\Text(175,60)[]
{$q_\mu=(k_3-k_1)_\mu$}
\Text(175,0)[]{$p_\nu=(k_1-k_2)_\nu$}\Text(175,47)[]{$\rightarrow$}\Text(175,13)[]{$\rightarrow$}
\ArrowLine(290,30)(340,50)\ArrowLine(340,10)(290,30)\ArrowLine(340,50)(340,10)
\Text(290,30)[]{$\bullet$}\Text(340,50)[]{$\bullet$}\Text(340,10)[]{$\bullet$}
\Text(280,35)[]{$\gamma_5$}\Text(350,45)[]{$\gamma_\mu$}\Text(350,15)[]{$\gamma_\nu$}
\Text(310,50)[]{$k+k_3$}\Text(360,30)[]{$k+k_1$}\Text(310,10)[]{$k+k_2$}
\DashArrowLine(240,30)(290,30){4}
\Photon(340,50)(380,50){2}{5}\Photon(340,10)(380,10){2}{5}\Text(365,60)[]{$q_\mu=(k_3-k_1)_\mu$}\Text(365,0)[]{$p_\nu=(k_1-k_2)_\nu$}
\Text(365,47)[]{$\rightarrow$}\Text(365,13)[]{$\rightarrow$}
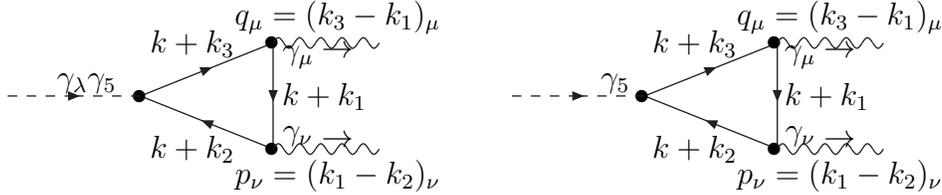
\end{picture}\begin{center}{{\sl Fig.3. One loop diagrammatical representation of correction to $T^{AVV}$ and
$T^{PVV}$.}}\end{center}

One can easily write down the  corresponding
$T^{AVV}_{\lambda\mu\nu}$ and $T^{PVV}_{\lambda\mu\nu}$ from the
above diagrams
\begin{eqnarray}
&&T^{(1),AVV}_{\lambda\mu\nu}=(-1)\int\frac{d^4k}{(2\pi)^4}{\rm
tr}
\{\gamma_\lambda\gamma_5\frac{i}{(k\hspace{-0.2cm}\slash+k\hspace{-0.2cm}\slash_2)-m}
\gamma_\nu\frac{i}{(k\hspace{-0.2cm}\slash+k\hspace{-0.2cm}\slash_1)-m}\gamma_\mu
\frac{i}{(k\hspace{-0.2cm}\slash+k\hspace{-0.2cm}\slash_3)-m}\}\\
&&T^{(1),PVV}_{\mu\nu}=(-1)\int\frac{d^4k}{(2\pi)^4}{\rm
tr}\{\gamma_5\frac{i}{(k\hspace{-0.2cm}\slash+k\hspace{-0.2cm}\slash_2)-m}
\gamma_\nu\frac{i}{(k\hspace{-0.2cm}\slash+k\hspace{-0.2cm}\slash_1)-m}
\gamma_\mu\frac{i}{(k\hspace{-0.2cm}\slash+k\hspace{-0.2cm}\slash_3)-m}\}
\end{eqnarray}

Repeating the same calculations done for massless QED, we yield
\begin{eqnarray}
T^{R,(1),\{AVV\}}_{\lambda\mu\nu}&=&T^{R,(1),\{AVV\}}_{0,\lambda\mu\nu}+T^{R,(1),\{AVV\}}_{-2,\lambda\mu\nu}\nonumber\\
T^{R,(1),\{AVV\}}_{0,\lambda\mu\nu}
&=&2\int_0^1dx[\epsilon_{\lambda\alpha\nu\mu}(k_2-k_1)_\alpha I_0^R(x,\mu_1)
+\epsilon_{\lambda\alpha\nu\mu}(2x-1)(k_3-k_2)_\alpha I_0^R(x,\mu_2)\nonumber\\
&&-\epsilon_{\lambda\mu\alpha\nu}(k_3-k_1)_\alpha I_0^R(x,\mu_3)]\nonumber\\
&&-2\epsilon_{\lambda\mu\nu\alpha}\int_0^1dx_1\int_0^{x_1}dx_2(-2k_3-2k_2+4k_1)_\alpha
I_0^R(x_i,\mu)\\
T^{R,(1),\{AVV\}}_{-2,\lambda\mu\nu}&=&-8\int_0^1dx_1\int_0^{x_1}dx_2\nonumber\\
&&\times\bigg\{\epsilon_{\lambda\alpha\nu\beta}(-\Delta+k_2)_\alpha(-\Delta+k_1)_\beta(-\Delta+k_3)_\mu\nonumber\\
&&+\epsilon_{\lambda\alpha\nu\rho}(-\Delta+k_2)_\alpha(-\Delta+k_1)_\mu(-\Delta+k_3)_\rho\nonumber\\
&&+\epsilon_{\lambda\alpha\beta\mu}(-\Delta+k_2)_\alpha(-\Delta+k_1)_\beta(-\Delta+k_3)_\nu\nonumber\\
&&+\epsilon_{\lambda\alpha\mu\rho}(-\Delta+k_2)_\alpha(-\Delta+k_1)_\nu(-\Delta+k_3)_\rho\nonumber\\
&&+\epsilon_{\lambda\nu\beta\rho}(-\Delta+k_2)_{\mu}(-\Delta+k_1)_\beta(-\Delta+k_3)_\rho\nonumber\\
&&+\epsilon_{\lambda\beta\mu\rho}(-\Delta+k_2)_\nu(-\Delta+k_1)_\beta(-\Delta+k_3)_\rho\nonumber\\
&&+\epsilon_{\alpha\nu\beta\mu}(-\Delta+k_2)_\alpha(-\Delta+k_1)_\beta(-\Delta+k_3)_\lambda\nonumber\\
&&+\epsilon_{\alpha\nu\mu\rho}(-\Delta+k_2)_\alpha(-\Delta+k_1)_\lambda(-\Delta+k_3)_\rho\nonumber\\
&&-\epsilon_{\nu\mu\beta\rho}(-\Delta+k_2)_\lambda(-\Delta+k_1)_\beta(-\Delta+k_3)_\rho\nonumber\\
&&+\frac{\epsilon_{\lambda\alpha\nu\mu}}{2}
(k_3-k_1)^2(-\Delta+k_2)_\alpha+\frac{\epsilon_{\lambda\nu\beta\mu}}{2}(k_3-k_2)^2(-\Delta+k_1)_\beta\nonumber\\
&&+\frac{\epsilon_{\lambda\nu\mu\rho}}{2}(k_2-k_1)^2(-\Delta+k_3)_\rho\bigg\}I_{-2}^R(x_i,\mu)
\end{eqnarray}
which has the same form as the massless QED case except the mass
of internal fermion is introduced. In above, $x$ and $x_i$ are
Feynman parameters and
\begin{eqnarray}
\mu_i^2 = \mu_s^2 + m^2 + M_i^{2}
\end{eqnarray}

By using the definitions
(\ref{defineI(-2ij)})-(\ref{defineI('00)}), we have
\begin{eqnarray}
(k_1-k_2)_\nu
T^{R,(1),\{AVV\}}_{\lambda\mu\nu}&=&4\epsilon_{\lambda\mu\nu\alpha}(k_1-k_2)_\nu(k_3-k_1)_\alpha
I_{0,(00)}\nonumber\\
&&-8\epsilon_{\lambda\mu\nu\alpha}(k_1-k_2)_\nu(k_3-k_1)_\alpha\nonumber\\
&&\times\{(k_1-k_2)^2[2I_{-2,(01)}-2I_{-2,(02)}-{1\over2}I_{-2,(00)}]\nonumber\\
&&+(k_1-k_2)\cdot(k_3-k_1)[2I_{-2,(11)}-I_{-2,(10)}]\}\label{massiveA-V-V-VWard1}\\
(k_3-k_1)_\mu
T^{R,(1),\{AVV\}}_{\lambda\mu\nu}&=&-4\epsilon_{\lambda\nu\alpha\mu}(k_1-k_2)_\alpha(k_3-k_1)_\mu
I_{0,(00)}^\prime\nonumber\\
&&+8\epsilon_{\lambda\nu\alpha\mu}(k_1-k_2)_\alpha(k_3-k_1)_\mu\nonumber\\
&&\times\{(k_1-k_2)\cdot(k_3-k_1)[2I_{-2,(11)}-I_{-2,(01)}]\nonumber\\
&&+(k_3-k_1)^2[2I_{-2,(10)}-2I_{-2,(20)}-{1\over2}I_{-2,(00)}]\}\label{massiveA-V-V-VWard2}\\
(k_3-k_2)_\lambda
T^{R,(1),\{AVV\}}_{\lambda\mu\nu}&=&-4\epsilon_{\mu\nu\lambda\alpha}(k_3-k_1)_\lambda(k_1-k_2)_\alpha
I_{0,(00)}^\prime\nonumber\\
&&-4\epsilon_{\mu\nu\lambda\alpha}(k_3-k_1)_\lambda(k_1-k_2)_\alpha
I_{0,(00)}\nonumber\\
&&-4\epsilon_{\mu\nu\lambda\alpha}(k_3-k_1)_\lambda(k_1-k_2)_\alpha\nonumber\\
&&\times\{(k_3-k_1)^2I_{-2,(00)}+(k_1-k_2)^2I_{-2,(00)}\nonumber\\
&&+2(k_1-k_2)\cdot(k_3-k_1)[I_{-2,(10)}+I_{-2,(01)}]\}
\end{eqnarray}
which have the same forms as
(\ref{A-V-V-VWard1})-(\ref{A-V-V-AWard}) except the mass of
internal fermion. In above the quantities $I_{-2,(ij)}$ and
$I_{0,(00)}$ are defined in App.~A. We now consider two cases with
$\mu_s^2\neq 0$ and $\mu_s^2 = 0$.

\subsection{Ward identities under the condition $\mu_s^2\neq0$
and $M_c^2\rightarrow\infty$}

\label{sec WDmuneq0}

Adopting the relations given in App.A., one can easily write the
above Ward identities as
\begin{eqnarray}
(k_1-k_2)_\nu
T^{R,(1),\{AVV\}}_{\lambda\mu\nu}&=&4\epsilon_{\lambda\mu\nu\alpha}(k_1-k_2)_\nu(k_3-k_1)_\alpha\nonumber\\
&&\times\bigg\{{i\over16\pi^2}[e^{-(m^2+\mu_s^2)/M_c^2}\int_0^1dx_1\int_0^{x_1}dx_2e^{-M^2/M_c^2}
-{1\over2}Y((p+q)^2,q^2)]\nonumber\\
&&-q^2I_{-2,(10)}-p^2I_{-2,(01)}+2(\mu_s^2+m^2)I_{-2,(00)}\nonumber\\
&&-2\{p^2I_{-2,(01)}-{1\over2}p^2I_{-2,(00)}\nonumber\\
&&-2[{-i\over64\pi^2}e^{-(m^2+\mu_s^2)/M_c^2}\int_0^1dx_1\int_0^{x_1}dx_2e^{-M^2/M_c^2}
-{\mu_s^2\over2}I_{-2,(00)}\nonumber\\
&&+{q^2\over4}I_{-2,(10)}+{3p^2\over4}I_{-2,(01)}]+[{-i\over64\pi^2}Y((p+q)^2,q^2)+{q^2\over2}
I_{-2,(00)}]\}\bigg\}\nonumber
\end{eqnarray}
Again the first part in the bracket comes from the difference of
two logarithemically divergent integrals and the others come from
the convergent integrals. It is seen that all terms cancel each
other, i.e.,
\begin{eqnarray}
(k_1-k_2)_\nu
T^{R,(1),\{AVV\}}_{\lambda\mu\nu}&=&0\label{massiveA-V-Vvector1}
\end{eqnarray}
Similarly, one has
\begin{eqnarray}
(k_3-k_1)_\mu
T^{R,(1),\{AVV\}}_{\lambda\mu\nu}&=&0\label{massiveA-V-Vvector2}
\end{eqnarray}
which shows that for massive QED the situation is same as the
massless case, the vector currents remain conserved with the
explicit use of the definition of $\gamma_5$ in the gamma trace.

For the axial-vector current, we yield
\begin{eqnarray}
(k_3-k_2)_\lambda
T^{R,(1),\{AVV\}}_{\lambda\mu\nu}&=&-16(\mu_s^2+m^2)
\epsilon_{\mu\nu\lambda\alpha}(k_3-k_1)_\lambda(k_1-k_2)_\alpha
I_{-2,(00)}(\mu_s^2)\nonumber\\
&&-{i\over2\pi^2}\epsilon_{\mu\nu\lambda\alpha}(k_3-k_1)_\lambda(k_1-k_2)_\alpha\nonumber\\
&&\times
e^{-(m^2+\mu_s^2)/M_c^2}\int_0^1dx_1\int_0^{x_1}dx_2e^{-M^2/M_c^2}\nonumber\\
&=&-16\mu_s^2\epsilon_{\mu\nu\lambda\alpha}(k_3-k_1)_\lambda(k_1-k_2)_\alpha
I_{-2,(00)}(\mu_s^2)\nonumber\\
&&-{i\over2\pi^2}\epsilon_{\mu\nu\lambda\alpha}(k_3-k_1)_\lambda(k_1-k_2)_\alpha\nonumber\\
&&\times
e^{-(m^2+\mu_s^2)/M_c^2}\int_0^1dx_1\int_0^{x_1}dx_2e^{-M^2/M_c^2}\nonumber\\
&&-16m^2\epsilon_{\mu\nu\lambda\alpha}(k_3-k_1)_\lambda(k_1-k_2)_\alpha
I_{-2,(00)}(\mu_s^2)\label{massiveA-V-Vaxialvector}
\end{eqnarray}

We now turn to evaluate the PVV diagram. Its loop amplitude is
given by
\begin{eqnarray}
T^{(1),PVV}_{\mu\nu}&=&4m\int\frac{d^4k}{(2\pi)^4}
\frac{1}{[(k+k_2)^2-m^2][(k+k_1)^2-m^2][(k+k_2)^2-m^2]}\nonumber\\
&&\times\{\epsilon_{\alpha\nu\beta\mu}(k+k_2)_\alpha(k+k_1)_\beta
+\epsilon_{\nu\alpha\mu\beta}(k+k_1)_\alpha(k+k_3)_\beta\nonumber\\
&&+\epsilon_{\alpha\nu\mu\beta}(k+k_2)_\alpha(k+k_3)_\beta\}
\end{eqnarray}
By using the Feynman parameterization method, it can be rewritten
as
\begin{eqnarray}
T^{(1),PVV}_{\mu\nu}&=&8m\int_0^1dx_1\int_0^{x_1}dx_2\int\frac{d^4k}{(2\pi)^4}
\epsilon_{\mu\nu\alpha\beta}\bigg\{\{(k+k_2)_\alpha(k+k_1)_\beta+(k+k_1)_\alpha(k+k_3)_\beta\nonumber\\
&&-(k+k_2)_\alpha(k+k_3)_\beta\}\frac{1}{[(k+\Delta)^2-M^2]^3}\bigg\}
\end{eqnarray}
Applying the loop regularization method and making a simple
algebraic calculation, we arrive at the following result
\begin{eqnarray}
T^{R,(1),PVV}_{\mu\nu}&=&8m\sum_{l=0}^Nc_l^N\int_0^1dx_1\int_0^{x_1}dx_2\int\frac{d^4k}{(2\pi)^4}
\epsilon_{\mu\nu\alpha\beta}\{(k_1-k_2)_\alpha(k_3-k_1)_\beta\}\frac{1}{[k^2-M^2_{l}]^3}\nonumber\\
&=&8m\epsilon_{\mu\nu\alpha\beta}(k_1-k_2)_\alpha(k_3-k_1)_\beta\int_0^1dx_1\int_0^{x_1}dx_2
I_{-2}^R(\mu)\nonumber\\
&=&8m\epsilon_{\mu\nu\alpha\beta}(k_1-k_2)_\alpha(k_3-k_1)_\beta
I_{-2,(00)}\label{PVVresult}
\end{eqnarray}
with the initial condition $M^2_{0}=M^2$.

In comparison with eq.$(\ref{SymmaxialWard})$, we obtain the
relation between two amplitudes
\begin{eqnarray}
(k_3-k_2)_\lambda
T^{R,(1),\{AVV\}}_{\lambda\mu\nu}&=&-16\mu_s^2\epsilon_{\mu\nu\lambda\alpha}(k_3-k_1)_\lambda(k_1-k_2)_\alpha
I_{-2,(00)}(\mu_s^2)\nonumber\\
&&-{i\over2\pi^2}\epsilon_{\mu\nu\lambda\alpha}(k_3-k_1)_\lambda(k_1-k_2)_\alpha\nonumber\\
&&\times
e^{-(m^2+\mu_s^2)/M_c^2}\int_0^1dx_1\int_0^{x_1}dx_2e^{-M^2/M_c^2}\nonumber\\
&&+2mT^{R,(1),PVV}_{\mu\nu}
\end{eqnarray}
which implies that the classical axial-vector Ward identity is in
general violated by quantum corrections.

Similar to the case in massless QED, if taking the two vector
states to be massless, i.e., $p^2=q^2=0$ and the axial-vector
state to be soft with $ m^2,p\cdot q \ll \mu_s^2 \ll M_c^2 \to
\infty$, as explained in the massless case, we have
\begin{eqnarray}
(k_3-k_1)_\mu T^{R,(1),\{AVV\}}_{\lambda\mu\nu} &=&0\\
(k_1-k_2)_\nu T^{R,(1),\{AVV\}}_{\lambda\mu\nu}&=&0\\
(k_3-k_2)_\lambda T^{R,(1),\{AVV\}}_{\lambda\mu\nu} & = &
2mT^{R,(1),PVV}_{\mu\nu}
\end{eqnarray}
which indicates that the Ward identities become preserved in this
case.

In the case that $\mu_s \gg m $, the quantum corrections of the
axial-vector Ward identity itself approaches to vanish
\begin{eqnarray}
(k_3-k_2)_\lambda T^{R,(1),\{AVV\}}_{\lambda\mu\nu} &=&
2mT^{R,(1),PVV}_{\mu\nu} \nonumber \\
& = & -{i\over4\pi^2}\frac{m^2}{\mu_s^2 + m^2}
\epsilon_{\mu\nu\alpha\beta}(k_3-k_1)_\alpha(k_1-k_2)_\beta \to 0
\end{eqnarray}

Generally speaking, in the limit $p^2$,$q^2$,$(p+q)^2$, $m^2
\ll\mu_s^2 \ll M_c^2\to \infty$, both the vector and axial-vector
currents become anomaly-free. Namely the Ward identities for both
currents are preserved in the quantum corrections. Obviously, to
arrive at this conclusion, the IR cut-off scale $\mu_s$ plays an
important role.

\subsection{Anomaly under the condition $\mu_s^2=0$ and
$M_c^2\rightarrow\infty$}

\label{subsec anomu0}

In the case that $\mu_s^2=0$, the relevant Ward identities have
the following forms
\begin{eqnarray}
(k_3-k_1)_\mu T^{R,(1),\{AVV\}}_{\lambda\mu\nu} &=&0\\
(k_1-k_2)_\nu T^{R,(1),\{AVV\}}_{\lambda\mu\nu}&=&0\\
(k_3-k_2)_\lambda
T^{R,(1),\{AVV\}}_{\lambda\mu\nu}&=&-{i\over2\pi^2}\epsilon_{\mu\nu\lambda\alpha}(k_3-k_1)_\lambda(k_1-k_2)_\alpha
\int_0^1dx_1\int_0^{x_1}dx_2e^{-(m^2+M^2)/M_c^2}\nonumber\\
&&+2mT^{R,(1),PVV}_{\mu\nu}
\end{eqnarray}
which indicates that in the loop regularization with $\mu_s = 0$
the vector Ward identities are preserved, while the axial-vector
current is no longer conserved.

Taking the limit $M_c^2\rightarrow\infty$, the above results are
simplified to be
\begin{eqnarray}
(k_3-k_1)_\mu T^{R,(1),\{AVV\}}_{\lambda\mu\nu} &=&0\\
(k_1-k_2)_\nu T^{R,(1),\{AVV\}}_{\lambda\mu\nu}&=&0\\
(k_3-k_2)_\lambda
T^{R,(1),\{AVV\}}_{\lambda\mu\nu}&=&2mT^{R,(1),PVV}_{\mu\nu}-{i\over4\pi^2}\epsilon_{\mu\nu\lambda\alpha}(k_3-k_1)_\lambda(k_1-k_2)_\alpha
\end{eqnarray}
which arrives at the well-known anomaly in the axial-vector Ward
identity.

The above Ward identities can in general be rewritten in form, by
considering eq.(\ref{PVVresult}), as
\begin{eqnarray}
(k_3-k_1)_\mu T^{R,(1),\{AVV\}}_{\lambda\mu\nu} &=&0\\
(k_1-k_2)_\nu T^{R,(1),\{AVV\}}_{\lambda\mu\nu}&=&0\\
(k_3-k_2)_\lambda
T^{R,(1),\{AVV\}}_{\lambda\mu\nu}&=&2mT^{R,(1),PVV}_{\mu\nu}-{i\over4\pi^2}
\epsilon_{\mu\nu\alpha\beta}(k_3-k_1)_\alpha(k_1-k_2)_\beta
\nonumber \\
& = & (I(m,p,q) - 1 ) {i\over4\pi^2}
\epsilon_{\mu\nu\alpha\beta}(k_3-k_1)_\alpha(k_1-k_2)_\beta
\end{eqnarray}
where the integral $I(m,p,q)$ is defined from eq.(\ref{PVVresult})
as
\begin{eqnarray}
I(m,p,q) = \int_0^1 dx_1 \int_0^{x_1} dx_2 \frac{2m^2}{m^2 +
M^2}\label{defineI(m,p,q)}
\end{eqnarray}

In the case that the external vector states are massless with
on-mass shell conditions: $p^2 = 0$ and $q^2 = 0$ and the
axial-vector state is soft with condition $(p+q)^2 = 2p\cdot q\ll
m^2$, to leading order electromagnetiam\cite{Zuber}, the above
integral can simply be performed
\begin{eqnarray}
I(m,p,q) = \int_0^1 dx_1 \int_0^{x_1} dx_2 \frac{2m^2}{m^2 + M^2}
= 2\int_0^1 dx_1 \int_0^{x_1} dx_2 = 1\label{special-I-1}
\end{eqnarray}
Thus
\begin{eqnarray}
 (k_3-k_2)_\lambda
T^{R,(1),\{AVV\}}_{\lambda\mu\nu} & = & (I(m,p,q) - 1 )
{i\over4\pi^2}
\epsilon_{\mu\nu\alpha\beta}(k_3-k_1)_\alpha(k_1-k_2)_\beta = 0
 \label{anomfree}
\end{eqnarray}
Namely the quantum corrections for both vector and axial-vector
become vanishing for massive QED when taking the general condition
$m^2\gg p^2, q^2, p\cdot q$.

Considering now the alternative case that $p^2 = 0$, $q^2 = 0$,
and $(p+q)^2 \gg m^2 $. In this case, the integral $I(m,p,q)$
approaches to zero
\begin{eqnarray} I(m,p,q) = \int_0^1 dx_1
\int_0^{x_1} dx_2 \frac{2m^2}{x_2(x_2 -x_1) (p+q)^2 + m^2} \to 0 ,
\quad  (p+q)^2 \gg m^2\label{special-I-2}
\end{eqnarray}
which coincides with massless QED at $\mu_s = 0$. In this case,
the axial-vector Ward identity gets anomaly.

 By including the cross diagrams, the Ward identities are
given by
\begin{eqnarray}
(k_3-k_1)_\mu
T^{R,\{AVV\}}_{\lambda\mu\nu}&=&0\\
(k_1-k_2)_\nu
T^{R,\{AVV\}}_{\lambda\mu\nu}&=&0\\
(k_3-k_2)_\lambda
T^{R,\{AVV\}}_{\lambda\mu\nu}&=&2mT^{R,PVV}_{\mu\nu}
-{i\over2\pi^2}\epsilon_{\mu\nu\alpha\beta}(k_3-k_1)_\alpha(k_1-k_2)_\beta
\end{eqnarray}
Their operator forms are
\begin{eqnarray}
\partial_\mu V_\mu(x)&=&0,\\
\partial_\mu
A_\mu(x)&=&2imP(x)+{e^2\over8\pi^2}F^{\mu\nu}(x)\widetilde{F}_{\mu\nu}(x)
\end{eqnarray}
which is the standard form of triangle anomaly.

It is seen that for massive QED the IR scale $\mu_s^2$ also plays
an important role in understanding the triangle anomaly.


\section{ Anomaly in Pauli-Villars regularization}

\label{sec PV}

\subsection{ Anomaly of massless QED in Pauli-Villars regularization}

\label{subsec masslessPV}

In spinor electrodynamics treated by Pauli-Villars regularization,
the fermionic part of the regularized lagrangian is\cite{PV}
\begin{eqnarray}
{\cal
L}^R_f=\bar{\psi}\gamma^\mu(i\partial_\mu-e\mathcal{A}_\mu)\psi
+\sum_{i=1}^2C_i\{\bar{\psi}_i\gamma^\mu(i\partial_\mu-e\mathcal{A}_\mu)\psi_i
-m_i\bar{\psi}_i\psi_i\}\label{regudL}
\end{eqnarray}
where $C_i$ and mass parameters $m_i$ are respectively specified
as
\begin{eqnarray}
&&C_1=1,~~~~C_2=-2\nonumber\\
&&m_1^2=2\Lambda^2,~~~~m_2^2=\Lambda^2
\end{eqnarray}

Extending the expressions in eqs.(\ref{divermasslessTAVV}) and
(\ref{convermasslessTAVV}) by adding the mass terms, one arrives
at the regularized amplitudes in the Pauli-Villars scheme
\begin{eqnarray}
T^{R,(1),\{AVV\}}_{\lambda\mu\nu}&=&T^{R,(1),\{AVV\}}_{0,\lambda\mu\nu}+T^{R,(1),\{AVV\}}_{-2,\lambda\mu\nu}\nonumber\\
T^{R,(1),\{AVV\}}_{0,\lambda\mu\nu}&=&2\sum_{i=0}^2C_i\int_0^1dx\{\epsilon_{\lambda\alpha\nu\mu}(k_2-k_1)_\alpha
 \int\frac{d^4k}{(2\pi)^4}\frac{1}{[k^2-M_{1i}^2]^2}\nonumber\\
&&+\epsilon_{\lambda\alpha\nu\mu}(2x-1)(k_3-k_2)_\alpha \int\frac{d^4k}{(2\pi)^4}\frac{1}{[k^2-M_{2i}^2]^2}\nonumber\\
&&+\epsilon_{\lambda\mu\alpha\nu}(k_3-k_1)_\alpha \int\frac{d^4k}{(2\pi)^4}\frac{1}{[k^2-M_{3i}^2]^2}\}\nonumber\\
&&-2\epsilon_{\lambda\mu\nu\alpha}\int_0^1dx_1\int_0^{x_1}dx_2(-2k_3-2k_2+4k_1)_\alpha
\int\frac{d^4k}{(2\pi)^4}\frac{k^2}{[k^2-M_i^2]^3}\\
T^{R,(1),\{AVV\}}_{-2,\lambda\mu\nu}&=&-8\sum_{i=0}^2C_i\int_0^1dx_1\int_0^{x_1}dx_2\nonumber\\
&&\times\bigg\{\epsilon_{\lambda\alpha\nu\beta}(-\Delta+k_2)_\alpha(-\Delta+k_1)_\beta(-\Delta+k_3)_\mu\nonumber\\
&&+\epsilon_{\lambda\alpha\nu\rho}(-\Delta+k_2)_\alpha(-\Delta+k_1)_\mu(-\Delta+k_3)_\rho\nonumber\\
&&+\epsilon_{\lambda\alpha\beta\mu}(-\Delta+k_2)_\alpha(-\Delta+k_1)_\beta(-\Delta+k_3)_\nu\nonumber\\
&&+\epsilon_{\lambda\alpha\mu\rho}(-\Delta+k_2)_\alpha(-\Delta+k_1)_\nu(-\Delta+k_3)_\rho\nonumber\\
&&+\epsilon_{\lambda\nu\beta\rho}(-\Delta+k_2)_{\mu}(-\Delta+k_1)_\beta(-\Delta+k_3)_\rho\nonumber\\
&&+\epsilon_{\lambda\beta\mu\rho}(-\Delta+k_2)_\nu(-\Delta+k_1)_\beta(-\Delta+k_3)_\rho\nonumber\\
&&+\epsilon_{\alpha\nu\beta\mu}(-\Delta+k_2)_\alpha(-\Delta+k_1)_\beta(-\Delta+k_3)_\lambda\nonumber\\
&&+\epsilon_{\alpha\nu\mu\rho}(-\Delta+k_2)_\alpha(-\Delta+k_1)_\lambda(-\Delta+k_3)_\rho\nonumber\\
&&-\epsilon_{\nu\mu\beta\rho}(-\Delta+k_2)_\lambda(-\Delta+k_1)_\beta(-\Delta+k_3)_\rho\nonumber\\
&&+\frac{\epsilon_{\lambda\alpha\nu\mu}}{2}
(k_3-k_1)^2(-\Delta+k_2)_\alpha+\frac{\epsilon_{\lambda\nu\beta\mu}}{2}(k_3-k_2)^2(-\Delta+k_1)_\beta\nonumber\\
&&+\frac{\epsilon_{\lambda\nu\mu\rho}}{2}(k_1-k_2)^2(-\Delta+k_3)_\rho\bigg\}
\int\frac{d^4k}{(2\pi)^4}\frac{1}{[k^2-M_i^2]^3}
\end{eqnarray}
with the initial condition $c_0=1$ and $m_0=0$ and
\begin{eqnarray}
-M_{i}^2&=&(x_1-x_2)(1-x_1)(k_2-k_1)^2+(k_1-k_3)^2x_2(1-x_1) \nonumber \\
& + & (k_3-k_2)^2x_2(x_1-x_2)-m_i^2  \\
-M_{1i}^2&=&x(1-x)(k_1-k_2)^2-m_i^2 \nonumber \\
-M_{2i}^2&=&x(1-x)(k_3-k_2)^2-m_i^2\\
-M_{3i}^2&=&x(1-x)(k_3-k_1)^2-m_i^2\nonumber
\end{eqnarray}

By adopting the useful relations given in App.~B and also in
App.~A, we have
\begin{eqnarray}
(k_3-k_1)_\mu
T^{R,(1),\{AVV\}}_{\lambda\mu\nu}&=&0\label{A-V-Vpaulivector1}\\
(k_1-k_2)_\nu
T^{R,(1),\{AVV\}}_{\lambda\mu\nu}&=&0\\
(k_3-k_2)_\lambda
T^{R,(1),\{AVV\}}_{\lambda\mu\nu}&=&{i\over4\pi^2}\sum_{i=0}^2C_i
\epsilon_{\mu\nu\lambda\alpha}(k_1-k_2)_\lambda(k_3-k_1)_\alpha
+\sum_{i=0}^2C_i[2m_iT_{\mu\nu}^{(1),PVV}(m_i)]\nonumber\\
&=&\sum_{i=0}^2C_i[2m_iT_{\mu\nu}^{(1),PVV}(m_i)]\label{A-V-Vpauliaxialvector1}
\end{eqnarray}
which shows that in Pauli-Villars regularization with the general
trace relation (\ref{trace10gamma}) from the definition of
$\gamma_5$, the vector currents are conserved. Note that such a
conclusion is not from the cancellation between the triangle
diagram of the original fermion and the regulator super-heavy
fermion, it is because both the vector currents in the original
triangle and the regulator triangle are conserved separately.
While partial cancellation occurs in the axial-vector Ward
identity.

For axial-vector Ward identity, it needs to further evaluate the
term $\sum_{i=0}^2C_i[2m_iT^{(1),PVV}_{\mu\nu}(m_i)]$. After some
algebra, we have
\begin{eqnarray}
\sum_{i=0}^2C_i[2m_iT^{(1),PVV}_{\mu\nu}(m_i)]&=&
16\epsilon_{\mu\nu\alpha\beta}(k_1-k_2)_\alpha(k_3-k_1)_\beta
\int_0^1dx_1\int_0^{x_1}dx_2\frac{i}{32\pi^2}\sum_{l=1}^2C_i\frac{m^2_i}{-M^2_i}\nonumber\\
&=&\frac{i}{4\pi^2}\epsilon_{\mu\nu\alpha\beta}(k_1-k_2)_\alpha(k_3-k_1)_\beta
\end{eqnarray}
where $m_1^2$ and $m_2^2$ have been taken to be infinity large, so
that
\begin{eqnarray}
\frac{m_i^2}{M_i^2}\rightarrow 1\label{limitPauli}
\end{eqnarray}
By considering the cross diagram, we finally obtain
\begin{eqnarray}
(k_3-k_2)_\lambda T^{R,\{AVV\}}_{\lambda\mu\nu}&=&
-\frac{i}{2\pi^2}\epsilon_{\mu\nu\alpha\beta}(k_1-k_2)_\alpha(k_3-k_1)_\beta
\end{eqnarray}

It is seen that the source of anomaly in Pauli-Villars
regularization actually arises from the heavy regulator fermion
loops. It is then unclear whether anomaly exists in the original
theory or is caused by the regularization scheme.

\subsection{ Anomaly of massive QED in Pauli-Villars regularization}

\label{sec massivePV}

In massive QED, the fermionic part of the regularized lagrangian
via Pauli-Villars regularization is\cite{PV}
\begin{eqnarray}
{\cal
L}^R_f=\bar{\psi}\gamma^\mu(i\partial_\mu-e\mathcal{A}_\mu)\psi-m\bar{\psi}\psi
+\sum_{i=1}^2C_i\{\bar{\psi}_i\gamma^\mu(i\partial_\mu-e\mathcal{A}_\mu)\psi_i-m_i\bar{\psi}_i\psi_i\}
\end{eqnarray}
where $C_i$ and mass parameters $m_i$ are respectively specified
as
\begin{eqnarray}
&&C_1=1,~~~~C_2=-2\nonumber\\
&&m_1^2=m^2+2\Lambda^2,~~~~m_2^2=m^2+\Lambda^2
\end{eqnarray}

With a similar evaluation as the massless case, the regularized
amplitudes are given by:
\begin{eqnarray}
T^{R,(1),\{AVV\}}_{\lambda\mu\nu}&=&T^{R,(1),\{AVV\}}_{0,\lambda\mu\nu}+T^{R,(1),\{AVV\}}_{-2,\lambda\mu\nu}\nonumber\\
T^{R,(1),\{AVV\}}_{0,\lambda\mu\nu}&=&2\sum_{i=0}^2C_i
\int_0^1dx[\epsilon_{\lambda\alpha\mu\nu}(k_2-k_1)_\alpha
\int\frac{d^4k}{(2\pi)^4}\frac{1}{[k^2-M_{1i}^2]^2}\nonumber\\
&&+\epsilon_{\lambda\alpha\nu\mu}(2x-1)(k_3-k_2)_\alpha \int\frac{d^4k}{(2\pi)^4}\frac{1}{[k^2-M_{2i}^2]^2}\nonumber\\
&&+\epsilon_{\lambda\mu\alpha\nu}(k_3-k_1)_\alpha \int\frac{d^4k}{(2\pi)^4}\frac{1}{[k^2-M_{3i}^2]^2}]\nonumber\\
&&-2\epsilon_{\lambda\mu\nu\alpha}\int_0^1dx_1\int_0^{x_1}dx_2(-2k_3-2k_2+4k_1)_\alpha
\int\frac{d^4k}{(2\pi)^4}\frac{k^2}{[k^2-M_i^2]^3}\\
T^{R,(1),\{AVV\}}_{-2,\lambda\mu\nu}&=&-8\sum_{i=0}^2C_i\int_0^1dx_1\int_0^{x_1}dx_2\nonumber\\
&&\times\bigg\{\epsilon_{\lambda\alpha\mu\beta}(-\Delta+k_2)_\alpha(-\Delta+k_1)_\beta(-\Delta+k_3)_\nu\nonumber\\
&&+\epsilon_{\lambda\alpha\mu\rho}(-\Delta+k_2)_\alpha(-\Delta+k_1)_\nu(-\Delta+k_3)_\rho\nonumber\\
&&+\epsilon_{\lambda\alpha\beta\nu}(-\Delta+k_2)_\alpha(-\Delta+k_1)_\beta(-\Delta+k_3)_\mu\nonumber\\
&&+\epsilon_{\lambda\alpha\nu\rho}(-\Delta+k_2)_\alpha(-\Delta+k_1)_\mu(-\Delta+k_3)_\rho\nonumber\\
&&+\epsilon_{\lambda\mu\beta\rho}(-\Delta+k_2)_{\nu}(-\Delta+k_1)_\beta(-\Delta+k_3)_\rho\nonumber\\
&&+\epsilon_{\lambda\beta\nu\rho}(-\Delta+k_2)_\mu(-\Delta+k_1)_\beta(-\Delta+k_3)_\rho\nonumber\\
&&+\epsilon_{\alpha\nu\beta\nu}(-\Delta+k_2)_\alpha(-\Delta+k_1)_\beta(-\Delta+k_3)_\lambda\nonumber\\
&&+\epsilon_{\alpha\mu\nu\rho}(-\Delta+k_2)_\alpha(-\Delta+k_1)_\lambda(-\Delta+k_3)_\rho\nonumber\\
&&-\epsilon_{\alpha\beta\nu\rho}(-\Delta+k_2)_\alpha(-\Delta+k_1)_\beta(-\Delta+k_3)_\rho\nonumber\\
&&-\epsilon_{\mu\nu\beta\rho}(-\Delta+k_2)_\lambda(-\Delta+k_1)_\beta(-\Delta+k_3)_\rho\nonumber\\
&&+\frac{\epsilon_{\lambda\alpha\mu\nu}}{2}
(k_3-k_1)^2(-\Delta+k_2)_\alpha+\frac{\epsilon_{\lambda\mu\beta\nu}}{2}(k_3-k_2)^2(-\Delta+k_1)_\beta\nonumber\\
&&+\frac{\epsilon_{\lambda\mu\nu\rho}}{2}(k_2-k_1)^2(-\Delta+k_3)_\rho\bigg\}
\int\frac{d^4k}{(2\pi)^4}\frac{1}{[k^2-M_i^2]^3}
\end{eqnarray}
which has the same form as the massless QED case except the mass
of internal fermion is introduced with the initial condition
$c_0=1$ and $m_0=m$.

Again adopting the useful relations presented in App. B and App.
A, we obtain the Ward identities as follows
\begin{eqnarray}
(k_3-k_1)_\mu
T^{R,(1),\{AVV\}}_{\lambda\mu\nu}&=&0\\
(k_1-k_2)_\nu
T^{R,(1),\{AVV\}}_{\lambda\mu\nu}&=&0\\
(k_3-k_2)_\lambda
T^{R,(1),\{AVV\}}_{\lambda\mu\nu}&=&{i\over4\pi^2}
\sum_{i=0}^2C_i\epsilon_{\mu\nu\lambda\alpha}(k_1-k_2)_\lambda(k_3-k_1)_\alpha
+\sum_{i=0}^2C_i[2m_iT_{\mu\nu}^{(1),PVV}(m_i)]\nonumber\\
&=&\sum_{i=0}^2C_i[2m_iT_{\mu\nu}^{(1),PVV}(m_i)]
\end{eqnarray}
which shows that, in the general relation of the trace of gamma
matrices (\ref{trace10gamma}), the situation is the same as the
massless case, the vector currents are conserved automatically,
while the axial-vector Ward identity is violated by the quantum
corrections.

Here $m_1^2$ and $m_2^2$ are the masses of heavy fields and taken
to be infinity large. By considering (\ref{limitPauli}), we have
\begin{eqnarray}
(k_3-k_2)_\lambda
T^{R,(1),\{AVV\}}_{\lambda\mu\nu}&=&2mT_{\mu\nu}^{(1),PVV}-{i\over4\pi^2}
\epsilon_{\mu\nu\alpha\beta}(k_3-k_1)_\alpha(k_1-k_2)_\beta
\end{eqnarray}

With the same discussions made in the loop regularization for
massive QED at $\mu_s =0$, the above Ward identity can generally
be written as
\begin{eqnarray}
 (k_3-k_2)_\lambda
T^{(1),\{AVV\}}_{\lambda\mu\nu} & = & (I(m,p,q) - 1 )
{i\over4\pi^2}
\epsilon_{\mu\nu\alpha\beta}(k_3-k_1)_\alpha(k_1-k_2)_\beta
\end{eqnarray}
where $I(m,p,q)$ is define in (\ref{defineI(m,p,q)}).

For two massless vector external states and one soft axial-vector
state: $p^2 = 0$, $q^2 = 0$, $(p+q)^2$ is soft, one see from
eq.(\ref{special-I-1}) that in this case both vector and
axial-vector receive no quantum corrections in massive QED.

For the alternative case that $p^2 = 0$, $q^2 = 0$, and $(p+q)^2
\gg m^2 $, thus eq.(\ref{special-I-2}) holds, which implies that
in this case the axial-vector becomes anomaly. Though
Pauli-Villars scheme leads to the same conclusions as the loop
regularization at $\mu_s = 0$, while the anomaly in Pauli-Villars
scheme is attribute to the regulator fermion loops rather than the
physical fermion loops in the original theory.

\section{Anomaly in dimensional regularization}

\label{sec DR}

Since $\gamma_5$ is an intrinsical four dimensional object, to
calculate anomaly with dimensional regularization, one can not
naively extend it to neither $n\neq4$ nor non-integer dimensions.
In dimensional regularization, $\gamma_5$ problem is usually
considered by using the definition in $n$ dimensions as
follows\cite{DR}
\begin{eqnarray}
\gamma_5=i\gamma_0\gamma_1\gamma_2\gamma_3
\end{eqnarray}
By this definition, then considering the extension of the
commutation relation of gamma matrix in $n$ dimension space
\begin{eqnarray}
\{\gamma_\mu, \gamma_\nu\}=2g_{\mu\nu}
\end{eqnarray}
It is not difficult to prove the following commutation relations
\begin{eqnarray}
&&\{\gamma_\mu^\parallel,\gamma_5\}=0,~~~~\gamma_\mu^\parallel ~~\mbox{is in the first four dimensions}\label{commu-4-dim}\\
&&[\gamma_\mu^\perp,\gamma_5]=0,~~~~\gamma_\mu^\perp ~~\mbox{is in
the extended dimensions}\label{commu-extra-dim}
\end{eqnarray}

To do the momentum integral with dimensional regularization, only
the integration variable is extended to $n$ dimensions while other
quantities still live in four dimensions. In the $n$ dimensions,
the loop momentum $k$ is decomposed into two parts
\begin{eqnarray}
k=k_\parallel+k_\perp\label{decompl}
\end{eqnarray}
where $k_\parallel$ and $k_\perp$ are the components of $k$ in
dimensions $0, 1, 2, 3$ and in other $n-4$ dimensions
respectively.

\subsection{Anomaly of massless QED in dimensional regularization}

\label{sec masslessDR}

Applying the above considerations to the amplitude defined in
(\ref{TAVV}), we then also divide the amplitude into two parts
corresponding to $k_\parallel$ and $k_\perp$
\begin{eqnarray}
T^{R,(1),AVV}_{\lambda\mu\nu}&=&T^{R,(1),AVV}_{\parallel,\lambda\mu\nu}
+
T^{R,(1),AVV}_{\perp,\lambda\mu\nu}\label{T=Tp+TP}\\
T^{R,(1),AVV}_{\parallel,\lambda\mu\nu}
&\equiv&-i\int\frac{d^nk}{(2\pi)^n}{\rm
tr}\{\frac{\gamma_5\gamma_\lambda(k\hspace{-0.2cm}\slash_\parallel+k\hspace{-0.2cm}\slash_2)
\gamma_\nu(k\hspace{-0.2cm}\slash_\parallel+k\hspace{-0.2cm}\slash_1)
\gamma_\mu(k\hspace{-0.2cm}\slash_\parallel
+k\hspace{-0.2cm}\slash_3)}{(k+k_2)^2(k+k_1)^2(k+k_3)^2}\}\label{Tparalell}\\
T^{R,(1),AVV}_{\perp,\lambda\mu\nu}&\equiv&-4\epsilon^{\mu\nu\lambda\alpha}\int\frac{d^nk}{(2\pi)^n}
\frac{(k_\parallel+k_3)_\alpha k_\perp^2}{(k+k_2)^2(k+k_1)^2(k+k_3)^2}\nonumber\\
&&-4\epsilon^{\mu\nu\lambda\alpha}\int\frac{d^nk}{(2\pi)^n}
\frac{k_\perp^2(k_\parallel+k_1)_\alpha}{(k+k_1)^2(k+k_1)^2(k+k_3)^2}\nonumber\\
&&-4\epsilon^{\mu\nu\lambda\alpha}\int\frac{d^nk}{(2\pi)^n}
\frac{k_\perp^2(k_\parallel+k_2)_\alpha}{(k+k_1)^2(k+k_1)^2(k+k_3)^2}\}
\end{eqnarray}
where we have used eqs.(\ref{commu-4-dim}),
(\ref{commu-extra-dim}) and (\ref{decompl}) and only kept the
non-zero terms.

For the terms depending on $k_\perp^2$, by using the Feynman
parametrization and App.~C., we find that
\begin{eqnarray}
T^{R,(1),AVV}_{\perp,\lambda\mu\nu}&=&{1\over3}{i\over4\pi^2}
\epsilon_{\mu\nu\lambda\alpha}(k_3+k_2-2k_1)_\alpha\label{masslessPerpend}
\end{eqnarray}
So that
\begin{eqnarray}
(k_3-k_1)_\mu
T^{R,(1),AVV}_{\perp,\lambda\mu\nu}&=&-{1\over3}{i\over4\pi^2}
\epsilon_{\mu\nu\lambda\alpha}(k_3-k_1)_\mu(k_1-k_2)_\alpha\\
(k_1-k_2)_\nu
T^{R,(1),AVV}_{\perp,\lambda\mu\nu}&=&{1\over3}{i\over4\pi^2}
\epsilon_{\mu\nu\lambda\alpha}(k_1-k_2)_\nu(k_3-k_1)_\alpha\\
(k_3-k_2)_\lambda
T^{R,(1),AVV}_{\perp,\lambda\mu\nu}&=&{2\over3}{i\over4\pi^2}
\epsilon_{\mu\nu\lambda\alpha}(k_3-k_1)_\lambda(k_2-k_1)_\alpha
\end{eqnarray}

For the terms independent of $k_\perp^2$, they are four
dimensional objects. By using the relation (\ref{trace10gamma})
for the trace of gamma matrices, we obtain
\begin{eqnarray}
T^{R,(1),\{AVV\}}_{\parallel,\lambda\mu\nu}
&=&T^{R,(1),\{AVV\}}_{\parallel,L,\lambda\mu\nu}+T^{R,(1),\{AVV\}}_{\parallel,C,\lambda\mu\nu}\nonumber\\
T^{R,(1),\{AVV\}}_{\parallel,L,\lambda\mu\nu}&=&4\int\frac{d^4k}{(2\pi)^4}
\bigg\{ \  \{-\epsilon_{\lambda\alpha\nu\beta}
(k_\parallel+k_2)_\alpha(k_\parallel+k_1)_\beta(k_\parallel+k_3)_\mu\nonumber\\
&&-\epsilon_{\lambda\alpha\nu\rho}
(k_\parallel+k_2)_\alpha(k_\parallel+k_1)_\mu(k_\parallel+k_3)_\rho\nonumber\\
&& -\epsilon_{\lambda\alpha\beta\mu}
(k_\parallel+k_2)_\alpha(k_\parallel+k_1)_\beta(k_\parallel+k_3)_\nu\nonumber\\
&&+\epsilon_{\lambda\alpha\beta\rho}
g_{\mu\nu}(k_\parallel+k_2)_\alpha(k_\parallel+k_1)_\beta(k_\parallel+k_3)_\rho\nonumber\\
&&-\epsilon_{\lambda\alpha\mu\rho}(k_\parallel+k_2)_\alpha(k_\parallel+k_1)_\nu(k_\parallel+k_3)_\rho\nonumber\\
&&-\epsilon_{\lambda\nu\beta\rho}(k_\parallel+k_2)_{\mu}(k_\parallel+k_1)_\beta(k_\parallel+k_3)_\rho\nonumber\\
&&-\epsilon_{\lambda\beta\mu\rho}(k_\parallel+k_2)_\nu(k_\parallel+k_1)_\beta(k_\parallel+k_3)_\rho\nonumber\\
&&-\epsilon_{\alpha\nu\beta\mu}(k_\parallel+k_2)_\alpha(k_\parallel+k_1)_\beta(k_\parallel+k_3)_\lambda\nonumber\\
&&+\epsilon_{\alpha\nu\beta\rho}
g_{\lambda\mu}(k_\parallel+k_2)_\alpha(k_\parallel+k_1)_\beta(k_\parallel+k_3)_\rho\nonumber\\
&&-\epsilon_{\alpha\nu\mu\rho}(k_\parallel+k_2)_\alpha(k_\parallel+k_1)_\lambda(k_\parallel+k_3)_\rho\nonumber\\
&&+\epsilon_{\alpha\beta\mu\rho}
g_{\lambda\nu}(k_\parallel+k_2)_\alpha(k_\parallel+k_1)_\beta(k_\parallel+k_3)_\rho\nonumber\\
&&+\epsilon_{\nu\mu\beta\rho}(k_\parallel+k_2)_\lambda(k_\parallel+k_1)_\beta(k_\parallel+k_3)_\rho\}\nonumber\\
&&\times\bigg[\frac{1}{(k+k_1)^2(k+k_2)^2(k+k_3)^2}\bigg]\nonumber\\
&&+\frac{\epsilon_{\lambda\alpha\nu\mu}}{2}
\bigg[\frac{(k_\parallel+k_2)_\alpha}{(k+k_2)^2(k+k_3)^2}
+\frac{(k_\parallel+k_2)_\alpha}{(k+k_2)^2(k+k_1)^2}\bigg]\nonumber\\
&&+\frac{\epsilon_{\lambda\nu\beta\mu}}{2}
\bigg[\frac{(k_\parallel+k_1)_\beta}{(k+k_1)^2(k+k_3)^2}
+\frac{(k_\parallel+k_1)_\beta}{(k+k_2)^2(k+k_1)^2}\bigg]\nonumber\\
&&+\frac{\epsilon_{\lambda\nu\mu\rho}}{2}
\bigg[\frac{(k_\parallel+k_3)_\rho}{(k+k_1)^2(k+k_3)^2}
+\frac{(k_\parallel+k_3)_\rho}{(k+k_2)^2(k+k_3)^2}\bigg]\bigg\}\label{dimTAVVL}\\
T^{R,(1),\{AVV\}}_{\parallel,C,\lambda\mu\nu}&=&-2\epsilon_{\lambda\alpha\nu\mu}\int\frac{d^4k}{(2\pi)^4}
\frac{(k_3-k_1)^2(k_\parallel+k_2)_\alpha}{(k+k_1)^2(k+k_2)^2(k+k_3)^2}\nonumber\\
&&-2\epsilon_{\lambda\nu\beta\mu}\int\frac{d^4k}{(2\pi)^4}
\frac{(k_3-k_2)^2(k_\parallel+k_1)_\beta}{(k+k_1)^2(k+k_2)^2(k+k_3)^2}\nonumber\\
&&-2\epsilon_{\lambda\nu\mu\rho}\int\frac{d^4k}{(2\pi)^4}
\frac{(k_1-k_2)^2(k_\parallel+k_3)_\rho}{(k+k_1)^2(k+k_2)^2(k+k_3)^2}\label{dimTAVVC}
\end{eqnarray}

In the dimensional regularization, we come to the following
relation
\begin{eqnarray}
\int\frac{d^nk}{(2\pi)^n}\frac{k_{\parallel\mu}k_{\parallel\nu}}{[k^2-M^2]^3}&=&{1\over
n}g_{\mu\nu}\int\frac{d^nk}{(2\pi)^n}\frac{k_{\parallel}^2}{[k^2-M^2]^3}\nonumber\\
&=&{1\over
n}g_{\mu\nu}\int\frac{d^nk}{(2\pi)^n}\frac{k^2-k_\perp^2}{[k^2-M^2]^3}\nonumber\\
&=&{1\over
n}g_{\mu\nu}\int\frac{d^nk}{(2\pi)^n}\{\frac{1}{[k^2-M^2]^2}
+\frac{M^2}{[k^2-M^2]^3}-\frac{k_\perp^2}{[k^2-M^2]^3}\}\nonumber\\
&=&{1\over
n}g_{\mu\nu}\int\frac{d^nk}{(2\pi)^n}\{\frac{1}{[k^2-M^2]^2}\}
\end{eqnarray}

Denoting the infinitesimal constant $\varepsilon$ in the
dimensional regularization as $\varepsilon=4-n$, we have
\begin{eqnarray}
\int\frac{d^nk}{(2\pi)^n}\frac{k_{\parallel\mu}k_{\parallel\nu}}{[k^2-M^2]^3}&=&{1\over
4-\varepsilon}g_{\mu\nu}\int\frac{d^nk}{(2\pi)^n}\{\frac{1}{[k^2-M^2]^2}\}\nonumber\\
&=&{1\over
4}g_{\mu\nu}\int\frac{d^nk}{(2\pi)^n}\{\frac{1}{[k^2-M^2]^2}\}+{\varepsilon\over
16}g_{\mu\nu}\int\frac{d^nk}{(2\pi)^n}\{\frac{1}{[k^2-M^2]^2}\}\nonumber\\
&=&{1\over
4}g_{\mu\nu}\int\frac{d^nk}{(2\pi)^n}\{\frac{1}{[k^2-M^2]^2}\}+{\varepsilon\over
16}g_{\mu\nu}{i\over16\pi^2}\{{2\over\varepsilon}+\cdots\}\nonumber\\
&=&{1\over
4}g_{\mu\nu}\int\frac{d^nk}{(2\pi)^n}\{\frac{1}{[k^2-M^2]^2}\}+{1\over
8}g_{\mu\nu}{i\over16\pi^2}+\mathcal{O}(\varepsilon)\label{paralellintegral}
\end{eqnarray}
We will not consider the last term in the following calculation
since it will vanish when one sets $\varepsilon \rightarrow 0$
which is required by the dimensional regularization.

After Feynman paramerization and some algebra, the parallel
momentum part of the regularized amplitude
$T^{R,(1),\{AVV\}}_{\parallel,\lambda\mu\nu}$ can be written as:
\begin{eqnarray}
T^{R,(1),\{AVV\}}_{\parallel,\lambda\mu\nu}&=&T^{R,(1),\{AVV\}}_{\parallel,0,\lambda\mu\nu}
+T^{R,(1),\{AVV\}}_{\parallel,-2,\lambda\mu\nu}\nonumber\\
T^{R,(1),\{AVV\}}_{\parallel,0,\lambda\mu\nu}
&=&2\int_0^1dx\{\epsilon_{\lambda\alpha\nu\mu}(k_2-k_1)_\alpha {i\over16\pi^2}[-\ln M_1^2]\nonumber\\
&&+\epsilon_{\lambda\alpha\nu\mu}(2x-1)(k_3-k_2)_\alpha{i\over16\pi^2}[-\ln M_2^2]\nonumber\\
&&+\epsilon_{\lambda\mu\alpha\nu}(k_3-k_1)_\alpha {i\over16\pi^2}[-\ln M_3^2]\}\nonumber\\
&&-2\epsilon_{\lambda\mu\nu\alpha}(-2k_3-2k_2+4k_1)_\alpha \int_0^1dx_1\int_0^{x_1}dx_2 {i\over16\pi^2}[-\ln M^2]\\
T^{R,(1),\{AVV\}}_{\parallel,-2,\lambda\mu\nu}&=&-8\int_0^1dx_1\int_0^{x_1}dx_2\nonumber\\
&&\times\bigg\{\epsilon_{\lambda\alpha\nu\beta}(-\Delta+k_2)_\alpha(-\Delta+k_1)_\beta(-\Delta+k_3)_\mu\nonumber\\
&&+\epsilon_{\lambda\alpha\nu\rho}(-\Delta+k_2)_\alpha(-\Delta+k_1)_\mu(-\Delta+k_3)_\rho\nonumber\\
&&+\epsilon_{\lambda\alpha\beta\mu}(-\Delta+k_2)_\alpha(-\Delta+k_1)_\beta(-\Delta+k_3)_\nu\nonumber\\
&&+\epsilon_{\lambda\alpha\mu\rho}(-\Delta+k_2)_\alpha(-\Delta+k_1)_\nu(-\Delta+k_3)_\rho\nonumber\\
&&+\epsilon_{\lambda\nu\beta\rho}(-\Delta+k_2)_{\mu}(-\Delta+k_1)_\beta(-\Delta+k_3)_\rho\nonumber\\
&&+\epsilon_{\lambda\beta\mu\rho}(-\Delta+k_2)_\nu(-\Delta+k_1)_\beta(-\Delta+k_3)_\rho\nonumber\\
&&+\epsilon_{\alpha\nu\beta\mu}(-\Delta+k_2)_\alpha(-\Delta+k_1)_\beta(-\Delta+k_3)_\lambda\nonumber\\
&&+\epsilon_{\alpha\nu\mu\rho}(-\Delta+k_2)_\alpha(-\Delta+k_1)_\lambda(-\Delta+k_3)_\rho\nonumber\\
&&-\epsilon_{\nu\mu\beta\rho}(-\Delta+k_2)_\lambda(-\Delta+k_1)_\beta(-\Delta+k_3)_\rho\nonumber\\
&&+\frac{\epsilon_{\lambda\alpha\nu\mu}}{2}
(k_3-k_1)^2(-\Delta+k_2)_\alpha+\frac{\epsilon_{\lambda\nu\beta\mu}}{2}(k_3-k_2)^2(-\Delta+k_1)_\beta\nonumber\\
&&+\frac{\epsilon_{\lambda\nu\mu\rho}}{2}(k_2-k_1)^2(-\Delta+k_3)_\rho\bigg\}{-i\over32\pi^2}\frac{1}{M^2}\nonumber\\
&&-{i\over16\pi^2}\epsilon_{\lambda\mu\nu\alpha}(-k_3-k_2+2k_1)_\alpha
\end{eqnarray}
Here $T^{(1),AVV}_{\parallel,0,\lambda\mu\nu}$ and
$T^{R,(1),AVV}_{\parallel,-2,\lambda\mu\nu}$ are corresponding to
the logarithemic and convergent terms in (\ref{dimTAVVL}) and
(\ref{dimTAVVC}). $x$ and $x_i$ are Feynman parameters. The last
term in above equation comes from the second term of
(\ref{paralellintegral}).

Using the relations given in App.~A. with $\mu_s^2=0$ and
$M_c^2\rightarrow\infty$, we find that the contributions from the
$k_\parallel$ have the following forms
\begin{eqnarray}
(k_3-k_1)_\mu T^{R,(1),\{AVV\}}_{\parallel,\lambda\mu\nu}
&=&-{1\over4}{i\over4\pi^2}\epsilon_{\lambda\mu\nu\alpha}(k_3-k_1)_\mu(k_1-k_2)_\alpha\\
(k_1-k_2)_\nu T^{R,(1),\{AVV\}}_{\parallel,\lambda\mu\nu}
&=&{1\over4}{i\over4\pi^2}\epsilon_{\lambda\mu\nu\alpha}(k_1-k_2)_\nu(k_3-k_1)_\alpha\\
(k_3-k_2)_\lambda T^{R,(1),\{AVV\}}_{\parallel,\lambda\mu\nu}
&=&{3\over2}{i\over4\pi^2}\epsilon_{\mu\nu\lambda\alpha}(k_1-k_2)_\lambda(k_3-k_1)_\alpha
\end{eqnarray}

Eventually, we yield by combining the two parts the following
results
\begin{eqnarray}
(k_3-k_1)_\mu
T^{R,(1),\{AVV\}}_{\lambda\mu\nu}&=&-{7\over12}\frac{i}{4\pi^2}
\epsilon_{\mu\nu\lambda\alpha}(k_3-k_1)_\mu(k_1-k_2)_\alpha\label{A-V-Vdimensionvector1}\\
(k_1-k_2)_\nu
T^{R,(1),\{AVV\}}_{\lambda\mu\nu}&=&{7\over12}\frac{i}{4\pi^2}
\epsilon_{\mu\nu\lambda\alpha}(k_1-k_2)_\nu(k_3-k_1)_\alpha\\
(k_3-k_2)_\lambda T^{R,(1),\{AVV\}}_{\lambda\mu
\nu}&=&{26\over12}\frac{i}{4\pi^2}
\epsilon_{\mu\nu\lambda\beta}(k_1-k_2)_\lambda(k_3-k_1)_\beta\label{A-V-Vdimensionaxialvector}
\end{eqnarray}
Including the cross diagrams, we arrive at the final results
\begin{eqnarray}
(k_3-k_1)_\mu
T^{R,\{AVV\}}_{\lambda\mu\nu}&=&-{7\over6}\frac{i}{4\pi^2}
\epsilon_{\nu\lambda\alpha\beta}(k_1-k_2)_\alpha(k_3-k_1)_\beta\nonumber\\
(k_1-k_2)_\nu
T^{R,\{AVV\}}_{\lambda\mu\nu}&=&{7\over6}\frac{i}{4\pi^2}
\epsilon_{\mu\lambda\alpha\beta}(k_1-k_2)_\alpha(k_3-k_1)_\beta\nonumber\\
(k_3-k_2)_\lambda
T^{R,\{AVV\}}_{\lambda\mu\nu}&=&{26\over6}\frac{i}{4\pi^2}
\epsilon_{\mu\nu\lambda\beta}(k_1-k_2)_\lambda(k_3-k_1)_\beta
\end{eqnarray}
Here both the vector Ward identity and axial-vector identity are
violated by quantum anomaly. To keep the vector current be
conserved, one can simply make the redefinition for the amplitudes
as follows
\begin{eqnarray}
\widetilde{T}^{(1),\{AVV\}}_{\lambda\mu\nu}(k_3-k_1,k_1-k_2)
=T^{(1),\{AVV\}}_{\lambda\mu\nu}(k_3-k_1,k_1-k_2)-T^{(1),\{AVV\}}_{\lambda\mu\nu}(0)\label{AVV-redefdimensionR-massless}
\end{eqnarray}
with
\begin{eqnarray}
T^{(1),\{AVV\}}_{\lambda\mu\nu}(0)={7\over6}{i\over4\pi^2}
\epsilon_{\mu\nu\lambda\alpha}[(k_3-k_1)_\alpha-(k_1-k_2)_\alpha]
\end{eqnarray}

As a consequence, the Ward identities of the redefined amplitudes
have the standard form of triangle anomaly appearing only in the
axial-vector current
\begin{eqnarray}
(k_3-k_1)_\mu
\widetilde{T}^{\{AVV\}}_{\lambda\mu\nu}&=&0\\
(k_1-k_2)_\nu
\widetilde{T}^{\{AVV\}}_{\lambda\mu\nu}&=&0\\
(k_3-k_2)_\lambda
\widetilde{T}^{\{AVV\}}_{\lambda\mu\nu}&=&{i\over2\pi^2}
\epsilon_{\mu\nu\lambda\alpha}(k_1-k_2)_\lambda(k_3-k_1)_\alpha
\end{eqnarray}

From the above explicit calculations in the dimensional
regularization, we see that, unlike in the loop regularization,
the divergences of both the vector and axial-vector currents are
violated by quantum corrections. The quantum corrections depend on
both the original four dimensions and the extended dimensions.
While as shown in\cite{DR}, if one multiplies the external
momentum of the axial-vector current on the AVV diagram before
evaluating the integrals, the final results only depend on the
extended dimensions and consequently only the axial-vector Ward
identity is violated by the anomaly. This implies that in the
dimensional regularization the anomaly of vector and axial-vector
currents due to quantum loop corrections depends on the procedure
of operation although the final anomaly, when normalizing to the
conserved vector current, has the same result of the standard
form. We shall further discuss the possible ambiguities arising
from the dimensional regularization.

\subsection{Anomaly of massive QED in dimensional regularization}

\label{subsec massiveDR}

In the massive QED, with the same reasoning as the massless case,
we can write the regularized parallel part of the amplitude as
\begin{eqnarray}
T^{R,(1),\{AVV\}}_{\parallel,\lambda\mu\nu}
&=&T^{R(1),\{AVV\}}_{\parallel,0,\lambda\mu\nu}+T^{R,(1),\{AVV\}}_{\parallel,-2,\lambda\mu\nu}\nonumber\\
T^{R,(1),\{AVV\}}_{\parallel,0,\lambda\mu\nu}
&=&2\int_0^1dx\{\epsilon_{\lambda\alpha\mu\nu}(k_2-k_1)_\alpha {i\over16\pi^2}[-\ln (M_1^{2}+m^2)]\nonumber\\
&&+\epsilon_{\lambda\alpha\nu\mu}(2x-1)(k_3-k_2)_\alpha{i\over16\pi^2}[-\ln (M_2^{2}+m^2)]\nonumber\\
&&+\epsilon_{\lambda\mu\alpha\nu}(k_3-k_1)_\alpha  {i\over16\pi^2}[-\ln (M_3^{2}+m^2)]\}\\
&&-2\epsilon_{\lambda\mu\nu\alpha}\int_0^1dx_1\int_0^{x_1}dx_2(-2k_3-2k_2+4k_1)_\alpha
{i\over16\pi^2}[-\ln (M^{2}+m^2)]\nonumber\\
T^{R,(1),\{AVV\}}_{\parallel,-2,\lambda\mu\nu}&=&-8\int_0^1dx_1\int_0^{x_1}dx_2\nonumber\\
&&\times\bigg\{\epsilon_{\lambda\alpha\mu\beta}(-\Delta+k_2)_\alpha(-\Delta+k_1)_\beta(-\Delta+k_3)_\nu\nonumber\\
&&+\epsilon_{\lambda\alpha\mu\rho}(-\Delta+k_2)_\alpha(-\Delta+k_1)_\nu(-\Delta+k_3)_\rho\nonumber\\
&&+\epsilon_{\lambda\alpha\beta\nu}(-\Delta+k_2)_\alpha(-\Delta+k_1)_\beta(-\Delta+k_3)_\mu\nonumber\\
&&+\epsilon_{\lambda\alpha\nu\rho}(-\Delta+k_2)_\alpha(-\Delta+k_1)_\mu(-\Delta+k_3)_\rho\nonumber\\
&&+\epsilon_{\lambda\mu\beta\rho}(-\Delta+k_2)_{\nu}(-\Delta+k_1)_\beta(-\Delta+k_3)_\rho\nonumber\\
&&+\epsilon_{\lambda\beta\nu\rho}(-\Delta+k_2)_\mu(-\Delta+k_1)_\beta(-\Delta+k_3)_\rho\nonumber\\
&&+\epsilon_{\alpha\nu\beta\nu}(-\Delta+k_2)_\alpha(-\Delta+k_1)_\beta(-\Delta+k_3)_\lambda\nonumber\\
&&+\epsilon_{\alpha\mu\nu\rho}(-\Delta+k_2)_\alpha(-\Delta+k_1)_\lambda(-\Delta+k_3)_\rho\nonumber\\
&&-\epsilon_{\alpha\beta\nu\rho}(-\Delta+k_2)_\alpha(-\Delta+k_1)_\beta(-\Delta+k_3)_\rho\nonumber\\
&&-\epsilon_{\mu\nu\beta\rho}(-\Delta+k_2)_\lambda(-\Delta+k_1)_\beta(-\Delta+k_3)_\rho\nonumber\\
&&+\frac{\epsilon_{\lambda\alpha\mu\nu}}{2}
(k_3-k_1)^2(-\Delta+k_2)_\alpha+\frac{\epsilon_{\lambda\mu\beta\nu}}{2}(k_3-k_2)^2(-\Delta+k_1)_\beta\nonumber\\
&&+\frac{\epsilon_{\lambda\mu\nu\rho}}{2}(k_2-k_1)^2(-\Delta+k_3)_\rho\bigg\}{-i\over32\pi^2}
\frac{1}{M^{2}+m^2}\nonumber\\
&&-{i\over16\pi^2}\epsilon_{\lambda\mu\nu\alpha}(-k_3-k_2+2k_1)_\alpha
\end{eqnarray}

Considering the relations given in App.A. and App.C., one can
easily get the following results
\begin{eqnarray}
(k_3-k_1)_\mu T^{R,(1),\{AVV\}}_{\parallel,\lambda\mu\nu}
&=&-{1\over4}{i\over4\pi^2}\epsilon_{\lambda\mu\nu\alpha}(k_3-k_1)_\mu(k_1-k_2)_\alpha\\
(k_1-k_2)_\nu T^{R,(1),\{AVV\}}_{\parallel,\lambda\mu\nu}
&=&{1\over4}{i\over4\pi^2}\epsilon_{\lambda\mu\nu\alpha}(k_1-k_2)_\nu(k_3-k_1)_\alpha\\
(k_3-k_2)_\lambda
T^{(1),\{AVV\}}_{\parallel,\lambda\mu\nu}
&=&-{3\over2}{i\over4\pi^2}\epsilon_{\mu\nu\lambda\alpha}(k_3-k_1)_\lambda(k_1-k_2)_\alpha\nonumber\\
&&-16m^2\epsilon_{\mu\nu\lambda\alpha}(k_3-k_1)_\lambda(k_1-k_2)_\alpha
I_{-2,(00)}
\end{eqnarray}
By considering the perpendicular contribution and the result of
the PVV diagram, the anomaly amplitude has the following form
\begin{eqnarray}
(k_3-k_1)_\mu
T^{(1),\{AVV\}}_{\lambda\mu\nu}&=&-{7\over12}{i\over4\pi^2}(k_3-k_1)_\mu(k_1-k_2)_\alpha\\
(k_1-k_2)_\nu
T^{(1),\{AVV\}}_{\lambda\mu\nu}&=&{7\over12}{i\over4\pi^2}(k_1-k_2)_\nu(k_3-k_1)_\alpha\\
(k_3-k_2)_\lambda
T^{(1),\{AVV\}}_{\lambda\mu\nu}&=&2mT_{\mu\nu}^{(1),PVV}-{26\over12}{i\over4\pi^2}
\epsilon_{\mu\nu\lambda\alpha}(k_3-k_1)_\lambda(k_1-k_2)_\alpha
\end{eqnarray}

By considering the cross diagram, we obtain the final results
\begin{eqnarray}
(k_3-k_1)_\mu
T^{\{AVV\}}_{\lambda\mu\nu}&=&-{7\over6}{i\over4\pi^2}(k_3-k_1)_\mu(k_1-k_2)_\alpha\\
(k_1-k_2)_\nu
T^{\{AVV\}}_{\lambda\mu\nu}&=&{7\over6}{i\over4\pi^2}(k_1-k_2)_\nu(k_3-k_1)_\alpha\\
(k_3-k_2)_\lambda
T^{\{AVV\}}_{\lambda\mu\nu}&=&2mT_{\mu\nu}^{PVV}-{26\over6}{i\over4\pi^2}
\epsilon_{\mu\nu\lambda\alpha}(k_3-k_1)_\lambda(k_1-k_2)_\alpha
\end{eqnarray}
which shows that in the direct calculation of the triangle diagram
by adopting the dimensional regularization, there are anomalies in
both vector and axial-vector Ward identities even when we treat
the gamma trace with the definition of $\gamma_5$.

Again to keep the vector currents conserved, one can make a
redefinition as done in (\ref{AVV-redefdimensionR-massless}).
After that, we have the Ward identities for the redefined
amplitude
\begin{eqnarray}
(k_3-k_1)_\mu
\tilde{T}^{\{AVV\}}_{\lambda\mu\nu}&=&0\\
(k_1-k_2)_\nu
\tilde{T}^{\{AVV\}}_{\lambda\mu\nu}&=&0\\
(k_3-k_2)_\lambda
\tilde{T}^{\{AVV\}}_{\lambda\mu\nu}
&=&2mT_{\mu\nu}^{PVV}-{i\over2\pi^2}\epsilon_{\mu\nu\lambda\alpha}(k_3-k_1)_\lambda(k_1-k_2)_\alpha
\end{eqnarray}

In conclusion, by adopting the dimensional regularization, the
direct calculation of the triangle diagram shows that even one
treat the gamma trace with the definition of $\gamma_5$
explicitly, both massless QED and massive QED, both vector and
axial-vector Ward identities are violated by quantum corrections.
But in this case, the magnitudes of the two vector Ward identities
are same and different from that of the axial-vector Ward
identity. It is then necessary to make a redefinition for the
amplitude in order to keep the vector currents conserved, so that
the anomaly lives in the axial-vector Ward identity. It is seen
that the situation is quite different from the case in the loop
regularization and Pauli-Villars scheme.


\section{Clarifications for the long-standing ambiguities in perturbation calculations of triangle anomaly}

\label{sec CLANOM}

In above sections, we have investigated anomaly by directly
calculating the triangle diagram with the most general trace
identity (\ref{trace10gamma}). As shown in section III that there
exist in general three independent treatments for carrying out the
trace of gamma matrices appearing in the triangle diagram. For
considerations of completeness and clarification, we shall
investigate other two treatments. In this section, we shall show
how the Ward identity anomaly depends on the treatment for the
trace of gamma matrices, which then allows us to make a
clarification for the long-standing ambiguities in the
perturbation calculations of triangle anomaly.

\subsection{Calculation of anomaly with (\ref{twovectorgamma}) in massless QED}

\label{subsec VVmassless}

In this subsection, we consider the case that the trace of gamma
matrix was evaluated by using (\ref{gammasymmetricvectors}). By
using that identity and performing the Dirac algebra, we can
express $T^{(1),A\{VV\}}_{\lambda\mu\nu}$ as
\begin{eqnarray}
T^{(1),A\{VV\}}_{\lambda\mu\nu}&=&T^{(1),A\{VV\}}_{L,\lambda\mu\nu}+T^{(1),A\{VV\}}_{C,\lambda\mu\nu}\nonumber\\
T^{(1),A\{VV\}}_{L,\lambda\mu\nu}&=&-4\int\frac{d^4k}{(2\pi)^4}
\bigg\{\{g_{\mu\nu}\epsilon_{\alpha\beta\lambda\xi}(k+k_1)_\alpha(k+k_2)_\beta(k+k_3)_\xi\nonumber\\
&&-\epsilon_{\nu\beta\lambda\xi}(k+k_1)_\mu(k+k_2)_\beta(k+k_3)_\xi
-\epsilon_{\mu\beta\lambda\xi}(k+k_1)_\nu(k+k_2)_\beta(k+k_3)_\xi\nonumber\\
&&-\epsilon_{\mu\alpha\nu\beta}(k+k_1)_\alpha(k+k_2)_\beta(k+k_3)_\lambda
-\epsilon_{\mu\alpha\nu\xi}(k+k_1)_\alpha(k+k_3)_\xi(k+k_2)_\lambda\}\nonumber\\
&&\times\bigg[\frac{1}{(k+k_1)^2(k+k_2)^2(k+k_3)^2}\bigg]\nonumber\\
&&+\frac{\epsilon_{\mu\alpha\nu\lambda}}{2}\bigg[\frac{(k+k_1)_\alpha}{(k+k_1)^2(k+k_2)^2}
+\frac{(k+k_1)_\alpha}{(k+k_1)^2(k+k_3)^2}\bigg]\bigg\}\label{linearTAVV}\\
T^{(1),A\{VV\}}_{C,\lambda\mu\nu}&=&-2\epsilon_{\mu\nu\alpha\lambda}\int\frac{d^4k}{(2\pi)^4}
\frac{(k_2-k_3)^2(k+k_1)_\alpha}{(k+k_1)^2(k+k_2)^2(k+k_3)^2}\label{converTAVV}
\end{eqnarray}

\subsubsection{Calculation of anomaly in loop regularization}

\label{subsubsec VVLRmassless}

After some basic algebra, the regularized amplitude is found to
have the following form
\begin{eqnarray}
T^{R(1),A\{VV\}}_{\lambda\mu\nu}&=&T^{R,(1),A\{VV\}}_{0
\lambda\mu\nu}+
T^{R,(1),A\{VV\}}_{-2\lambda\mu\nu} \nonumber \\
T^{R,(1),A\{VV\}}_{0\lambda\mu\nu}& = &
2\int^1_0dx_1\int_0^{x_1}dx_2
\{\epsilon_{\mu\lambda\nu\beta}(k_2+k_3-2k_1)_\beta I_0^R(x_i,\mu)\}\nonumber\\
&&-2\epsilon_{\mu\alpha\nu\lambda}\{\int_0^1dx(-\Delta_1+k_1)_\alpha
I_0^R(x,\mu_1)+\int_0^1dx(-\Delta_3+k_1)_\alpha
I_0^R(x,\mu_3)\}\nonumber\\
T^{R,(1),A\{VV\}}_{-2\lambda\mu\nu} & = & -4\int^1_0dx_1\int_0^{x_1}dx_2\nonumber\\
&&\times\bigg\{-2\{\epsilon_{\nu\beta\lambda\xi}[(-\Delta_1+k_1)_\mu(-\Delta_1+k_2)_\beta(-\Delta_1+k_3)_\xi]\nonumber\\
&&+\epsilon_{\mu\beta\lambda\alpha}[(-\Delta_1+k_3)_\alpha(-\Delta_1+k_2)_\beta(-\Delta_1+k_1)_\nu]\nonumber\\
&&+\epsilon_{\mu\alpha\nu\beta}[(-\Delta_1+k_1)_\alpha(-\Delta_1+k_2)_\beta(-\Delta_1+k_3)_\lambda]\nonumber\\
&&+\epsilon_{\mu\alpha\nu\beta}[(-\Delta_1+k_1)_\alpha(-\Delta_1+k_3)_\beta(-\Delta_1+k_2)_\lambda]\}\nonumber\\
&&+\epsilon_{\mu\nu\alpha\lambda}\{(k_2-k_3)^2(-\Delta_1+k_1)_\alpha\}\bigg\}I_{-2}^R(x_i,\mu)
\end{eqnarray}

We now first check the Ward identity $(k_1-k_2)_\nu
T^{R,A\{VV\}}_{\lambda\mu\nu}$. By using the definitions of
$\Delta_i$, we have
\begin{eqnarray}
(k_1-k_2)_\nu T^{R,(1),A\{VV\}}_{\lambda\mu\nu}&=&-2\epsilon_{\mu\nu\lambda\beta}(k_1-k_2)_\nu(k_3-k_1)_\beta\nonumber\\
&&\times\bigg\{4(k_3-k_1)\cdot(k_1-k_2)[I_{-2,(11)}+I_{-2,(20)}-I_{-2,10}]\nonumber\\
&&-4(k_2-k_1)^2[I_{-2,(02)}+I_{-2,(11)}-I_{-2,(01)}]+2(k_2-k_3)^2I_{-2,(10)}\bigg\}\nonumber\\
&&-2\epsilon_{\mu\nu\lambda\beta}(k_1-k_2)_\nu(k_3-k_1)_\beta
I_{0,(00)}
\end{eqnarray}

By adopting the explicit relations given in App.~A, we then obtain
\begin{eqnarray}
(k_1-k_2)_\nu
T^{R,(1),A\{VV\}}_{\lambda\mu\nu}&=&-2\epsilon_{\mu\nu\lambda\beta}(k_1-k_2)_\nu(k_3-k_1)_\beta\nonumber\\
&&\times\bigg\{\{{i\over16\pi^2}[e^{-\mu_s^2/M_c^2}\int_0^1dx_1\int_0^{x_1}dx_2e^{-M_1^2/M_c^2}
-{1\over2}Y((p+q)^2,q^2)]\nonumber\\
&&-q^2I_{-2,(10)}-p^2I_{-2,(01)}+2\mu_s^2I_{-2,(00)}\}+2(p+q)^2I_{-2,(10)}\nonumber\\
&&+4\{p^2I_{-2,(01)}-p\cdot qI_{-2,(10)}\nonumber\\
&&+{i\over64\pi^2}e^{-\mu_s^2/M_c^2}\int_0^1dx_1\int_0^{x_1}dx_2e^{-M_1^2/M_c^2}+{\mu_s^2\over2}
I_{-2,(00)}-{q^2\over4}I_{-2,(10)}\nonumber\\
&&-{3p^2\over4}I_{-2,(01)}+{i\over128\pi^2}Y((p+q)^2,q^2)-{p^2\over2}I_{-2,(10)}\}\bigg\}
\end{eqnarray}
where the first part in the bracket comes from the difference of
two logarithmically divergent integrals and the others come from
the convergent integrals. Many terms cancel each other, the final
expression gets the following simple form
\begin{eqnarray}
(k_1-k_2)_\nu
T^{R,(1),A\{VV\}}_{\lambda\mu\nu}&=&8\mu_s^2\epsilon_{\mu\lambda\alpha\beta}(k_1-k_2)_\alpha(k_3-k_1)_\beta
I_{-2,(00)}(0,\mu_s^2)\label{V-Vvector1}\\
&&+\frac{i}{4\pi^2}\epsilon_{\mu\lambda\alpha\beta}(k_1-k_2)_\alpha(k_3-k_1)_\beta
e^{-\mu_s^2/M_c^2}\int_0^1dx_1\int_0^{x_1}dx_2e^{-M^2/M_c^2}\nonumber
\end{eqnarray}
which is different from the result (\ref{A-V-Vvector1}) obtained
from the general treatment of the gamma trace, where the vector
current is found to be conserved automatically.

Similarly, we have the Ward identity for the second vector current
\begin{eqnarray}
(k_3-k_1)_\mu
T^{R,(1),A\{VV\}}_{\lambda\mu\nu}&=&-8\mu_s^2\epsilon_{\nu\lambda\alpha\beta}(k_1-k_2)_\alpha(k_3-k_1)_\beta
I_{-2,(00)}(0,\mu_s^2)\label{V-V-vector2}\\
&&-\frac{i}{4\pi^2}\epsilon_{\nu\lambda\alpha\beta}(k_1-k_2)_\alpha(k_3-k_1)_\beta
e^{-\mu_s^2/M_c^2}\int_0^1dx_1\int_0^{x_1}dx_2e^{-M_1^2/M_c^2}\nonumber
\end{eqnarray}
which has the same result as the first vector current except a
total sign that is expected from the treatment in which the two
gamma matrices with the Lorentz indices of vector currents are
grouped to reduce the number of gamma matrices.

For the divergence of the axial-vector current, we find that
\begin{eqnarray}
(k_3-k_2)_\lambda T^{R,(1),A\{VV\}}_{\lambda\mu\nu}
&=&4\epsilon_{\mu\nu\alpha\beta}(k_2-k_1)_\alpha(k_3-k_1)_\beta\nonumber\\
&&\times\bigg\{2(k_2-k_1)\cdot(k_3-k_2)[I_{-2,(20)}-I_{-2,(02)}+I_{-2,(01)}-I_{-2,(10)}]\nonumber\\
&&+2(k_3-k_2)^2[I_{-2,(10)}-I_{-2,(20)}+I_{-2,(11)}]\nonumber\\
&&-(k_2-k_3)^2[I_{-2,(01)}+I_{-2,(10)}]\bigg\}\nonumber\\
&&-2\epsilon_{\mu\nu\lambda\beta}(k_1-k_2)_\lambda(k_3-k_1)_\beta I^\prime_{0,(00)}\nonumber\\
&&-2\epsilon_{\mu\nu\lambda\beta}(k_1-k_2)_\lambda(k_3-k_1)_\beta I_{0,(00)}\nonumber\\
&=&0\label{divAVV}
\end{eqnarray}
which is identically equal to zero, namely there is no anomaly for
axial-vector current in this treatment.  It is obviously different
from the conclusion give by the general treatment
(\ref{SymmaxialWard}).

As before, we shall consider two interesting cases with
$\mu_s\neq0$ and $\mu_s=0$.

Firstly, considering the case that $\mu_s \neq 0$. Like section
III.A, for on-mass shell condition in the massless QED, i.e., $p^2
= (k_1 - k_2 )^2 = 0$, $q^2 = (k_3 - k_1 )^2 = 0$, $(p+q)^2 =
2p\cdot q$ is soft with $\mu_s^2 \gg (p+q)^2$ and $\mu_s^2 \ll
M_c^2\to \infty$, the vector current becomes conserved
\begin{eqnarray}
(k_3-k_1)_\mu T^{R,(1),A\{VV\}}_{\lambda\mu\nu}&=& 0\\
(k_1-k_2)_\nu T^{R,(1),A\{VV\}}_{\lambda\mu\nu}&=& 0
\end{eqnarray}
which shows that in loop regularization with the introduction of
nonzero IR cut-off scale $\mu_s\neq 0$ for massless QED, the Ward
identities of both vector and axial-vector currents become
conserved in the conditions that $p^2$, $q^2$, $(p+q)^2 \ll
\mu_s^2  \ll M_c^2 \to \infty$, which comes to the same statement
made in section III.A. in which the gamma trace was treated with
the definition of $\gamma_5$.

Secondly, considering the case that $\mu_s =0$, we then arrive at
the following results
\begin{eqnarray}
(k_3-k_1)_\mu
T^{R,(1),A\{VV\}}_{\lambda\mu\nu}&=&-\frac{i}{4\pi^2}\epsilon_{\nu\lambda\alpha\beta}(k_1-k_2)_\alpha(k_3-k_1)_\beta
\int_0^1dx_1\int_0^{x_1}dx_2e^{-M_1^2/M_c^2}\label{loopmasslessV-V-1}\\
(k_1-k_2)_\nu
T^{R,(1),A\{VV\}}_{\lambda\mu\nu}&=&\frac{i}{4\pi^2}\epsilon_{\mu\lambda\alpha\beta}(k_1-k_2)_\alpha(k_3-k_1)_\beta
\int_0^1dx_1\int_0^{x_1}dx_2e^{-M^2/M_c^2}\\
(k_3-k_2)_\lambda
T^{R,(1),A\{VV\}}_{\lambda\mu\nu}&=&0\label{loopmasslessV-V-3}
\end{eqnarray}
Taking $M_c\to \infty$, we yield
\begin{eqnarray}
(k_3-k_1)_\mu
T^{R,(1),A\{VV\}}_{\lambda\mu\nu}&=&-\frac{i}{8\pi^2}\epsilon_{\nu\lambda\alpha\beta}(k_1-k_2)_\alpha(k_3-k_1)_\beta\\
(k_1-k_2)_\nu
T^{R,(1),A\{VV\}}_{\lambda\mu\nu}&=&\frac{i}{8\pi^2}\epsilon_{\mu\lambda\alpha\beta}(k_1-k_2)_\alpha(k_3-k_1)_\beta\\
(k_3-k_2)_\lambda T^{R,(1),A\{VV\}}_{\lambda\mu\nu}&=&0
\end{eqnarray}
which is again different from the results
(\ref{symmevector1}-\ref{symmeaxialvector}) which were deduced
from the general treatment of the gamma trace. It is explicitly
seen that in the loop regularization with $\mu_s = 0$, when the
two gamma matrices with the Lorentz indices of the vector current
and one gamma between them are grouped to reduce the number of
gamma matrices in the trace, the vector currents in massless QED
are no longer conserved. To keep the vector Ward identities be
conserved, one can simply redefine the physical amplitude via
\begin{eqnarray}
\widetilde{T}^{(1),A\{VV\}}_{\lambda\mu\nu}(k_3-k_1,k_1-k_2)
=T^{R,(1),A\{VV\}}_{\lambda\mu\nu}(k_3-k_1,k_1-k_2)-T^{(1),A\{VV\}}_{\lambda\mu\nu}(0)\label{massiveV-Vredefi}
\end{eqnarray}
with
\begin{eqnarray}
T^{(1),A\{VV\}}_{\lambda\mu\nu}(0)&=&-{i\over8\pi^2}\epsilon_{\mu\nu\lambda\alpha}[(k_3-k_1)_\alpha-(k_1-k_2)_\alpha]
\end{eqnarray}
Thus the redefined amplitude has the following Ward identities
\begin{eqnarray}
(k_3-k_1)_\mu
\widetilde{T}^{(1),A\{VV\}}_{\lambda\mu\nu}&=&0\\
(k_1-k_2)_\nu
\widetilde{T}^{(1),A\{VV\}}_{\lambda\mu\nu}&=&0\\
(k_3-k_2)_\lambda
\widetilde{T}^{(1),A\{VV\}}_{\lambda\mu\nu}
&=&-{i\over4\pi^2}\epsilon_{\mu\nu\alpha\beta}(k_3-k_1)_\alpha(k_1-k_2)_\beta
\end{eqnarray}

By including the cross diagrams, we finally obtain the Ward
identities with anomaly of axial-vector current
\begin{eqnarray}
(k_3-k_1)_\mu
\widetilde{T}^{A\{VV\}}_{\lambda\mu\nu}&=&0\\
(k_1-k_2)_\nu
\widetilde{T}^{A\{VV\}}_{\lambda\mu\nu}&=&0\\
(k_3-k_2)_\lambda
\widetilde{T}^{A\{VV\}}_{\lambda\mu\nu}&=&-{i\over2\pi^2}
\epsilon_{\mu\nu\alpha\beta}(k_3-k_1)_\alpha(k_1-k_2)_\beta
\end{eqnarray}
Its operator form can be expressed as
\begin{eqnarray}
\partial_\mu V_\mu(x)&=&0,\\
\partial_\mu
A_\mu(x)&=&{e^2\over8\pi^2}F^{\mu\nu}(x)\widetilde{F}_{\mu\nu}(x)
\end{eqnarray}

In conclusion, it is seen from the above analysis that the IR-cut
off $\mu_s$ still plays an important role for an anomaly-free
treatment in the present case. While the Ward identity anomaly of
currents with treating the trace of gamma matrices with
(\ref{twovectorgamma}) is different from the case with the general
relation (\ref{trace10gamma}). When we treat using the relation
(\ref{twovectorgamma}), the two vector Ward identities are
violated by quantum corrections, the axial-vector Ward identity
becomes conserved. To keep the vector current conserved, one needs
to make a redefinition for the amplitude, so that the anomaly
appears in the axial-vector Ward identity.

\subsubsection{Calculation of anomaly by using Pauli-Villars regularization}

\label{subsubsec VVPVmassless}

When applying for the Pauli-Villars regularization, the
regularized amplitude reads
\begin{eqnarray}
T^{R,(1),A\{VV\}}_{L,\lambda\mu\nu}&=&-4\sum_{i=0}^2C_i\int\frac{d^4k}{(2\pi)^4}
\bigg\{\{g_{\mu\nu}\epsilon_{\alpha\beta\lambda\xi}(k+k_1)_\alpha(k+k_2)_\beta(k+k_3)_\xi\nonumber\\
&&-\epsilon_{\nu\beta\lambda\xi}(k+k_1)_\mu(k+k_2)_\beta(k+k_3)_\xi
-\epsilon_{\mu\beta\lambda\xi}(k+k_1)_\nu(k+k_2)_\beta(k+k_3)_\xi\nonumber\\
&&-\epsilon_{\mu\alpha\nu\beta}(k+k_1)_\alpha(k+k_3)_\beta(k+k_2)_\lambda
-\epsilon_{\mu\alpha\nu\xi}(k+k_1)_\alpha(k+k_2)_\xi(k+k_3)_\lambda\}\nonumber\\
&&\times\bigg[\frac{1}{[(k+k_1)^2-m_i^2][(k+k_2)^2-m_i^2][(k+k_3)^2-m_i^2]}\bigg]\nonumber\\
&&+\frac{\epsilon_{\mu\alpha\nu\lambda}}{2}
\bigg[\frac{(k+k_1)_\alpha}{[(k+k_1)^2-m_i^2][(k+k_2)^2-m_i^2]}\nonumber\\
&&+\frac{(k+k_1)_\alpha}{[(k+k_1)^2-m_i^2][(k+k_3)^2-m_i^2]}\bigg]\bigg\}\nonumber\\
T^{(1),A\{VV\}}_{C,\lambda\mu\nu}
&=&-2\epsilon_{\mu\nu\alpha\lambda}\sum_{i=0}^2C_i\int\frac{d^4k}{(2\pi)^4}\nonumber\\
&&\times\frac{2m_i^2(k+k_2)_\alpha-[4m_i^2-(k_2-k_3)^2](k+k_1)_\alpha
+2m_i^2(k+k_3)_\alpha}{[(k+k_1)^2-m_i^2][(k+k_2)^2-m_i^2][(k+k_3)^2-m_i^2]}\nonumber
\end{eqnarray}
with the initial condition $c_0=1$ and $m_0=0$. After adopting
Feynman parametrization and shifting the integration variables, we
obtain the similar expressions as those in loop regularization
\begin{eqnarray}
T^{(1),A\{VV\}}_{\lambda\mu\nu}&=&T^{R,(1),A\{VV\}}_{0
\lambda\mu\nu}+
T^{R,(1),A\{VV\}}_{-2\lambda\mu\nu} \nonumber \\
T^{R,(1),A\{VV\}}_{0\lambda\mu\nu}&=&-2\sum_{i=0}^2C_i\int^1_0dx_1\int_0^{x_1}dx_2
\{\epsilon_{\mu\lambda\nu\beta}(k_2+k_3-2k_1)_\beta\int\frac{d^4k}{(2\pi)^2}\frac{k^2}{[k^2-M_{i}^2]^3}\}\nonumber\\
&&-2\epsilon_{\mu\alpha\nu\lambda}\{\sum_{i=0}^2C_i\int_0^1dx
(-\Delta_2+k_1)_\alpha\int\frac{d^4k}{(2\pi)^2}\frac{1}{[k^2-M_{1i}^2]^2}\nonumber\\
&&+\sum_{i=0}^2C_i\int_0^1dx(-\Delta_3+k_1)_\alpha \int\frac{d^4k}{(2\pi)^2}\frac{1}{[k^2-M_{3i}^2]^2}\}\nonumber\\
T^{R,(1),A\{VV\}}_{-2\lambda\mu\nu}&=&4\sum_{i=0}^2C_i\int^1_0dx_1\int_0^{x_1}dx_2\nonumber\\
&&\times\bigg\{2\{\epsilon_{\mu\beta\lambda\xi}[(-\Delta_1+k_1)_\nu(-\Delta_1+k_2)_\beta(-\Delta_1+k_3)_\xi]\nonumber\\
&&+\epsilon_{\nu\beta\lambda\alpha}[(-\Delta_1+k_3)_\alpha(-\Delta_1+k_2)_\beta(-\Delta_1+k_1)_\mu]\nonumber\\
&&+\epsilon_{\mu\alpha\nu\beta}[(-\Delta_1+k_1)_\alpha(-\Delta_1+k_2)_\beta(-\Delta_1+k_3)_\lambda]\nonumber\\
&&+\epsilon_{\mu\alpha\nu\beta}[(-\Delta_1+k_1)_\alpha(-\Delta_1+k_3)_\beta(-\Delta_1+k_2)_\lambda ]\}\nonumber\\
&&-2\epsilon_{\mu\nu\alpha\lambda}m_i^2(k_2+k_3-2k_1)_\alpha\nonumber\\
&&-\epsilon_{\mu\nu\alpha\lambda}(k_2-k_3)^2(-\Delta_1+k_1)_\alpha\bigg\}
\int\frac{d^4k}{(2\pi)^2}\frac{1}{[k^2-M_{i}^2]^3}
\end{eqnarray}

By taking the useful relations given in App.~B and also in App.~A
with $\mu_s=0, m=0$ and $M_c\rightarrow\infty$, we have
\begin{eqnarray}
(k_3-k_1)_\mu
T^{R,(1),A\{VV\}}_{\lambda\mu\nu}&=&\sum_{i=0}^2C_i(\frac{-i}{8\pi^2})
\epsilon_{\nu\lambda\alpha\beta}(k_1-k_2)_\alpha(k_3-k_1)_\beta=0\\
(k_1-k_2)_\nu
T^{R,(1),A\{VV\}}_{\lambda\mu\nu}&=&\sum_{i=0}^2C_i(\frac{i}{8\pi^2})
\epsilon_{\mu\lambda\alpha\beta}(k_1-k_2)_\alpha(k_3-k_1)_\beta=0\\
(k_3-k_2)_\lambda
T^{R,(1),A\{VV\}}_{\lambda\mu\nu}&=&\sum_{i=0}^2C_i[2m_iT^{(1),PVV}_{\mu\nu}(m_i)]
\end{eqnarray}
which shows that in Pauli-Villars regularization the vector Ward
identities are preserved automatically. Although this is the same
as eqs.(\ref{A-V-Vpaulivector1}-\ref{A-V-Vpauliaxialvector1})
obtained by using the general trace relation (\ref{trace10gamma}),
while the mechanism becomes different. In the present case, there
are anomalies in both the original vector currents and the
regulator vector currents of super-heavy fields, but the sum of
their anomalous terms cancel each other. In the previous case, the
vector currents of both the original fermion and regulator
fermions are conserved separately.

For axial-vector Ward identity, by using the same condition as
(\ref{limitPauli}), we have
\begin{eqnarray}
\sum_{i=0}^2C_i[2m_iT^{(1),PVV}_{\mu\nu}(m_i)]
&=&\frac{i}{4\pi^2}\epsilon_{\mu\nu\alpha\beta}(k_1-k_2)_\alpha(k_3-k_1)_\beta
\end{eqnarray}
It is seen again that the source of anomaly in Pauli-Villars
regularization arises from the heavy regulator fermion loops.

\subsubsection{Calculation of anomaly by using dimensional regularization}

\label{sec VVDRmassless}

By using (\ref{gammasymmetricvectors}) and considering the
amplitude (\ref{T=Tp+TP}), one can write the parallel part of the
amplitude (\ref{Tparalell}) as
\begin{eqnarray}
T^{R,(1),A\{VV\}}_{\parallel,L,\lambda\mu\nu}&=&-4\int\frac{d^nk}{(2\pi)^n}
\bigg\{\{g_{\mu\nu}\epsilon_{\alpha\beta\lambda\xi}
(k_\parallel+k_1)_\alpha(k_\parallel+k_2)_\beta(k_\parallel+k_3)_\xi\nonumber\\
&&-\epsilon_{\nu\beta\lambda\xi}(k_\parallel+k_1)_\mu(k_\parallel+k_2)_\beta(k_\parallel+k_3)_\xi
-\epsilon_{\mu\beta\lambda\xi}(k_\parallel+k_1)_\nu(k_\parallel+k_2)_\beta(k_\parallel+k_3)_\xi\nonumber\\
&&-\epsilon_{\mu\alpha\nu\beta}(k_\parallel+k_1)_\alpha(k_\parallel+k_2)_\beta(k_\parallel+k_3)_\lambda
-\epsilon_{\mu\alpha\nu\xi}(k_\parallel+k_1)_\alpha(k_\parallel+k_3)_\xi(k_\parallel+k_2)_\lambda\}\nonumber\\
&&\times\bigg[\frac{1}{(k+k_1)^2(k+k_2)^2(k+k_3)^2}\bigg]\nonumber\\
&&+\frac{\epsilon_{\mu\alpha\nu\lambda}}{2}\bigg[\frac{(k_\parallel+k_1)_\alpha}{(k+k_1)^2(k+k_2)^2}
+\frac{(k_\parallel+k_1)_\alpha}{(k+k_1)^2(k+k_3)^2}\bigg]\bigg\}\label{2dimTAVVL}\\
T^{(1),A\{VV\}}_{\parallel,C,\lambda\mu\nu}&=&-2\epsilon_{\mu\nu\alpha\lambda}
\int\frac{d^nk}{(2\pi)^n}\frac{(k_2-k_3)^2(k_\parallel+k_1)_\alpha}{(k+k_1)^2(k+k_2)^2(k+k_3)^2}\label{2dimTAVVC}
\end{eqnarray}

By adopting Feynman parametrization, shifting the integration
variables and using App.~C., we obtain
\begin{eqnarray}
T^{(1),A\{VV\}}_{\parallel,\lambda\mu\nu}&=&T^{(1),A\{VV\}}_{\parallel,0,
\lambda\mu\nu}+
T^{(1),A\{VV\}}_{\parallel,-2,\lambda\mu\nu} \nonumber \\
T^{(1),A\{VV\}}_{\parallel,0,\lambda\mu\nu}&=&2\int^1_0dx_1\int_0^{x_1}dx_2
\epsilon_{\mu\lambda\nu\beta}(k_2+k_3-2k_1)_\beta{i\over16\pi^2}[-\ln M^2]\nonumber\\
&&-2\epsilon_{\mu\lambda\nu\alpha}\{\int_0^1dxx(k_2-k_1)_\alpha {i\over16\pi^2}[-\ln M_1^2]\nonumber\\
&&+\int_0^1dxx(k_3-k_1)_\alpha {i\over16\pi^2}[-\ln M_3^2]\} \nonumber \\
T^{R,(1),A\{VV\}}_{\parallel,-2,\lambda\mu\nu}&=&-4\int^1_0dx_1\int_0^{x_1}dx_2\nonumber\\
&&\times\bigg\{-2\{\epsilon_{\nu\beta\lambda\xi}[(-\Delta_1+k_1)_\mu(-\Delta_1+k_2)_\beta(-\Delta_1+k_3)_\xi]\nonumber\\
&&+\epsilon_{\mu\beta\lambda\alpha}[(-\Delta_1+k_3)_\alpha(-\Delta_1+k_2)_\beta(-\Delta_1+k_1)_\nu]\nonumber\\
&&+\epsilon_{\mu\alpha\nu\beta}[(-\Delta_1+k_1)_\alpha(-\Delta_1+k_2)_\beta(-\Delta_1+k_3)_\lambda]\nonumber\\
&&+\epsilon_{\mu\alpha\nu\beta}[(-\Delta_1+k_1)_\alpha(-\Delta_1+k_3)_\beta(-\Delta_1+k_2)_\lambda ]\}\nonumber\\
&&+\epsilon_{\mu\nu\alpha\lambda}\{(k_2-k_3)^2(-\Delta_1+k_1)_\alpha\}\bigg\}\frac{-i}{32\pi^2}{1\over
M^2}\nonumber\\
&&+{i\over32\pi^2}\epsilon_{\mu\lambda\nu\beta}(k_2+k_3-2k_1)_\beta
\end{eqnarray}
where $T^{(1),AVV}_{\parallel,0,\lambda\mu\nu}$ and
$T^{R,(1),AVV}_{\parallel,-2,\lambda\mu\nu}$ are corresponding to
the logarithemic and convergent terms in (\ref{2dimTAVVL}) and
(\ref{2dimTAVVC}). $x$ and $x_i$ are Feynman parameters. By using
the relations given in App.~A. with $\mu_s^2=0$ and
$M_c^2\rightarrow\infty$, we find that the contributions from the
$k_\parallel$ have the following forms
\begin{eqnarray}
(k_3-k_1)_\mu
T^{R,(1),A\{VV\}}_{\parallel,\lambda\mu\nu}&=&{10\over16}\frac{-i}{4\pi^2}
\epsilon_{\nu\lambda\alpha\beta}(k_1-k_2)_\alpha(k_3-k_1)_\beta\nonumber\\
(k_1-k_2)_\nu
T^{R,(1),A\{VV\}}_{\parallel,\lambda\mu\nu}&=&{10\over16}\frac{i}{4\pi^2}
\epsilon_{\mu\lambda\alpha\beta}(k_1-k_2)_\alpha(k_3-k_1)_\beta\nonumber\\
(k_3-k_2)_\lambda T^{R,(1),A\{VV\}}_{\parallel,\lambda\mu
\nu}&=&{4\over16}{i\over4\pi^2}\epsilon_{\mu\lambda\nu\beta}(k_1-k_2)_\lambda(k_3-k_1)_\beta
\end{eqnarray}

By including the perpendicular part, we yield the following
results
\begin{eqnarray}
(k_3-k_1)_\mu
T^{R,(1),A\{VV\}}_{\lambda\mu\nu}&=&{14\over48}\frac{-i}{4\pi^2}
\epsilon_{\nu\lambda\alpha\beta}(k_1-k_2)_\alpha(k_3-k_1)_\beta\nonumber\\
(k_1-k_2)_\nu
T^{R,(1),A\{VV\}}_{\lambda\mu\nu}&=&{14\over48}\frac{i}{4\pi^2}
\epsilon_{\mu\lambda\alpha\beta}(k_1-k_2)_\alpha(k_3-k_1)_\beta\nonumber\\
(k_3-k_2)_\lambda T^{R,(1),A\{VV\}}_{\lambda\mu
\nu}&=&{20\over48}\frac{i}{4\pi^2}
\epsilon_{\mu\nu\lambda\beta}(k_1-k_2)_\lambda(k_3-k_1)_\beta
\end{eqnarray}
which shows that it is similar to the previous case indicated in
(\ref{A-V-Vdimensionvector1}-\ref{A-V-Vdimensionaxialvector}),
namely all three currents have anomalies, but their magnitudes are
different as seen from the numerical coefficients.

Including the cross diagrams, we arrive at the final results
\begin{eqnarray}
(k_3-k_1)_\mu
T^{R,A\{VV\}}_{\lambda\mu\nu}&=&{14\over24}\frac{-i}{4\pi^2}
\epsilon_{\nu\lambda\alpha\beta}(k_1-k_2)_\alpha(k_3-k_1)_\beta\nonumber\\
(k_1-k_2)_\nu
T^{R,A\{VV\}}_{\lambda\mu\nu}&=&{14\over24}\frac{i}{4\pi^2}
\epsilon_{\mu\lambda\alpha\beta}(k_1-k_2)_\alpha(k_3-k_1)_\beta\nonumber\\
(k_3-k_2)_\lambda
T^{R,A\{VV\}}_{\lambda\mu\nu}&=&{20\over24}\frac{i}{4\pi^2}
\epsilon_{\mu\nu\lambda\beta}(k_1-k_2)_\lambda(k_3-k_1)_\beta
\end{eqnarray}
Again both the vector Ward identities and axial-vector identity
are violated. To keep the vector current be conserved, the
following redefinition for the amplitudes is needed
\begin{eqnarray}
\widetilde{T}^{(1),A\{VV\}}_{\lambda\mu\nu}(k_3-k_1,k_1-k_2)
=T^{(1),A\{VV\}}_{\lambda\mu\nu}(k_3-k_1,k_1-k_2)-T^{(1),A\{VV\}}_{\lambda\mu\nu}(0)\label{VV-redefdimensionR-massless}
\end{eqnarray}
with
\begin{eqnarray}
T^{(1),A\{VV\}}_{\lambda\mu\nu}(0)={14\over24}{-i\over4\pi^2}
\epsilon_{\mu\nu\lambda\alpha}[(k_3-k_1)_\alpha-(k_1-k_2)_\alpha]
\end{eqnarray}

One can check that the redefined amplitude has the Ward identity
anomaly only in the axial-vector current
\begin{eqnarray}
(k_3-k_1)_\mu
\widetilde{T}^{A\{VV\}}_{\lambda\mu\nu}&=&0\\
(k_1-k_2)_\nu
\widetilde{T}^{A\{VV\}}_{\lambda\mu\nu}&=&0\\
(k_3-k_2)_\lambda
\widetilde{T}^{A\{VV\}}_{\lambda\mu\nu}&=&-{i\over2\pi^2}
\epsilon_{\mu\nu\alpha\beta}(k_3-k_1)_\alpha(k_1-k_2)_\beta
\end{eqnarray}

From the above explicit calculations, it is seen that in the
dimensional regularization the divergences of both the vector and
axial-vector currents are violated by quantum corrections. The
quantum corrections also depend on both the original four
dimensions and the extended dimensions. This is similar to the
case where the trace was evaluated by using the definition of
$\gamma_5$ explicitly. However, the magnitudes of anomalous terms
are different in these two treatments, so that the redefinitions
of the amplitudes are also different, although the final form is
the same when normalizing to the conserved vector currents.

\subsection{Calculation of anomaly by using (\ref{vector-axialvectorgamma}) in massless QED}

\label{subsec AVmassless}

 Instead of treating the trace of gamma matrix with the relation
(\ref{vector-axialvectorgamma}), we now consider in this
subsection the case that the trace of gamma matrix is evaluated
with the relation (\ref{vector-axialvectorgamma}).

After Dirac algebra, we can write $T^{(1),AVV}_{\lambda\mu\nu}$ as
\begin{eqnarray}
T^{(1),\{AV\}V}_{\lambda\mu\nu}&=&T^{(1),\{AV\}V}_{L,\lambda\mu\nu}+T^{(1),\{AV\}V}_{C,\lambda\mu\nu}\nonumber\\
T^{(1),\{AV\}V}_{L,\lambda\mu\nu}&=&-4\int\frac{d^4k}{(2\pi)^4}
\bigg\{\{g_{\nu\lambda}\epsilon_{\alpha\beta\mu\xi}(k+k_1)_\alpha(k+k_2)_\beta(k+k_3)_\xi\nonumber\\
&&+\epsilon_{\nu\beta\mu\xi}(k+k_1)_\beta(k+k_2)_\lambda(k+k_3)_\xi
+\epsilon_{\lambda\beta\mu\xi}(k+k_1)_\beta(k+k_2)_\nu(k+k_3)_\xi\nonumber\\
&&+\epsilon_{\lambda\alpha\nu\beta}(k+k_1)_\beta(k+k_2)_\alpha(k+k_3)_\mu
+\epsilon_{\lambda\alpha\nu\xi}(k+k_1)_\mu(k+k_2)_\alpha(k+k_3)_\xi\}\nonumber\\
&&\times\bigg[\frac{1}{(k+k_1)^2(k+k_2)^2(k+k_3)^2}\bigg]\nonumber\\
&&-\frac{\epsilon_{\lambda\alpha\nu\mu}}{2}\bigg[\frac{(k+k_2)_\alpha}{(k+k_2)^2(k+k_3)^2}
+\frac{(k+k_2)_\alpha}{(k+k_2)^2(k+k_1)^2}\bigg]\bigg\}\label{linearTAVV-1}\\
T^{(1),\{AV\}V}_{C,\lambda\mu\nu}&=&-2\epsilon_{\lambda\alpha\nu\mu}\int\frac{d^4k}{(2\pi)^4}
\frac{(k_3-k_1)^2(k+k_2)_\alpha}{(k+k_1)^2(k+k_2)^2(k+k_3)^2}\label{converTAVV-1}
\end{eqnarray}
Again we will check the Ward identities by using three
regularization schemes.

\subsubsection{Calculation of anomaly in the loop regularization}

\label{subsubsec AVLRmassless}

 Applying for the loop regularization, we have after some basic
 algebra that
\begin{eqnarray}
T^{R,(1),\{AV\}V}_{\lambda\mu\nu}&=&T^{(1),\{AV\}V}_{0
\lambda\mu\nu}+T^{(1),\{AV\}V}_{-2\lambda\mu\nu} \nonumber \\
T^{R,(1),\{AV\}V}_{0\lambda\mu\nu}& = &
2\int^1_0dx_1\int_0^{x_1}dx_2
\{\epsilon_{\mu\lambda\nu\beta}(k_3+k_1-2k_2)_\beta I_0^R(x_i,\mu)\}\nonumber\\
&&-2\epsilon_{\mu\alpha\nu\lambda}\{\int_0^1dx(-\Delta_2+k_2)_\alpha
I_0^R(x,\mu_2)+\int_0^1dx(-\Delta_1+k_2)_\alpha
I_0^R(x,\mu_1)\}\nonumber\\
T^{R,(1),\{AV\}V}_{-2\lambda\mu\nu} & = &
-4\int^1_0dx_1\int_0^{x_1}dx_2 \times\bigg\{-2\{\epsilon_{\mu\beta\nu\xi}
[(-\Delta+k_1)_\beta(-\Delta+k_2)_\lambda(-\Delta+k_3)_\xi]\nonumber\\
&&+\epsilon_{\mu\beta\lambda\xi}[(-\Delta+k_1)_\beta(-\Delta+k_2)_\nu(-\Delta+k_3)_\xi]\nonumber\\
&&+\epsilon_{\nu\alpha\lambda\beta}[(-\Delta+k_1)_\beta(-\Delta+k_2)_\alpha(-\Delta+k_3)_\mu]\nonumber\\
&&+\epsilon_{\nu\alpha\lambda\xi}[(-\Delta+k_1)_\mu(-\Delta+k_3)_\xi(-\Delta+k_2)_\alpha]\}\nonumber\\
&&+\epsilon_{\lambda\alpha\nu\mu}\{(k_3-k_1)^2(-\Delta+k_2)_\alpha\}\bigg\}I_{-2}^R(x_i,\mu)
\end{eqnarray}
From the definitions of $\Delta$ and $\Delta_i$, we have
\begin{eqnarray}
(k_1-k_2)_\nu
T^{R,(1),\{AV\}V}_{\lambda\mu\nu}&=&-2\epsilon_{\nu\lambda\mu\xi}(k_1-k_2)_\nu(k_3-k_1)_\xi\nonumber\\
&&\times\bigg\{4[(k_1-k_2)^2I_{-2,(01)}-(k_1-k_2)^2I_{-2,(02)}\nonumber\\
&&+(k_3-k_1)\cdot(k_1-k_2)I_{-2,(11)}]+2(k_3-k_1)^2I_{-2,(10)}\bigg\}\label{vectorA-V}\\
&&-2\epsilon_{\nu\lambda\mu\xi}(k_1-k_2)_\nu(k_3-k_1)_\xi\{I_{0,(00)}+{i\over32\pi^2}Y((p+q)^2,q^2)\}\nonumber\\
(k_3-k_2)_\lambda T^{R,(1),\{AV\}V}_{\lambda\mu\nu}&=&-2\epsilon_{\nu\lambda\mu\xi}(k_1-k_2)_\nu(k_3-k_1)_\xi\nonumber\\
&&\times\bigg\{4\{(k_1-k_2)^2[I_{-2,01}-I_{-2,(02)}]\nonumber\\
&&+(k_1-k_2)\cdot(k_3-k_1)[I_{-2,(01)}-I_{-2,(02)}+I_{-2,(11)}]\nonumber\\
&&+(k_3-k_1)^2I_{-2,(11)}\}\nonumber\\
&&+2\{(k_3-k_1)^2[I_{-2,(00)}-I_{-2,(01)}-I_{-2,(10)}\}\bigg\}\nonumber\\
&&-2\epsilon_{\nu\lambda\mu\xi}(k_1-k_2)_\nu(k_3-k_1)_\xi
I^\prime_{0,(00)}
\end{eqnarray}
where the factor $(i/32\pi^2)Y((p+q)^2,q^2)$ in the last term of
$(\ref{vectorA-V})$ comes from:
\begin{eqnarray}
\int_0^1
dxx[I_0^R(x,\mu_3)-I_0^R(x,\mu_2)]={i\over16\pi^2}{1\over2}Y((p+q)^2,q^2)
\end{eqnarray}

With the help of the relations given in App.A. at $m^2=0$, the
above results are simplified to be
\begin{eqnarray}
(k_1-k_2)_\nu
T^{R,(1),\{AV\}V}_{\lambda\mu\nu}&=&-8\mu_s^2\epsilon_{\nu\lambda\mu\xi}(k_1-k_2)_\nu(k_3-k_1)_\xi I_{-2,(00)}\\
&&-{i\over4\pi^2}\epsilon_{\nu\lambda\mu\xi}(k_1-k_2)_\nu(k_3-k_1)_\xi
e^{-\mu_s^2/M_c^2}\int_0^1dx_1\int_0^{x_1}dx_2e^{-M^2/M_c^2}\nonumber\\
(k_3-k_2)_\lambda
T^{R,(1),\{AV\}V}_{\lambda\mu\nu}&=&8\mu_s^2\epsilon_{\nu\lambda\mu\xi}(k_1-k_2)_\lambda(k_3-k_1)_\xi I_{-2,(00)}\\
&&+{i\over4\pi^2}\epsilon_{\nu\lambda\mu\xi}(k_1-k_2)_\lambda(k_3-k_1)_\xi
e^{-\mu_s^2/M_c^2}\int_0^1dx_1\int_0^{x_1}dx_2e^{-M^2/M_c^2}\nonumber
\end{eqnarray}
It is seen that the first identity is the same as
(\ref{V-Vvector1}) except a total sign, but different from
(\ref{A-V-Vvector1}). The second identity is different from either
eq.(\ref{divAVV}) or eq.(\ref{SymmaxialWard}).

By making the similar evaluation, it is easy to check the Ward
identity of the remaining vector current. We find that
\begin{eqnarray}
(k_3-k_1)_\mu T^{R,(1),\{AV\}V}_{\lambda\mu\nu}&=&0
\end{eqnarray}
which shows that the Ward identity of this vector current is
preserved automatically, which is the same as eq.
(\ref{A-V-Vvector2}) but different from eq. (\ref{V-V-vector2}).

Following the previous subsection, we first consider the case with
$\mu_s\neq 0$ in massless QED. As in section III.A. with the
conditions $p^2=(k_1-k_2)^2=0, q^2=(k_3-k_1)^2=0$ and $(p+q)^2=
2p\cdot q $ is soft with $\mu_s^2\gg (p+q^2)$ and $\mu_s^2\ll
M_c^2\to \infty$, the vector current and axial-vector current
become conserved
\begin{eqnarray}
(k_1-k_2)_\nu
T^{R,(1),\{AV\}V}_{\lambda\mu\nu}&=&0\\
(k_3-k_2)_\lambda T^{R,(1),\{AV\}V}_{\lambda\mu\nu}&=&0
\end{eqnarray}
which means that in the calculation with the trace relation
(\ref{vector-axialvectorgamma}), all the three Ward identities,
under the condition that $p^2, q^2, (p+q)^2 \ll \mu_s^2 \ll
M_c^2$, are conserved by the quantum corrections.

It is noticed that the conclusions under such a condition are the
same for all the three treatments. Nevertheless, one may see that
there exist subtle differences among the three treatments: in the
treatment with trace relation (\ref{twovectorgamma}), the
axial-vector is conserved automatically; in the treatment with
relation (\ref{vector-axialvectorgamma}), one of the vector
current becomes conserved automatically; in the treatment with
relation (\ref{trace10gamma}), two vector currents are conserved
automatically.

We now consider the second case $\mu_s=0$. In this case, three
Ward identities of the vector and axial-vector currents are found
to have the following forms
\begin{eqnarray}
(k_3-k_1)_\mu T^{R,(1),\{AV\}V}_{\lambda\mu\nu}&=&0\\
(k_1-k_2)_\nu
T^{R,(1),\{AV\}V}_{\lambda\mu\nu}&=&-{i\over4\pi^2}\epsilon_{\nu\lambda\mu\xi}(k_1-k_2)_\nu(k_3-k_1)_\xi
\int_0^1dx_1\int_0^{x_1}dx_2e^{-M^2/M_c^2}\\
(k_3-k_2)_\lambda
T^{R,(1),\{AV\}V}_{\lambda\mu\nu}&=&{i\over4\pi^2}\epsilon_{\nu\lambda\mu\xi}(k_1-k_2)_\lambda(k_3-k_1)_\xi
\int_0^1dx_1\int_0^{x_1}dx_2e^{-M^2/M_c^2}
\end{eqnarray}
Taking the UV cut-off $M_c$ to be infinity, i.e.,
$M_c\rightarrow\infty$, the Ward identities become
\begin{eqnarray}
(k_3-k_1)_\mu T^{R,(1),\{AV\}V}_{\lambda\mu\nu}&=&0\\
(k_1-k_2)_\nu
T^{R,(1),\{AV\}V}_{\lambda\mu\nu}&=&-{i\over8\pi^2}\epsilon_{\nu\lambda\mu\xi}(k_1-k_2)_\nu(k_3-k_1)_\xi\\
(k_3-k_2)_\lambda
T^{R,(1),\{AV\}V}_{\lambda\mu\nu}&=&{i\over8\pi^2}\epsilon_{\nu\lambda\mu\xi}(k_1-k_2)_\lambda(k_3-k_1)_\xi
\end{eqnarray}
which indicates that when taking the IR cut-off scale $\mu_s=0$
and UV cut-off scale $M_c\to \infty$ in the loop regularization,
and grouping the gamma with the Lorentz indices of one of vector
currents and axial-vector current to reduce the number of gamma in
the trace, the anomalies appear in the two grouped currents, the
remaining vector current is conserved.

To have two vectors conserved, the physical amplitude should be
redefined as
\begin{eqnarray}
\tilde{T}^{(1),\{AV\}V}_{\lambda\mu\nu}(k_3-k_1,k_1-k_2)
&=&T^{R,(1),\{AV\}V}_{\lambda\mu\nu}(k_3-k_1,k_1-k_2)-T^{(1),\{AV\}V}_{\lambda\mu\nu}(0)\label{redefA-V}
\end{eqnarray}
with
\begin{eqnarray}
T^{(1),\{AV\}V}_{\lambda\mu\nu}(0)=-{i\over
8\pi^2}\epsilon_{\nu\lambda\mu\xi}(k_3-k_1)_\xi
\end{eqnarray}
Then the redefined amplitude satisfies the following Ward
identities
\begin{eqnarray}
(k_3-k_1)_\mu \tilde{T}^{(1),\{AV\}V}_{\lambda\mu\nu}&=&0\\
(k_1-k_2)_\nu
\tilde{T}^{(1),\{AV\}V}_{\lambda\mu\nu}&=&0\\
(k_3-k_2)_\lambda
\tilde{T}^{(1),\{AV\}V}_{\lambda\mu\nu}&=&{i\over4\pi^2}\epsilon_{\nu\lambda\mu\xi}(k_1-k_2)_\lambda(k_3-k_1)_\xi
\end{eqnarray}
By including the cross diagram, we get the standard form of Ward
identity anomaly
\begin{eqnarray}
(k_3-k_1)_\mu \tilde{T}^{(1),\{AV\}V}_{\lambda\mu\nu}&=&0\\
(k_1-k_2)_\nu
\tilde{T}^{(1),\{AV\}V}_{\lambda\mu\nu}&=&0\\
(k_3-k_2)_\lambda
\tilde{T}^{(1),\{AV\}V}_{\lambda\mu\nu}
&=&-{i\over2\pi^2}\epsilon_{\mu\nu\alpha\beta}(k_3-k_1)_\alpha(k_1-k_2)_\beta
\end{eqnarray}

It is seen that in the treatment with relation
(\ref{vector-axialvectorgamma}), the IR scale $\mu_s$ plays the
same role as the other two cases. Comparing with the treatment
with relation (\ref{twovectorgamma}), one observes that the Ward
identity anomalies always appear in the currents with the Lorentz
indices classified in the group which was treated with relation
(\ref{gammaidentity}), in the perturbative calculations of
triangle anomaly, how the Ward identity anomalies depend on the
treatments for the trace of gamma matrices. After comparing all
the three treatments for the trace of gamma matrices in the loop
regularization, it is not difficult to arrive at the conclusion:
In order to obtain a unique and right solution for the Ward
identity anomaly, one shall make the most general treatment for
the trace of gamma matrices, namely, by using the definition of
$\gamma_5$.

\subsubsection{Calculation of anomaly by using Pauli-Villars regularization}

\label{subsubsec AVPVmassless}

In Pauli-Villars regularization, the regularized version of
$(\ref{linearTAVV-1})$ and $(\ref{converTAVV-1})$ are given by
\begin{eqnarray}
T^{R,(1),\{AV\}V}_{L,\lambda\mu\nu}&=&-4\sum_{i=0}^2C_i\int\frac{d^4k}{(2\pi)^4}
\bigg\{\{g_{\nu\lambda}\epsilon_{\alpha\beta\mu\xi}(k+k_1)_\alpha(k+k_2)_\beta(k+k_3)_\xi\nonumber\\
&&+\epsilon_{\nu\beta\mu\xi}(k+k_1)_\beta(k+k_2)_\lambda(k+k_3)_\xi
+\epsilon_{\lambda\beta\mu\xi}(k+k_1)_\beta(k+k_2)_\nu(k+k_3)_\xi\nonumber\\
&&+\epsilon_{\lambda\alpha\nu\beta}(k+k_1)_\beta(k+k_2)_\alpha(k+k_3)_\mu
+\epsilon_{\lambda\alpha\nu\xi}(k+k_1)_\mu(k+k_2)_\alpha(k+k_3)_\xi\}\nonumber\\
&&\times\bigg[\frac{1}{[(k+k_1)^2-m_i^2][(k+k_2)^2-m_i^2][(k+k_3)^2-m_i^2]}\bigg]\nonumber\\
&&-\frac{\epsilon_{\lambda\alpha\nu\mu}}{2}\bigg[\frac{(k+k_2)_\alpha}{[(k+k_2)^2-m_i^2][(k+k_3)^2-m_i^2]}\nonumber\\
&&~~~~~~~~~~+\frac{(k+k_2)_\alpha}{[(k+k_2)^2-m_i^2][(k+k_1)^2-m_i^2]}\bigg]\bigg\}\nonumber\\
T^{R,(1),\{AV\}V}_{C,\lambda\mu\nu}&=&-2\epsilon_{\lambda\alpha\nu\mu}\sum_{i=0}^2C_i\int\frac{d^4k}{(2\pi)^4}
\frac{1}{[(k+k_1)^2-m_i^2][(k+k_2)^2-m_i^2][(k+k_3)^2-m_i^2]}\nonumber\\
&&~~~~~~~~\times\{2m_i^2(k+k_3)_\alpha-[4m_i^2-(k_3-k_1)^2](k+k_2)_\alpha+2m_i^2(k+k_1)_\alpha\}\nonumber
\end{eqnarray}

Following the same evaluation as done in the previous subsection,
we obtain the following Ward identities
\begin{eqnarray}
(k_3-k_1)_\mu T^{R,(1),\{AV\}V}_{\lambda\mu\nu}&=&0\\
(k_1-k_2)_\nu T^{R,(1),\{AV\}V}_{\lambda\mu\nu}
&=&-{i\over8\pi^2}\sum_{i=0}^2C_i\epsilon_{\nu\lambda\mu\xi}(k_1-k_2)_\nu(k_3-k_1)_\xi=0\nonumber\\
(k_3-k_2)_\lambda T^{R, (1),\{AV\}V}_{\lambda\mu\nu}
&=&-{i\over8\pi^2}\sum_{i=0}^2C_i\epsilon_{\nu\lambda\mu\xi}(k_1-k_2)_\xi(k_3-k_1)_\lambda
+\sum_{i=0}^2C_i[2m_iT_{\mu\nu}^{(1),PVV}(m_i)]\nonumber\\
&=&\sum_{i=0}^2C_i[2m_iT_{\mu\nu}^{(1),PVV}(m_i)]
\end{eqnarray}
which shows that in the treatment with relation
(\ref{vector-axialvectorgamma}), the Ward identities are the same
as the ones in the treatment with relation (\ref{twovectorgamma}).
Although the final results are the same in these two treatments,
while the anomalies in the two vector currents are cancelled by
the heavy regulator field in latter case. Here the remaining
vector current is conserved automatically.

Taking the masses of regulator fermions to be infinity large, we
have
\begin{eqnarray}
\sum_{i=0}^2C_i[2m_iT^{(1),PVV}_{\mu\nu}(m_i)]
&=&-\frac{i}{4\pi^2}\epsilon_{\mu\nu\alpha\beta}(k_3-k_1)_\alpha(k_1-k_2)_\beta
\end{eqnarray}
Here the source of anomaly remains arising from the heavy
regulator fermion loops.

\subsubsection{Calculation of anomaly by using dimensional regularization}

\label{subsubsec AVDRmassless}

By using eq. (\ref{vector-axialvectorgamma}), the regularized
version for the parallel part of the amplitude can be written as
\begin{eqnarray}
T^{R,(1),\{AV\}V}_{\parallel,\lambda\mu\nu}&=&T^{R,(1),\{AV\}V}_{\parallel,0
\lambda\mu\nu}+T^{R,(1),\{AV\}V}_{\parallel,-2\lambda\mu\nu} \nonumber \\
T^{R,(1),\{AV\}V}_{\parallel,0\lambda\mu\nu}& = &
2\int^1_0dx_1\int_0^{x_1}dx_2
\{\epsilon_{\mu\lambda\nu\beta}(k_3+k_1-2k_2)_\beta {i\over16\pi^2}[-\ln M^2]\}\nonumber\\
&&-2\epsilon_{\mu\alpha\nu\lambda}\{\int_0^1dx(-\Delta_2+k_2)_\alpha
{i\over16\pi^2}[-\ln
M_2^2]\nonumber\\
&&+\int_0^1dx(-\Delta_1+k_2)_\alpha
{i\over16\pi^2}[-\ln M_1^2]\}\nonumber\\
T^{R,(1),\{AV\}V}_{\parallel,-2\lambda\mu\nu} & = & -4\int^1_0dx_1\int_0^{x_1}dx_2\nonumber\\
&&\times\bigg\{-2\{\epsilon_{\mu\beta\nu\xi}[(-\Delta+k_1)_\beta(-\Delta+k_2)_\lambda(-\Delta+k_3)_\xi]\nonumber\\
&&+\epsilon_{\mu\beta\lambda\xi}[(-\Delta+k_1)_\beta(-\Delta+k_2)_\nu(-\Delta+k_3)_\xi]\nonumber\\
&&+\epsilon_{\nu\alpha\lambda\beta}[(-\Delta+k_1)_\beta(-\Delta+k_2)_\alpha(-\Delta+k_3)_\mu]\nonumber\\
&&+\epsilon_{\nu\alpha\lambda\xi}[(-\Delta+k_1)_\mu(-\Delta+k_3)_\xi(-\Delta+k_2)_\alpha]\}\nonumber\\
&&+\epsilon_{\lambda\alpha\nu\mu}\{(k_3-k_1)^2(-\Delta+k_2)_\alpha\}\bigg\}{-i\over32\pi^2}\frac{1}{M^2}\nonumber\\
&&+{i\over32\pi^2}\epsilon_{\mu\lambda\nu\beta}(k_3+k_1-2k_2)_\beta
\end{eqnarray}
From the definitions of $\Delta$ and $\Delta_i$ and the relations
given in App.A. and App.C., we obtain the following Ward
identities
\begin{eqnarray}
(k_3-k_1)_\mu T^{(1),\{AV\}V}_{\parallel,\lambda\mu\nu}
&=&-{4\over16}{i\over4\pi^2}\epsilon_{\mu\nu\lambda\alpha}(k_3-k_1)_\mu(k_1-k_2)_\alpha\nonumber\\
(k_1-k_2)_\nu T^{(1),\{AV\}V}_{\parallel,\lambda\mu\nu}
&=&-{10\over16}{i\over4\pi^2}\epsilon_{\nu\lambda\mu\xi}(k_1-k_2)_\nu(k_3-k_1)_\xi\nonumber\\
(k_3-k_2)_\lambda T^{(1),\{AV\}V}_{\parallel,\lambda\mu\nu}
&=&{10\over16}{i\over4\pi^2}\epsilon_{\nu\lambda\mu\xi}(k_1-k_2)_\lambda(k_3-k_1)_\xi
\end{eqnarray}
Combining the result in eq. $(\ref{masslessPerpend})$, we have
\begin{eqnarray}
(k_3-k_1)_\mu T^{(1),\{AV\}V}_{\lambda\mu\nu}
&=&-{14\over24}{i\over4\pi^2}\epsilon_{\mu\nu\lambda\alpha}(k_3-k_1)_\mu(k_1-k_2)_\alpha\nonumber\\
(k_1-k_2)_\nu T^{(1),\{AV\}V}_{\lambda\mu\nu}
&=&-{7\over24}{i\over4\pi^2}\epsilon_{\nu\lambda\mu\xi}(k_1-k_2)_\nu(k_3-k_1)_\xi\nonumber\\
(k_3-k_2)_\lambda T^{(1),\{AV\}V}_{\lambda\mu\nu}
&=&{31\over24}{i\over4\pi^2}\epsilon_{\nu\lambda\mu\xi}(k_1-k_2)_\lambda(k_3-k_1)_\xi
\end{eqnarray}

By including the cross diagram, we arrive at the final results
\begin{eqnarray}
(k_3-k_1)_\mu T^{R,\{AV\}V}_{\lambda\mu\nu}
&=&-{14\over12}{i\over4\pi^2}\epsilon_{\nu\lambda\mu\xi}(k_1-k_2)_\xi(k_3-k_1)_\mu\nonumber\\
(k_1-k_2)_\nu T^{R,\{AV\}V}_{\lambda\mu\nu}
&=&-{7\over12}{i\over4\pi^2}\epsilon_{\nu\lambda\mu\xi}(k_1-k_2)_\nu(k_3-k_1)_\xi\nonumber\\
(k_3-k_2)_\lambda T^{R,\{AV\}V}_{\lambda\mu\nu}
&=&{31\over12}{i\over4\pi^2}\epsilon_{\nu\lambda\mu\xi}(k_1-k_2)_\lambda(k_3-k_1)_\xi
\end{eqnarray}
which indicates that when treating the gamma trace with relation
(\ref{vector-axialvectorgamma}) in the dimensional regularization,
anomalies remain appearing in all the three currents. But the
explicit forms of anomaly are different from the case in which the
gamma trace is treated with relation (\ref{twovectorgamma}) and
all the amplitudes are different from the conclusion based on the
relation (\ref{twovectorgamma}). In addition, in the present case,
the magnitudes of all the three currents are different while in
the latter case the magnitudes of the two vector currents are
same. To require two vectors conserved, the physical amplitude
should be redefined as
\begin{eqnarray}
\tilde{T}^{R,\{AV\}V}_{\lambda\mu\nu}(k_3-k_1,k_1-k_2)
&=&T^{R,\{AV\}V}_{\lambda\mu\nu}(k_3-k_1,k_1-k_2)-T^{\{AV\}V}_{\lambda\mu\nu}(0)
\end{eqnarray}
with
\begin{eqnarray}
T^{\{AV\}V}_{\lambda\mu\nu}(0)=-{7\over12}{i\over
4\pi^2}\epsilon_{\nu\lambda\mu\xi}(k_3-k_1)_\xi-{14\over12}{i\over4\pi^2}
\epsilon_{\nu\lambda\mu\xi}(k_1-k_2)_\xi\label{redefinitionV-Adimension}
\end{eqnarray}

With the above redefinition, one then gets the standard form of
anomaly
\begin{eqnarray}
(k_3-k_1)_\mu \tilde{T}^{R,\{AV\}V}_{\lambda\mu\nu}&=&0\\
(k_1-k_2)_\nu
\tilde{T}^{R,\{AV\}V}_{\lambda\mu\nu}&=&0\\
(k_3-k_2)_\lambda
\tilde{T}^{R,\{AV\}V}_{\lambda\mu\nu}&=&-{i\over2\pi^2}
\epsilon_{\mu\nu\alpha\beta}(k_3-k_1)_\alpha(k_1-k_2)_\beta
\end{eqnarray}

In conclusion, in the treatment of the gamma trace with relation
$(\ref{vector-axialvectorgamma})$ in dimensional regularization,
the anomalies still appear in both vector and axial-vector Ward
identities. But the anomalies in the three Ward identities are
different from that in treatment with relation
$(\ref{twovectorgamma})$, so that the redefinition of the
amplitude is also different although both treatments can make the
vector currents conserved and the axial-vector Ward identity is
violated by quantum corrections.

\subsection{Calculation of anomaly with relation (\ref{twovectorgamma}) in massive QED}

\label{subsec VVmassive}

\subsubsection{Calculation of anomaly in the loop regularization}

\label{subsubsec VVLRmassive}

As in the massless case, we first consider the case that the trace
of gamma matrix is calculated by using the relation
(\ref{gammasymmetricvectors}) and repeating the same calculations
as done for massless QED, we yield
\begin{eqnarray}
T^{R,(1),A\{VV\}}_{0\lambda\mu\nu}&=&2\int^1_0dx_1\int_0^{x_1}dx_2\{\epsilon_{\mu\lambda\nu\beta}
(k_2+k_3-2k_1)_\beta I_0^R(x_i,\mu)\}\nonumber\\
&&-2\epsilon_{\mu\alpha\nu\lambda}\{\int_0^1dx(-\Delta_1+k_1)_\alpha
I_0^R(x,\mu_1)+\int_0^1dx(-\Delta_3+k_1)_\alpha
I_0^R(x,\mu_3)\}\nonumber\\
T^{R,(1),A\{VV\}}_{-2\lambda\mu\nu}&=&-4\int^1_0dx_1\int_0^{x_1}dx_2\nonumber\\
&&\times\bigg\{-2\{\epsilon_{\nu\beta\lambda\xi}[(-\Delta_1+k_1)_\mu(-\Delta_1+k_2)_\beta(-\Delta_1+k_3)_\xi]\nonumber\\
&&+\epsilon_{\mu\beta\lambda\alpha}[(-\Delta_1+k_3)_\alpha(-\Delta_1+k_2)_\beta(-\Delta_1+k_1)_\nu]\nonumber\\
&&+\epsilon_{\mu\alpha\nu\beta}[(-\Delta_1+k_1)_\alpha(-\Delta_1+k_2)_\beta(-\Delta_1+k_3)_\lambda]\nonumber\\
&&+\epsilon_{\mu\alpha\nu\beta}[(-\Delta_1+k_1)_\alpha(-\Delta_1+k_3)_\beta(-\Delta_1+k_2)_\lambda
]\}\nonumber\\
&&+2\epsilon_{\mu\nu\alpha\lambda}\{m^2(k_2+k_3-2k_1)_\alpha\}\nonumber\\
&&+\epsilon_{\mu\nu\alpha\lambda}\{(k_2-k_3)^2(-\Delta_1+k_1)_\alpha\}\bigg\}I_{-2}^R(x_i,\mu)
\end{eqnarray}

The Ward identity for vector current is given by
\begin{eqnarray}
(k_1-k_2)_\nu T^{R,(1),A\{VV\}}_{\lambda\mu\nu}&=&-2\epsilon_{\mu\nu\lambda\beta}(k_1-k_2)_\nu(k_3-k_1)_\beta\nonumber\\
&&\times\bigg\{4(k_3-k_1)\cdot(k_1-k_2)[I_{-2,(11)}+I_{-2,(20)}-I_{-2,10}]\nonumber\\
&&-4(k_2-k_1)^2[I_{-2,(02)}+I_{-2,(11)}-I_{-2,(01)}]\nonumber\\
&&-4m^2I_{-2,(00)}+2(k_2-k_3)^2I_{-2,(10)}\bigg\}\nonumber\\
&&-2\epsilon_{\mu\nu\lambda\beta}(k_1-k_2)_\nu(k_3-k_1)_\beta
I_{0,(00)}
\end{eqnarray}

By adopting the relations given in App.~A, we have
\begin{eqnarray}
(k_1-k_2)_\nu
T^{R,(1),A\{VV\}}_{\lambda\mu\nu}&=&-2\epsilon_{\mu\nu\lambda\beta}(k_1-k_2)_\nu(k_3-k_1)_\beta\nonumber\\
&&\times\bigg\{\{{i\over16\pi^2}[e^{-(m^2+\mu_s^2)/M_c^2}\int_0^1dx_1\int_0^{x_1}dx_2
e^{-M^2/M_c^2}-{1\over2}Y((p+q)^2,q^2)]\nonumber\\
&&-q^2I_{-2,(10)}-p^2I_{-2,(01)}+2(m^2+\mu_s^2)I_{-2,(00)}\}\nonumber\\
&&-4m^2I_{-2,(00)}+2(p+q)^2I_{-2,(10)}\\
&&+4\{p^2I_{-2,(01)}-p\cdot qI_{-2,(10)}\nonumber\\
&&+{i\over64\pi^2}e^{-(m^2+\mu_s^2)/M_c^2}\int_0^1dx_1\int_0^{x_1}dx_2
e^{-M^2/M_c^2}+{m^2+\mu_s^2\over2}I_{-2,(00)}\nonumber\\
&&-{q^2\over4}I_{-2,(10)}-{3p^2\over4}I_{-2,(01)}+{i\over128\pi^2}Y((p+q)^2,q^2)-{p^2\over2}I_{-2,(10)}\}\bigg\}\nonumber
\end{eqnarray}
With simple algebraic evaluation, the above expression is
simplified to be
\begin{eqnarray}
(k_1-k_2)_\nu
T^{R,(1),A\{VV\}}_{\lambda\mu\nu}&=&8\mu_s^2\epsilon_{\mu\lambda\alpha\beta}(k_1-k_2)_\alpha(k_3-k_1)_\beta
I_{-2,(00)}\nonumber\\
&&+\frac{i}{4\pi^2}\epsilon_{\mu\lambda\alpha\beta}(k_1-k_2)_\alpha(k_3-k_1)_\beta\nonumber\\
&&\times e^{-(m^2 + \mu_s^2)/M_c^2} \int_0^1dx_1\int_0^{x_1}dx_2
e^{-M^{2}/M_c^2}\nonumber
\end{eqnarray}
which is different from (\ref{massiveA-V-V-VWard1}) and the two
superficially logarithemically divergent integrals cancel each
other. Similarly, one has
\begin{eqnarray}
(k_3-k_1)_\mu
T^{R,(1),A\{VV\}}_{\lambda\mu\nu}&=&-8\mu_s^2\epsilon_{\nu\lambda\alpha\beta}(k_1-k_2)_\alpha(k_3-k_1)_\beta
I_{-2,(00)}\nonumber\\
&&-\frac{i}{4\pi^2}\epsilon_{\nu\lambda\alpha\beta}(k_1-k_2)_\alpha(k_3-k_1)_\beta\nonumber\\
&&\times e^{-(m^2 + \mu_s^2)/M_c^2}
\int_0^1dx_1\int_0^{x_1}dx_2e^{-M^{2}/M_c^2}\nonumber
\end{eqnarray}
which is different from (\ref{massiveA-V-V-VWard2}).

For the Ward identity of axial-vector current, we have
\begin{eqnarray}
(k_3-k_2)_\lambda T^{R,(1),A\{VV\}}_{\lambda\mu\nu}&=&
4\epsilon_{\mu\nu\alpha\beta}(k_2-k_1)_\alpha(k_3-k_1)_\beta\nonumber\\
&&\times\bigg\{2(k_2-k_1)\cdot(k_3-k_2)[I_{-2,(20)}-I_{-2,(02)}+I_{-2,(01)}-I_{-2,(10)}]\nonumber\\
&&+2(k_3-k_2)^2[I_{-2,(10)}-I_{-2,(20)}+I_{-2,(11)}]\nonumber\\
&&+4m^2I_{-2,(00)}-(k_2-k_3)^2[I_{-2,(01)}+I_{-2,(10)}]\bigg\}\nonumber\\
&&-2\epsilon_{\mu\nu\lambda\beta}(k_1-k_2)_\lambda(k_3-k_1)_\beta I^\prime_{0,(00)}\nonumber\\
&&-2\epsilon_{\mu\nu\lambda\beta}(k_1-k_2)_\lambda(k_3-k_1)_\beta I_{0,(00)}\nonumber\\
&=&16m^2\epsilon_{\mu\nu\alpha\beta}(k_2-k_1)_\alpha(k_3-k_1)_\beta
I_{-2,(00)}
\end{eqnarray}
which is different from (\ref{massiveA-V-Vaxialvector}).

In comparison with eq.(\ref{PVVresult}), we obtain the following
relation between two amplitudes
\begin{eqnarray}
(k_3-k_2)_\lambda
T^{R,(1),A\{VV\}}_{\lambda\mu\nu}&=&2mT^{R,(1),PVV}_{\mu\nu}
\end{eqnarray}
which shows that the classical axial-vector Ward identity is still
preserved by quantum corrections.

To further evaluate the Ward identity of vector current, we take
the limit $M_c^2\rightarrow\infty$ allowed for the renormalizable
massive QED. Under this limit, the Ward identities are simplified
to be
\begin{eqnarray}
(k_3-k_1)_\mu
T^{R,(1),A\{VV\}}_{\lambda\mu\nu}&=&-8\mu_s^2\epsilon_{\nu\lambda\alpha\beta}(k_1-k_2)_\alpha(k_3-k_1)_\beta
I_{-2,(00)}\nonumber\\
&&-\frac{i}{8\pi^2}\epsilon_{\nu\lambda\alpha\beta}(k_1-k_2)_\alpha(k_3-k_1)_\beta\\
(k_1-k_2)_\nu
T^{R,(1),A\{VV\}}_{\lambda\mu\nu}&=&8\mu_s^2\epsilon_{\mu\lambda\alpha\beta}(k_1-k_2)_\alpha(k_3-k_1)_\beta
I_{-2,(00)}\nonumber\\
&&+\frac{i}{8\pi^2}\epsilon_{\mu\lambda\alpha\beta}(k_1-k_2)_\alpha(k_3-k_1)_\beta\\
(k_3-k_2)_\lambda
T^{R,(1),A\{VV\}}_{\lambda\mu\nu}&=&2mT^{R,(1),PVV}_{\mu\nu} =
16m^2\epsilon_{\mu\nu\alpha\beta}(k_1-k_2)_\alpha(k_3-k_1)_\beta
I_{-2,(00)}
\end{eqnarray}

Firstly, consider the case $\mu_s\neq0$. Similar to the case in
massless QED, if taking the two vector states state to be massless
with the conditions $p^2=q^2=0$ and the axial-vector state
$(p+q)^2$ to be soft, we have
\begin{eqnarray}
(k_3-k_1)_\mu T^{R,(1),A\{VV\}}_{\lambda\mu\nu} &=&0\\
(k_1-k_2)_\nu T^{R,(1),A\{VV\}}_{\lambda\mu\nu}&=&0\\
(k_3-k_2)_\lambda T^{R,(1),A\{VV\}}_{\lambda\mu\nu} & = &
2mT^{R,(1),PVV}_{\mu\nu}
\end{eqnarray}
which indicates that the Ward identities are well preserved in
this case.

In the case that $\mu_s \gg m $, the axial-vector current also
approaches to anomaly free
\begin{eqnarray}
(k_3-k_2)_\lambda T^{R,(1),A\{VV\}}_{\lambda\mu\nu} &=&
2mT^{(1),PVV}_{\mu\nu} \nonumber \\
& = & -{i\over4\pi^2}\frac{m^2}{\mu_s^2 + m^2}
\epsilon_{\mu\nu\alpha\beta}(k_3-k_1)_\alpha(k_1-k_2)_\beta \to 0
\label{massive-mu>>m}
\end{eqnarray}

Generally speaking, in the limit $p^2$,$q^2$,$(p+q)^2$, $m^2
\ll\mu_s^2 \ll M_c^2\to \infty$, the Ward identities for both
currents are preserved in the quantum corrections. Obviously, to
arrive at this conclusion, the IR cut-off scale $\mu_s$ plays an
important role.

Secondly, consider the case that $\mu_s^2=0$, the relevant Ward
identities have the following forms
\begin{eqnarray}
(k_3-k_1)_\mu T^{R,(1),A\{VV\}}_{\lambda\mu\nu}
&=&-\frac{i}{4\pi^2}\epsilon_{\nu\lambda\alpha\beta}(k_1-k_2)_\alpha(k_3-k_1)_\beta
\int_0^1dx_1\int_0^{x_1}dx_2e^{-(m^2 + M^{2})/M_c^2}\\
(k_1-k_2)_\nu T^{R,(1),A\{VV\}}_{\lambda\mu\nu}
&=&\frac{i}{4\pi^2}\epsilon_{\mu\lambda\alpha\beta}(k_1-k_2)_\alpha(k_3-k_1)_\beta
\int_0^1dx_1\int_0^{x_1}dx_2e^{-(m^2 +M^{2})/M_c^2}\\
(k_3-k_2)_\lambda T^{R,(1),A\{VV\}}_{\lambda\mu\nu}
&=&2mT^{R,(1),PVV}_{\mu\nu}
\end{eqnarray}
which indicates that in the loop regularization with $\mu_s = 0$
the vector current is no longer conserved, while the axial-vector
Ward identity is still preserved.

Taking the limit $M_c^2\rightarrow\infty$, the above results are
simplified to be
\begin{eqnarray}
(k_3-k_1)_\mu T^{R,(1),A\{VV\}}_{\lambda\mu\nu}
&=&-\frac{i}{4\pi^2}\epsilon_{\nu\lambda\alpha\beta}(k_1-k_2)_\alpha(k_3-k_1)_\beta\\
(k_1-k_2)_\nu T^{R,(1),A\{VV\}}_{\lambda\mu\nu}
&=&\frac{i}{4\pi^2}\epsilon_{\mu\lambda\alpha\beta}(k_1-k_2)_\alpha(k_3-k_1)_\beta\\
(k_3-k_2)_\lambda T^{R,(1),A\{VV\}}_{\lambda\mu\nu}
&=&2mT^{R,(1),PVV}_{\mu\nu}
\end{eqnarray}
which shows that in this case the well-known anomaly appears in
the vector Ward identity.

Again, by using the redefinition (\ref{massiveV-Vredefi}), we get
the amplitude which makes the vector current conserved and anomaly
lives in the axial-vector Ward identity
\begin{eqnarray}
(k_3-k_1)_\mu
\widetilde{T}^{(1),A\{VV\}}_{\lambda\mu\nu}&=&0\\
(k_1-k_2)_\nu
\widetilde{T}^{(1),A\{VV\}}_{\lambda\mu\nu}&=&0\\
(k_3-k_2)_\lambda \widetilde{T}^{(1),A\{VV\}}_{\lambda\mu\nu}& = &
(I(m,p,q) - 1 ) {i\over4\pi^2}
\epsilon_{\mu\nu\alpha\beta}(k_3-k_1)_\alpha(k_1-k_2)_\beta
\end{eqnarray}
where the integral $I(m,p,q)$ is defined by
(\ref{defineI(m,p,q)}).

In the case that the external vector states are massless with
conditions: $p^2 = 0$, $q^2 = 0$ and the axial-vector state be
soft, using (\ref{special-I-1}), we have
\begin{eqnarray}
 (k_3-k_2)_\lambda
\widetilde{T}^{(1),A\{VV\}}_{\lambda\mu\nu}  = 0
 \label{VV-anomfree}
\end{eqnarray}
Namely both vector and axial-vector receive no contribution from
quantum corrections for massive QED with general conditions
$m^2\gg p^2, q^2, p\cdot q$.

In an alternative case that $p^2 = 0$, $q^2 = 0$, and $(p+q)^2 \gg
m^2 $, by using (\ref{special-I-2}), the axial-vector gets
anomaly.

By including the cross diagrams, the Ward identities are given by
\begin{eqnarray}
(k_3-k_1)_\mu
\widetilde{T}^{A\{VV\}}_{\lambda\mu\nu}&=&0\\
(k_1-k_2)_\nu
\widetilde{T}^{A\{VV\}}_{\lambda\mu\nu}&=&0\\
(k_3-k_2)_\lambda
\widetilde{T}^{A\{VV\}}_{\lambda\mu\nu}&=&2mT^{R,PVV}_{\mu\nu}
-{i\over2\pi^2}\epsilon_{\mu\nu\alpha\beta}(k_3-k_1)_\alpha(k_1-k_2)_\beta
\end{eqnarray}
with the operator forms
\begin{eqnarray}
\partial_\mu V_\mu(x)&=&0,\\
\partial_\mu
A_\mu(x)&=&2imP(x)+{e^2\over8\pi^2}F^{\mu\nu}(x)\widetilde{F}_{\mu\nu}(x)
\end{eqnarray}
which is the standard form of triangle anomaly.

It is seen that for massive QED the IR scale $\mu_s^2$ plays the
same role for two treatments.

\subsubsection{Calculation of anomaly by using Pauli-Villars regularization}

\label{subsubsec VVPVmassive}

With a similar evaluation as the massless case, the regularized
versions are given by
\begin{eqnarray}
T^{R,(1),A\{VV\}}_{L,\lambda\mu\nu}&=&-4\sum_{i=0}^2C_i\int\frac{d^4k}{(2\pi)^4}
\bigg\{\{g_{\mu\nu}\epsilon_{\alpha\beta\lambda\xi}(k+k_1)_\alpha(k+k_2)_\beta(k+k_3)_\xi\nonumber\\
&&-\epsilon_{\nu\beta\lambda\xi}(k+k_1)_\mu(k+k_2)_\beta(k+k_3)_\xi
-\epsilon_{\mu\beta\lambda\xi}(k+k_1)_\nu(k+k_2)_\beta(k+k_3)_\xi\nonumber\\
&&-\epsilon_{\mu\alpha\nu\beta}(k+k_1)_\alpha(k+k_3)_\beta(k+k_2)_\lambda
-\epsilon_{\mu\alpha\nu\xi}(k+k_1)_\alpha(k+k_2)_\xi(k+k_3)_\lambda\}\nonumber\\
&&\times\bigg[\frac{1}{[(k+k_1)^2-m_i^2][(k+k_2)^2-m_i^2][(k+k_3)^2-m_i^2]}\bigg]\nonumber\\
&&+\frac{\epsilon_{\mu\alpha\nu\lambda}}{2}\bigg[\frac{(k+k_1)_\alpha}{[(k+k_1)^2-m_i^2][(k+k_2)^2-m_i^2]}\nonumber\\
&&+\frac{(k+k_1)_\alpha}{[(k+k_1)^2-m_i^2][(k+k_3)^2-m_i^2]}\bigg]\bigg\}\nonumber\\
T^{(1),A\{VV\}}_{C,\lambda\mu\nu}&=&-2\epsilon_{\mu\nu\alpha\lambda}\sum_{i=0}^2C_i\int\frac{d^4k}{(2\pi)^4}\nonumber\\
&&\times\frac{2m_i^2(k+k_2)_\alpha-[4m_i^2-(k_2-k_3)^2](k+k_1)_\alpha
+2m_i^2(k+k_3)_\alpha}{[(k+k_1)^2-m_i^2][(k+k_2)^2-m_i^2][(k+k_3)^2-m_i^2]}\nonumber
\end{eqnarray}
with the initial condition $c_0=1$ and $m_0=m$. After adopting
Feynman parametrization and shifting the integration variables, we
have
\begin{eqnarray}
T^{R,(1),A\{VV\}}_{\lambda\mu\nu} &=&
T^{R,(1),A\{VV\}}_{0\lambda\mu\nu}
+ T^{R,(1),A\{VV\}}_{-2\lambda\mu\nu} \nonumber \\
T^{R,(1),A\{VV\}}_{0\lambda\mu\nu}&=&-2\sum_{i=0}^2C_i\int^1_0dx_1\int_0^{x_1}dx_2
\{\epsilon_{\mu\lambda\nu\beta}(k_2+k_3-2k_1)_\beta\int\frac{d^4k}{(2\pi)^2}\frac{k^2}{[k^2-M_{i}^2]^3}\}\nonumber\\
&&-2\epsilon_{\mu\alpha\nu\lambda}\{\sum_{i=0}^2C_i\int_0^1dx(-\Delta_2+k_1)_\alpha
\int\frac{d^4k}{(2\pi)^4}\frac{1}{[k^2-M_{1i}^2]^2}\nonumber\\
&&+\sum_{i=0}^2C_i\int_0^1dx(-\Delta_3+k_1)_\alpha
\int\frac{d^4k}{(2\pi)^4}\frac{1}{[k^2-M_{3i}^2]^2}\}\nonumber\\
T^{R,(1),A\{VV\}}_{-2\lambda\mu\nu}&=&4\sum_{i=0}^2C_i\int^1_0dx_1\int_0^{x_1}dx_2\nonumber\\
&&\times\bigg\{2\{\epsilon_{\mu\beta\lambda\xi}
[(-\Delta_1+k_1)_\nu(-\Delta_1+k_2)_\beta(-\Delta_1+k_3)_\xi]\nonumber\\
&&+\epsilon_{\nu\beta\lambda\alpha}[(-\Delta_1+k_3)_\alpha(-\Delta_1+k_2)_\beta(-\Delta_1+k_1)_\mu]\nonumber\\
&&+\epsilon_{\mu\alpha\nu\beta}[(-\Delta_1+k_1)_\alpha(-\Delta_1+k_2)_\beta(-\Delta_1+k_3)_\lambda]\nonumber\\
&&+\epsilon_{\mu\alpha\nu\beta}[(-\Delta_1+k_1)_\alpha(-\Delta_1+k_3)_\beta(-\Delta_1+k_2)_\lambda ]\}\nonumber\\
&&-2\epsilon_{\mu\nu\alpha\lambda}m_i^2(k_2+k_3-2k_1)_\alpha\nonumber\\
&&-\epsilon_{\mu\nu\alpha\lambda}(k_2-k_3)^2(-\Delta_1+k_1)_\alpha\bigg\}
\int\frac{d^4k}{(2\pi)^4}\frac{1}{[k^2-M_{i}^2]^3}
\end{eqnarray}

After some algebra and again adopting the useful relations
presented in App.~B and App.~A, we have
\begin{eqnarray}
(k_3-k_1)_\mu
T^{R,(1),A\{VV\}}_{\lambda\mu\nu}&=&\sum_{i=0}^2C_i(\frac{-i}{8\pi^2})
\epsilon_{\lambda\alpha\beta}(k_1-k_2)_\alpha(k_3-k_1)_\beta=0\\
(k_1-k_2)_\nu
T^{R,(1),A\{VV\}}_{\lambda\mu\nu}&=&\sum_{i=0}^2C_i(\frac{i}{8\pi^2})
\epsilon_{\mu\lambda\alpha\beta}(k_1-k_2)_\alpha(k_3-k_1)_\beta=0\\
(k_3-k_2)_\lambda
T^{R,(1),A\{VV\}}_{\lambda\mu\nu}&=&\sum_{i=0}^2C_i[2m_iT^{(1),PVV}_{\mu\nu}(m_i)]
\end{eqnarray}
which is the same as the massless case that the vector ward
identity is automatically preserved. For axial-vector current,
using (\ref{limitPauli}), we have
\begin{eqnarray}
 (k_3-k_2)_\lambda
\widetilde{T}^{(1),A\{VV\}}_{\lambda\mu\nu} & = & (I(m,p,q) - 1 )
{i\over4\pi^2}
\epsilon_{\mu\nu\alpha\beta}(k_3-k_1)_\alpha(k_1-k_2)_\beta
\end{eqnarray}

For the external states with conditions: $p^2 = 0$, $q^2 = 0$, and
$(p+q)^2=2p\cdot q$ being soft with $m^2\gg p\cdot q$, one has
from eq.(\ref{special-I-1}) both vector and axial-vector become
anomaly free in massive QED.

Considering the alternative case that $p^2 = 0$, $q^2 = 0$, and
$(p+q)^2 \gg m^2 $ and eq.(\ref{special-I-2}), then the
axial-vector gets anomaly. It is seen that Pauli-Villars scheme
leads to the same conclusions as loop regularization with $\mu_s =
0$.

\subsubsection{Calculation of anomaly by using dimensional regularization}

\label{subsubsec VVDRmassive}

As seen from eq. (\ref{T=Tp+TP}), in the dimensional
regularization the amplitude $T^{(1),A\{VV\}}_{\lambda\mu\nu}$ can
be decomposed in terms of two parts
$T^{(1),A\{VV\}}_{\parallel,\lambda\mu\nu}$ and
$T^{(1),A\{VV\}}_{\perp,\lambda\mu\nu}$. For the part
$T^{(1),A\{VV\}}_{\perp,\lambda\mu\nu}$, the result is the same as
the one given in eq.(\ref{masslessPerpend}). To evaluate the part
$T^{(1),A\{VV\}}_{\parallel,\lambda\mu\nu}$, we can adopt the
integration relations presented in App.~C. and consider the
relations given in App.~A.. The final results with the cross
diagram are found to be
\begin{eqnarray}
(k_3-k_1)_\mu
T^{R,A\{VV\}}_{\lambda\mu\nu}&=&{14\over24}\frac{-i}{4\pi^2}
\epsilon_{\nu\lambda\alpha\beta}(k_1-k_2)_\alpha(k_3-k_1)_\beta\\
(k_1-k_2)_\nu
T^{R,A\{VV\}}_{\lambda\mu\nu}&=&{14\over24}\frac{i}{4\pi^2}
\epsilon_{\mu\lambda\alpha\beta}(k_1-k_2)_\alpha(k_3-k_1)_\beta\nonumber\\
(k_3-k_2)_\lambda
T^{R,A\{VV\}}_{\lambda\mu\nu}&=&32m^2\epsilon_{\alpha\mu\beta\nu}(k_3-k_1)_\alpha(k_1-k_2)_\beta
I_{-2,(00)}+{20\over24}\frac{i}{4\pi^2}\epsilon_{\mu\nu\lambda\beta}(k_1-k_2)_\lambda(k_3-k_1)_\beta\nonumber
\end{eqnarray}

Consider the result eq.(\ref{PVVresult}), the three Ward
identities can be rewritten as
\begin{eqnarray}
(k_3-k_1)_\mu T^{R,A\{VV\}}_{\lambda\mu\nu}
&=&{14\over24}\frac{-i}{4\pi^2}\epsilon_{\nu\lambda\alpha\beta}(k_1-k_2)_\alpha(k_3-k_1)_\beta\\
(k_1-k_2)_\nu T^{R,A\{VV\}}_{\lambda\mu\nu}
&=&{14\over24}\frac{i}{4\pi^2} \epsilon_{\mu\lambda\alpha\beta}(k_1-k_2)_\alpha(k_3-k_1)_\beta\nonumber\\
(k_3-k_2)_\lambda T^{R,A\{VV\}}_{\lambda\mu\nu}
&=&2mT^{R,PVV}_{\mu\nu}+{20\over24}\frac{i}{4\pi^2}
\epsilon_{\mu\nu\lambda\beta}(k_1-k_2)_\lambda(k_3-k_1)_\beta\nonumber
\end{eqnarray}

To keep the conservation of vector current, making a similar
redefinition for the physical amplitude as in
eq.(\ref{VV-redefdimensionR-massless}), we then obtain for the
redefined amplitude the standard form of Ward identities
\begin{eqnarray}
(k_3-k_1)_\mu\widetilde{T}^{A\{VV\}}_{\lambda\mu\nu}(k_3-k_1,k_1-k_2)&=&0\\
(k_1-k_2)_\nu\widetilde{T}^{A\{VV\}}_{\lambda\mu\nu}(k_3-k_1,k_1-k_2)&=&0\\
(k_3-k_2)_\nu\widetilde{T}^{A\{VV\}}_{\lambda\mu\nu}(k_3-k_1,k_1-k_2)&=&2mT^{R,PVV}_{\mu\nu}
-{i\over2\pi^2}\epsilon_{\mu\nu\alpha\beta}(k_3-k_1)_\alpha(k_1-k_2)_\beta
\end{eqnarray}
The discussions given in the previous sections are applicable to
the above results.

\subsection{Calculation of anomaly with relation (\ref{vector-axialvectorgamma}) in massive QED}

\label{subsec AVmassive}

\subsubsection{Calculation of anomaly in the loop regularization}

\label{subsubsec AVLRmassive}

In this subsection, we will investigate the anomaly by evaluate
the trace of gamma matix by using the relation
(\ref{vector-axialvectorgamma}). Repeating the calculations done
in the massless case, the massive AVV amplitude can be expressed
as
\begin{eqnarray}
T^{R,(1),\{AV\}V}_{\lambda\mu\nu}&=&T^{R,(1),\{AV\}V}_{0
\lambda\mu\nu}+T^{R,(1),\{AV\}V}_{-2\lambda\mu\nu} \nonumber \\
T^{R,(1),\{AV\}V}_{0\lambda\mu\nu}& = &
2\int^1_0dx_1\int_0^{x_1}dx_2
\{\epsilon_{\mu\lambda\nu\beta}(k_3+k_1-2k_2)_\beta I_0^R(x_i,\mu)\}\nonumber\\
&&-2\epsilon_{\mu\alpha\nu\lambda}\{\int_0^1dx(-\Delta_2+k_2)_\alpha
I_0^R(x,\mu_2)+\int_0^1dx(-\Delta_1+k_2)_\alpha
I_0^R(x,\mu_1)\}\nonumber\\
T^{R,(1),\{AV\}V}_{-2\lambda\mu\nu} & = & -4\int^1_0dx_1\int_0^{x_1}dx_2\nonumber\\
&&\times\bigg\{-2\{\epsilon_{\mu\beta\nu\xi}[(-\Delta+k_1)_\beta(-\Delta+k_2)_\lambda(-\Delta+k_3)_\xi]\nonumber\\
&&+\epsilon_{\mu\beta\lambda\xi}[(-\Delta+k_1)_\beta(-\Delta+k_2)_\nu(-\Delta+k_3)_\xi]\nonumber\\
&&+\epsilon_{\nu\alpha\lambda\beta}[(-\Delta+k_1)_\beta(-\Delta+k_2)_\alpha(-\Delta+k_3)_\mu]\nonumber\\
&&+\epsilon_{\nu\alpha\lambda\xi}[(-\Delta+k_1)_\mu(-\Delta+k_3)_\xi(-\Delta+k_2)_\alpha]\}\nonumber\\
&&+\epsilon_{\lambda\alpha\nu\mu}[(k_3-k_1)^2(-\Delta+k_2)_\alpha]\nonumber\\
&&+2m^2\epsilon_{\lambda\nu\mu\alpha}(k_3-k_1)_\alpha\bigg\}I_{-2}^R(x_i,\mu)
\end{eqnarray}

For three Ward identities, by using the definitions of $\Delta$
and $\Delta_i$, we have
\begin{eqnarray}
(k_1-k_2)_\nu
T^{R,(1),\{AV\}V}_{\lambda\mu\nu}&=&-2\epsilon_{\nu\lambda\mu\xi}(k_1-k_2)_\nu(k_3-k_1)_\xi\nonumber\\
&&\times\bigg\{4[(k_1-k_2)^2I_{-2,(01)}-(k_1-k_2)^2I_{-2,(02)}\nonumber\\
&&+(k_3-k_1)\cdot(k_1-k_2)I_{-2,(11)}]+2(k_3-k_1)^2I_{-2,(10)}\nonumber\\
&&+2m^2\epsilon_{\lambda\nu\mu\alpha}(k_1-k_2)_\nu(k_3-k_1)_\alpha I_{-2,(00)}\bigg\}\\
&&-2\epsilon_{\nu\lambda\mu\xi}(k_1-k_2)_\nu(k_3-k_1)_\xi\{I_{0,(00)}+{i\over32\pi^2}Y((p+q)^2,q^2)\}\nonumber\\
(k_3-k_2)_\lambda
T^{R,(1),\{AV\}V}_{\lambda\mu\nu}&=&-2\epsilon_{\nu\lambda\mu\xi}(k_1-k_2)_\xi(k_3-k_1)_\lambda\nonumber\\
&&\times\bigg\{4[(k_1-k_2)^2I_{-2,(01)}-(k_1-k_2)^2I_{-2,(02)}\nonumber\\
&&+(k_1-k_2)\cdot (k_3-k_1)I_{-2,(01)}-(k_1-k_2)\cdot
(k_3-k_1)I_{-2,(02)}\nonumber\\
&&+(k_1-k_2)\cdot
(k_3-k_1)I_{-2,(11)}+(k_3-k_1)^2I_{-2,(11)}]\nonumber\\
&&+2[(k_3-k_1)^2I_{-2,(00)}-(k_3-k_1)^2I_{-2,(01)}-(k_3-k_1)^2I_{-2,(10)}]\nonumber\\
&&++2m^2\epsilon_{\lambda\nu\mu\alpha}(k_1-k_2)_\lambda(k_3-k_1)_\alpha\bigg\}\nonumber\\
&&-2\epsilon_{\nu\lambda\mu\xi}(k_1-k_2)_\xi(k_3-k_1)_\lambda
I_{0,(00)}^\prime\\
(k_3-k_1)_\mu
T^{R,(1),\{AV\}V}_{\lambda\mu\nu}&=&-2\epsilon_{\nu\lambda\mu\xi}(k_3-k_1)_\mu(k_1-k_2)_\xi\nonumber\\
&&\times\bigg\{-4[-(k_1-k_2)\cdot(k_3-k_1)I_{-2,(01)}+(k_1-k_2)\cdot(k_3-k_1)I_{-2,(02)}\nonumber\\
&&+2(k_1-k_2)\cdot(k_3-k_1)I_{-2,(11)}+(k_3-k_1)^2I_{-2,(10)}-(k_3-k_1)^2I_{-2,(11)}\nonumber\\
&&-2(k_3-k_1)^2I_{-2,(20)}+(k_3-k_1)^2I_{-2,(10)}]\nonumber\\
&&+2[(k_3-k_1)^2I_{-2,(00)}-(k_3-k_1)^2I_{-2,(01)}]\bigg\}\nonumber\\
&&-2\epsilon_{\nu\lambda\mu\xi}(k_3-k_1)_\mu(k_1-k_2)_\xi\{I_{0,(00)}+{i\over32\pi^2}Y((p+q)^2,q^2)\}\nonumber\\
&&-2\epsilon_{\nu\lambda\mu\xi}(k_3-k_1)_\mu(k_1-k_2)_\xi
I_{0,(00)}^\prime
\end{eqnarray}

Using the massive relations given in App.A., we can rewrite the
above Ward identities as
\begin{eqnarray}
(k_3-k_1)_\mu
T^{R,(1),\{AV\}V}_{\lambda\mu\nu}&=&0\label{massiveWard2-1}\\
(k_1-k_2)_\nu T^{R,(1),\{AV\}V}_{\lambda\mu\nu}
&=&-8\mu_s^2\epsilon_{\nu\lambda\mu\xi}(k_1-k_2)_\nu(k_3-k_1)_\xi I_{-2,(00)}\\
&&-{i\over4\pi^2}\epsilon_{\nu\lambda\mu\xi}(k_1-k_2)_\nu(k_3-k_1)_\xi
e^{-(m^2+\mu_s^2)/M_c^2}\int_0^1dx_1\int_0^{x_1}dx_2e^{-M^2/M_c^2}\nonumber\\
(k_3-k_2)_\lambda T^{R,(1),\{AV\}V}_{\lambda\mu\nu}
&=&-8\mu_s^2\epsilon_{\nu\lambda\mu\xi}(k_1-k_2)_\xi(k_3-k_1)_\lambda  I_{-2,(00)}\nonumber\\
&&-{i\over4\pi^2}\epsilon_{\nu\lambda\mu\xi}(k_1-k_2)_\xi(k_3-k_1)_\lambda
e^{-(m^2+\mu_s^2)/M_c^2}\int_0^1dx_1\int_0^{x_1}dx_2e^{-M^2/M_c^2}\nonumber\\
&&+16m^2\epsilon_{\nu\lambda\mu\xi}(k_1-k_2)_\nu(k_3-k_1)_\xi
I_{-2,(00)}\label{massiveWard2-3}
\end{eqnarray}

It is noticed that in the treatment with the trace relation
(\ref{vector-axialvectorgamma}), one of the vectors is conserved.
This result is not the same as the one considered in the previous
subsection, where the relation (\ref{twovectorgamma}) was used in
the calculation and anomalies appear in two vector Ward
identities.

By considering the PVV amplitude result (\ref{PVVresult}), the
axial-vector Ward identity can be reexpressed as:
\begin{eqnarray}
(k_3-k_2)_\lambda T^{R,(1),\{AV\}V}_{\lambda\mu\nu}
&=&-8\mu_s^2\epsilon_{\nu\lambda\mu\xi}(k_1-k_2)_\xi(k_3-k_1)_\lambda  I_{-2,(00)}\nonumber\\
&&-{i\over4\pi^2}\epsilon_{\nu\lambda\mu\xi}(k_1-k_2)_\xi(k_3-k_1)_\lambda
e^{-(m^2+\mu_s^2)/M_c^2}\int_0^1dx_1\int_0^{x_1}dx_2e^{-M^2/M_c^2}\nonumber\\
&&+2m T_{\mu\nu}^{R,(1),PVV}
\end{eqnarray}

As the massive QED is renormalizable, we take the limit
$M_c\rightarrow\infty$, the Ward identities become
\begin{eqnarray}
(k_3-k_1)_\mu T^{R,(1),\{AV\}V}_{\lambda\mu\nu}&=&0\\
(k_1-k_2)_\nu T^{R,(1),\{AV\}V}_{\lambda\mu\nu}
&=&-8\mu_s^2\epsilon_{\nu\lambda\mu\xi}(k_1-k_2)_\nu(k_3-k_1)_\xi I_{-2,(00)}\\
&&-{i\over4\pi^2}\epsilon_{\nu\lambda\mu\xi}(k_1-k_2)_\nu(k_3-k_1)_\xi\nonumber\\
(k_3-k_2)_\lambda T^{R,(1),\{AV\}V}_{\lambda\mu\nu}
&=&-8\mu_s^2\epsilon_{\nu\lambda\mu\xi}(k_1-k_2)_\xi(k_3-k_1)_\lambda  I_{-2,(00)}\nonumber\\
&&-{i\over4\pi^2}\epsilon_{\nu\lambda\mu\xi}(k_1-k_2)_\xi(k_3-k_1)_\lambda\nonumber\\
&&+2m T_{\mu\nu}^{R,(1),PVV}
\end{eqnarray}

As in section III.B, we now consider two cases. Firstly, consider
the case that $\mu_s\neq0$, taking the external vector states be
massless and on their mass shell $p^2=(k_1-k_2)^2=0,
q^2=(k_3-k_1)^2=0$ and the axial-vector state be soft, we then
have
\begin{eqnarray}
(k_3-k_1)_\mu T^{R,(1),\{AV\}V}_{\lambda\mu\nu}&=&0\\
(k_1-k_2)_\nu
T^{R,(1),\{AV\}V}_{\lambda\mu\nu}&=&0\\
(k_3-k_2)_\lambda T^{R,(1),\{AV\}V}_{\lambda\mu\nu}&=&2m
T_{\mu\nu}^{R,(1),PVV}
\end{eqnarray}
which means that with the above conditions both the vector and
axial-vector Ward identities are preserved. In fact, in the case
$\mu_s\gg m$, the quantum corrections to the axial-vector current
Ward identity also approach to zero as shown in eq.
(\ref{massive-mu>>m}).

In conclusion, in the treatment with relation
(\ref{vector-axialvectorgamma}), the anomaly free conditions are
the same as the case in the treatment with relation
(\ref{twovectorgamma}).

We now come to consider the case with $\mu_s=0$, the massive Ward
identities eqs.(\ref{massiveWard2-1}-\ref{massiveWard2-3}) can be
rewritten as
\begin{eqnarray}
(k_3-k_1)_\mu T^{R,(1),\{AV\}V}_{\lambda\mu\nu}&=&0\\
(k_1-k_2)_\nu T^{R,(1),\{AV\}V}_{\lambda\mu\nu}
&=&-{i\over4\pi^2}\epsilon_{\nu\lambda\mu\xi}(k_1-k_2)_\nu(k_3-k_1)_\xi
\int_0^1dx_1\int_0^{x_1}dx_2e^{-(M^2+m^2)/M_c^2}\nonumber\\
(k_3-k_2)_\lambda T^{R,(1),\{AV\}V}_{\lambda\mu\nu}
&=&-{i\over4\pi^2}\epsilon_{\nu\lambda\mu\xi}(k_1-k_2)_\xi(k_3-k_1)_\lambda
\int_0^1dx_1\int_0^{x_1}dx_2e^{-(M^2+m^2)/M_c^2}\nonumber\\
&&+2mT_{\mu\nu}^{R,(1),PVV}
\end{eqnarray}

Taking the limit $M_c\rightarrow\infty$ since the massive QED is
renormalizable, the results can be expressed as
\begin{eqnarray}
(k_3-k_1)_\mu T^{R,(1),\{AV\}V}_{\lambda\mu\nu}&=&0\\
(k_1-k_2)_\nu T^{R,(1),\{AV\}V}_{\lambda\mu\nu}
&=&-{i\over8\pi^2}\epsilon_{\nu\lambda\mu\xi}(k_1-k_2)_\nu(k_3-k_1)_\xi\\
(k_3-k_2)_\lambda T^{R,(1),\{AV\}V}_{\lambda\mu\nu}
&=&2mT_{\mu\nu}^{R,(1),PVV}-{i\over8\pi^2}\epsilon_{\nu\lambda\mu\xi}(k_1-k_2)_\xi(k_3-k_1)_\lambda
\end{eqnarray}
which are the Ward identities when the trace of gamma matrices is
manipulated with relation (\ref{vector-axialvectorgamma}) where
the gamma matrices of the Lorentz indices of the axial-vector
current and one of the vector currents are grouped to reduce the
number of gamma matrices.

Again to keep both the vector currents conserved, we can use eq.
(\ref{redefA-V}) to redefine the physical amplitude. The Ward
identities for the redefined amplitude become
\begin{eqnarray}
(k_3-k_1)_\mu \tilde{T}^{(1),\{AV\}V}_{\lambda\mu\nu}&=&0\\
(k_1-k_2)_\nu
\tilde{T}^{(1),\{AV\}V}_{\lambda\mu\nu}&=&0\\
(k_3-k_2)_\lambda \tilde{T}^{(1),\{AV\}V}_{\lambda\mu\nu}&=&
(I(m,p,q) - 1 ) {i\over4\pi^2}
\epsilon_{\mu\nu\alpha\beta}(k_3-k_1)_\alpha(k_1-k_2)_\beta
\end{eqnarray}
where the function $I(m,p,q)$ was defined in eq.
(\ref{defineI(m,p,q)}). In the case that the external vector are
massless with conditions: $p^2 = 0$, $q^2 = 0$, and the
axial-vector state $(p+q)^2 = 2p\cdot q$ is soft and using eq.
(\ref{special-I-2}), we have
\begin{eqnarray}
 (k_3-k_2)_\lambda
\widetilde{T}^{(1),\{AV\}V}_{\lambda\mu\nu} = 0
 \label{AV-anomfree}
\end{eqnarray}
Namely both vector and axial-vector become anomaly free for
massive QED with the general conditions $m^2 \gg p^2, q^2,
(p+q)^2$.

While in an alternative case that $p^2 = 0$, $q^2 = 0$ and
$(p+q)^2\gg m^2$, we have
\begin{eqnarray}
(k_3-k_1)_\mu
\widetilde{T}^{(1)\{AV\}V}_{\lambda\mu\nu}&=&0\\
(k_1-k_2)_\nu
\widetilde{T}^{(1),\{AV\}V}_{\lambda\mu\nu}&=&0\\
(k_3-k_2)_\lambda
\widetilde{T}^{(1),\{AV\}V}_{\lambda\mu\nu}&=&2mT^{R,(1),PVV}_{\mu\nu}
-{i\over4\pi^2}\epsilon_{\mu\nu\alpha\beta}(k_3-k_1)_\alpha(k_1-k_2)_\beta
\end{eqnarray}
which can be regarded as a consistent check to the massless QED at
$\mu_s = 0$. In this case, the divergence of the axial-vector
current gets anomaly.

When including the cross diagrams, the Ward identities are given
by
\begin{eqnarray}
(k_3-k_1)_\mu
\widetilde{T}^{\{AV\}V}_{\lambda\mu\nu}&=&0\\
(k_1-k_2)_\nu
\widetilde{T}^{\{AV\}V}_{\lambda\mu\nu}&=&0\\
(k_3-k_2)_\lambda
\widetilde{T}^{\{AV\}V}_{\lambda\mu\nu}&=&2mT^{R,PVV}_{\mu\nu}
-{i\over2\pi^2}\epsilon_{\mu\nu\alpha\beta}(k_3-k_1)_\alpha(k_1-k_2)_\beta
\end{eqnarray}

\subsubsection{Calculation of anomaly by using Pauli-Villars regularization}

\label{subsubsec AVPVmassive}

In Pauli-Villars regularization, the regularized version of
(\ref{linearTAVV-1}) and (\ref{converTAVV-1}) have the following
forms
\begin{eqnarray}
T^{R,(1),\{AV\}V}_{L,\lambda\mu\nu}&=&-4\sum_{i=0}^2C_i\int\frac{d^4k}{(2\pi)^4}
\bigg\{\{g_{\nu\lambda}\epsilon_{\alpha\beta\mu\xi}(k+k_1)_\alpha(k+k_2)_\beta(k+k_3)_\xi\nonumber\\
&&+\epsilon_{\nu\beta\mu\xi}(k+k_1)_\beta(k+k_2)_\lambda(k+k_3)_\xi
+\epsilon_{\lambda\beta\mu\xi}(k+k_1)_\beta(k+k_2)_\nu(k+k_3)_\xi\nonumber\\
&&+\epsilon_{\lambda\alpha\nu\beta}(k+k_1)_\beta(k+k_2)_\alpha(k+k_3)_\mu
+\epsilon_{\lambda\alpha\nu\xi}(k+k_1)_\mu(k+k_2)_\alpha(k+k_3)_\xi\}\nonumber\\
&&\times\bigg[\frac{1}{[(k+k_1)^2-m_i^2][(k+k_2)^2-m_i^2][(k+k_3)^2-m_i^2]}\bigg]\nonumber\\
&&-\frac{\epsilon_{\lambda\alpha\nu\mu}}{2}\bigg[\frac{(k+k_2)_\alpha}{[(k+k_2)^2-m_i^2][(k+k_3)^2-m_i^2]}\nonumber\\
&&~~~~~~~~~~+\frac{(k+k_2)_\alpha}{[(k+k_2)^2-m_i^2][(k+k_1)^2-m_i^2]}\bigg]\bigg\}\nonumber\\
T^{R,(1),\{AV\}V}_{C,\lambda\mu\nu}&=&-2\epsilon_{\lambda\alpha\nu\mu}\sum_{i=0}^2C_i\int\frac{d^4k}{(2\pi)^4}
\frac{1}{[(k+k_1)^2-m_i^2][(k+k_2)^2-m_i^2][(k+k_3)^2-m_i^2]}\nonumber\\
&&~~~~~~~~\times\{2m_i^2(k+k_3)_\alpha-[4m_i^2-(k_3-k_1)^2](k+k_2)_\alpha+2m_i^2(k+k_1)_\alpha\}\nonumber
\end{eqnarray}

By shifting the integration variable and making some algebra, we
have
\begin{eqnarray}
T^{R,(1),\{AV\}V}_{\lambda\mu\nu}&=&T^{R,(1),\{AV\}V}_{0
\lambda\mu\nu}+T^{R,(1),\{AV\}V}_{-2\lambda\mu\nu} \nonumber \\
T^{R,(1),\{AV\}V}_{0\lambda\mu\nu}& = &
2\sum_{i=0}^2C_i\int^1_0dx_1\int_0^{x_1}dx_2
\{\epsilon_{\mu\lambda\nu\beta}(k_3+k_1-2k_2)_\beta \int\frac{d^4k}{(2\pi)^4}\frac{k^2}{[k^2-M_i^2]^3}\}\nonumber\\
&&-2\epsilon_{\mu\alpha\nu\lambda}\{\sum_{i=0}^2C_i\int_0^1dx(-\Delta_2+k_2)_\alpha
\int\frac{d^4k}{(2\pi)^4}\frac{1}{[k^2-M_{2i}^2]^2}\nonumber\\
&&+\sum_{i=0}^2C_i\int_0^1dx(-\Delta_1+k_2)_\alpha
\int\frac{d^4k}{(2\pi)^4}\frac{1}{[k^2-M_{1i}^2]^2}\}\nonumber\\
T^{R,(1),\{AV\}V}_{-2\lambda\mu\nu} & = & -4\sum_{i=0}^2C_i\int^1_0dx_1\int_0^{x_1}dx_2\nonumber\\
&&\times\bigg\{-2\{\epsilon_{\mu\beta\nu\xi}[(-\Delta+k_1)_\beta(-\Delta+k_2)_\lambda(-\Delta+k_3)_\xi]\nonumber\\
&&+\epsilon_{\mu\beta\lambda\xi}[(-\Delta+k_1)_\beta(-\Delta+k_2)_\nu(-\Delta+k_3)_\xi]\nonumber\\
&&+\epsilon_{\nu\alpha\lambda\beta}[(-\Delta+k_1)_\beta(-\Delta+k_2)_\alpha(-\Delta+k_3)_\mu]\nonumber\\
&&+\epsilon_{\nu\alpha\lambda\xi}[(-\Delta+k_1)_\mu(-\Delta+k_3)_\xi(-\Delta+k_2)_\alpha]\}\nonumber\\
&&+\epsilon_{\lambda\alpha\nu\mu}[(k_3-k_1)^2(-\Delta+k_2)_\alpha]\nonumber\\
&&+2m^2\epsilon_{\lambda\nu\mu\alpha}(k_3-k_1)_\alpha\bigg\}\int\frac{d^4k}{(2\pi)^4}\frac{1}{[k^2-M_i^2]^3}
\end{eqnarray}

From the relations given in App.B and also in App.A., we arrive at
the following results
\begin{eqnarray}
(k_3-k_1)_\mu T^{(1),\{AV\}V}_{\lambda\mu\nu}&=&0\\
(k_1-k_2)_\nu T^{R,(1),\{AV\}V}_{\lambda\mu\nu}
&=&-{i\over8\pi^2}\sum_{i=0}^2C_i\epsilon_{\nu\lambda\mu\xi}(k_1-k_2)_\nu(k_3-k_1)_\xi=0\nonumber\\
(k_3-k_2)_\lambda T^{R, (1),\{AV\}V}_{\lambda\mu\nu}
&=&-{i\over8\pi^2}\sum_{i=0}^2C_i\epsilon_{\nu\lambda\mu\xi}(k_1-k_2)_\xi(k_3-k_1)_\lambda
+\sum_{i=0}^2C_i[2m_iT_{\mu\nu}^{(1),PVV}(m_i)]\nonumber\\
&=&\sum_{i=0}^2C_i[2m_iT_{\mu\nu}^{(1),PVV}(m_i)]
\end{eqnarray}
Note that as in the massless case, although both two vectors are
conserved, while one is caused by the cancellation of the heavy
fermion fields and another is conserved automatically. In the
infinity mass limit for regulator heavy fermions, we have
\begin{eqnarray}
(k_3-k_2)_\lambda T^{R,
(1),\{AV\}V}_{\lambda\mu\nu}&=&2mT_{\mu\nu}^{(1),PVV}-\frac{i}{4\pi^2}\epsilon_{\mu\nu\alpha\beta}(k_3-k_1)_\alpha(k_2-k_1)_\beta
\end{eqnarray}

\subsubsection{Calculation of anomaly by using dimensional regularization}

\label{subsubsec AVDRmassive}

In the massive QED, the parallel part of the amplitude is given by
\begin{eqnarray}
T^{R,(1),\{AV\}V}_{\parallel,\lambda\mu\nu}&=&T^{R,(1),\{AV\}V}_{\parallel,0
\lambda\mu\nu}+T^{R,(1),\{AV\}V}_{\parallel,-2\lambda\mu\nu} \nonumber \\
T^{R,(1),\{AV\}V}_{\parallel,0\lambda\mu\nu}& = &
2\int^1_0dx_1\int_0^{x_1}dx_2
\{\epsilon_{\mu\lambda\nu\beta}(k_3+k_1-2k_2)_\beta {i\over16\pi^2}[-\ln (M^{2}+m^2)]\}\nonumber\\
&&-2\epsilon_{\mu\alpha\nu\lambda}\{\int_0^1dx(-\Delta_1+k_2)_\alpha
{i\over16\pi^2}[-\ln
(M_2^{2}+m^2)]\nonumber\\
&&+\int_0^1dx(-\Delta+k_2)_\alpha
{i\over16\pi^2}[-\ln (M_1^{2}+m^2)]\}\nonumber\\
T^{R,(1),\{AV\}V}_{\parallel,-2\lambda\mu\nu} & = & -4\int^1_0dx_1\int_0^{x_1}dx_2\nonumber\\
&&\times\bigg\{-2\{\epsilon_{\mu\beta\nu\xi}[(-\Delta+k_1)_\beta(-\Delta+k_2)_\lambda(-\Delta+k_3)_\xi]\nonumber\\
&&+\epsilon_{\mu\beta\lambda\xi}[(-\Delta+k_1)_\beta(-\Delta+k_2)_\nu(-\Delta+k_3)_\xi]\nonumber\\
&&+\epsilon_{\nu\alpha\lambda\beta}[(-\Delta+k_1)_\beta(-\Delta+k_2)_\alpha(-\Delta+k_3)_\mu]\nonumber\\
&&+\epsilon_{\nu\alpha\lambda\xi}[(-\Delta+k_1)_\mu(-\Delta+k_3)_\xi(-\Delta+k_2)_\alpha]\}\nonumber\\
&&+\epsilon_{\lambda\alpha\nu\mu}[(k_3-k_1)^2(-\Delta+k_2)_\alpha]\nonumber\\
&&+2m^2\epsilon_{\lambda\nu\mu\alpha}(k_3-k_1)_\alpha\bigg\}{-i\over32\pi^2}\frac{1}{M^{2}+m^2}
\end{eqnarray}

Using the definitions of $\Delta$ and $\Delta_i$, as well as the
relations given in App.C. and App.A., we can express the Ward
identities as
\begin{eqnarray}
(k_3-k_1)_\mu T^{R,(1),\{AV\}V}_{\lambda\mu\nu}&=&0\\
(k_1-k_2)_\nu T^{R,(1),\{AV\}V}_{\parallel,\lambda\mu\nu}
&=&-{10\over16}{i\over4\pi^2}\epsilon_{\nu\lambda\mu\xi}(k_1-k_2)_\nu(k_3-k_1)_\xi\nonumber\\
(k_3-k_2)_\lambda T^{R,(1),\{AV\}V}_{\parallel,\lambda\mu\nu}
&=&{10\over16}{i\over4\pi^2}\epsilon_{\nu\lambda\mu\xi}(k_1-k_2)_\lambda(k_3-k_1)_\xi\nonumber\\
&&+16m^2\epsilon_{\nu\lambda\mu\xi}(k_1-k_2)_\nu(k_3-k_1)_\xi
I_{-2,(00)}
\end{eqnarray}

Noticing the result for the PVV diagram and the conclusion of the
perpendicular part of the amplitude, we have
\begin{eqnarray}
(k_3-k_1)_\mu T^{(1),\{AV\}V}_{\lambda\mu\nu}
&=&-{14\over24}{i\over4\pi^2}\epsilon_{\mu\nu\lambda\alpha}(k_3-k_1)_\mu(k_1-k_2)_\alpha\nonumber\\
(k_1-k_2)_\nu T^{(1),\{AV\}V}_{\lambda\mu\nu}
&=&-{7\over24}{i\over4\pi^2}\epsilon_{\nu\lambda\mu\xi}(k_1-k_2)_\nu(k_3-k_1)_\xi\nonumber\\
(k_3-k_2)_\lambda T^{(1),\{AV\}V}_{\lambda\mu\nu}
&=&{31\over24}{i\over4\pi^2}\epsilon_{\nu\lambda\mu\xi}(k_1-k_2)_\lambda(k_3-k_1)_\xi+2mT_{\mu\nu}^{(1),PVV}
\end{eqnarray}

To keep the vector currents conserved, making again the
redefinition (\ref{redefinitionV-Adimension}), we then yield the
familiar anomaly
\begin{eqnarray}
(k_3-k_1)_\mu \tilde{T}^{(1),\{AV\}V}_{\lambda\mu\nu}&=&0\nonumber\\
(k_1-k_2)_\nu \tilde{T}^{(1),\{AV\}V}_{\lambda\mu\nu}&=&0\nonumber\\
(k_3-k_2)_\lambda \tilde{T}^{(1),\{AV\}V}_{\lambda\mu\nu}
&=&2mT_{\mu\nu}^{(1),PVV}-{i\over4\pi^2}\epsilon_{\mu\nu\alpha\beta}(k_3-k_1)_\alpha(k_1-k_2)_\beta
\end{eqnarray}
Including the cross diagram, we arrive at the final results
\begin{eqnarray}
(k_3-k_1)_\mu \tilde{T}^{\{AV\}V}_{\lambda\mu\nu}&=&0\nonumber\\
(k_1-k_2)_\nu \tilde{T}^{\{AV\}V}_{\lambda\mu\nu}&=&0\nonumber\\
(k_3-k_2)_\lambda \tilde{T}^{\{AV\}V}_{\lambda\mu\nu}
&=&2mT_{\mu\nu}^{PVV}-{i\over2\pi^2}\epsilon_{\mu\nu\alpha\beta}(k_3-k_1)_\alpha(k_1-k_2)_\beta
\end{eqnarray}

\subsection{On ambiguities of reducing the triangle diagram by acting external momentum }

\label{sunsec AB}

In some literatures, it is often to demonstrate the calculations
of anomaly via the effective PVV amplitude instead of calculating
the AVV triangle diagram directly. Here the PVV amplitude is
obtained by acting the external momentum of currents on the AVV
amplitude. In the following we will show within the framework of
massless QED how some ambiguities arise from such kind of
approaches in the dimensional regularization and Pauli-Villars
scheme. In the loop regularization, as the regularization
prescription operates on the irreducible loop integrals (ILIs)
that are evaluated from Feynman loop integrals, the action of
external momentum on the ILIs does not change the structure of
Ward identities, so that there exist no such ambiguities.

\subsubsection{The calculation with dimensional regularization}

\label{subsubsec ABDR}

Following ref. \cite{DR}, acting the external momentum of
axial-vector current $(k_3-k_2)_\lambda$ on the amplitude
(\ref{TAVV}) and using the relation
\begin{eqnarray}
(k_3-k_2)_\lambda\gamma_\lambda\gamma_5
=(k\hspace{-0.2cm}\slash+k_3\hspace{-0.3cm}\slash)\gamma_5
+\gamma_5(k\hspace{-0.2cm}\slash+k_2\hspace{-0.3cm}\slash)-2\gamma_5k_\perp\hspace{-0.35cm}\slash
\end{eqnarray}
we then have
\begin{eqnarray}
(k_3-k_2)_\lambda
T^{(1),AVV}_{\lambda\mu\nu}&=&i\int\frac{d^nk}{(2\pi)^n}{\rm
tr}\{\gamma_5\frac{1}{(k\hspace{-0.2cm}\slash+k\hspace{-0.2cm}\slash_2)}
\gamma_\nu\frac{1}{(k\hspace{-0.2cm}\slash+k\hspace{-0.2cm}\slash_1)}
\gamma_\mu\}\nonumber\\
&&+i\int\frac{d^nk}{(2\pi)^n}{\rm tr}\{\gamma_5
\gamma_\nu\frac{1}{(k\hspace{-0.2cm}\slash+k\hspace{-0.2cm}\slash_1)}
\gamma_\mu\frac{1}{(k\hspace{-0.2cm}\slash+k\hspace{-0.2cm}\slash_3)}\}\nonumber\\
&&-2i\int\frac{d^nk}{(2\pi)^n}{\rm
tr}\{\gamma_5k_\perp\hspace{-0.35cm}\slash\frac{1}{(k\hspace{-0.2cm}\slash+k\hspace{-0.2cm}\slash_2)}
\gamma_\nu\frac{1}{(k\hspace{-0.2cm}\slash+k\hspace{-0.2cm}\slash_1)}
\gamma_\mu\frac{1}{(k\hspace{-0.2cm}\slash+k\hspace{-0.2cm}\slash_3)}\}
\end{eqnarray}

An explicit calculation shows that the first two terms vanish
separately, then the above amplitude is simplified to be
\begin{eqnarray}
(k_3-k_2)_\lambda
T^{(1),AVV}_{\lambda\mu\nu}&=&-2i\int\frac{d^nk}{(2\pi)^n}{\rm
tr}\{\gamma_5k_\perp\hspace{-0.35cm}\slash\frac{1}{(k\hspace{-0.2cm}\slash+k\hspace{-0.2cm}\slash_2)}
\gamma_\nu\frac{1}{(k\hspace{-0.2cm}\slash+k\hspace{-0.2cm}\slash_1)}
\gamma_\mu\frac{1}{(k\hspace{-0.2cm}\slash+k\hspace{-0.2cm}\slash_3)}\}
\end{eqnarray}

Using the relation in the dimensional regularization that
\begin{eqnarray}
(k\hspace{-0.2cm}\slash_\perp)^2=k_\perp^2\rightarrow{(d-4)\over
d}l^2
\end{eqnarray}
and considering the results in App.C, we have
\begin{eqnarray}
(k_3-k_2)_\lambda
T^{(1),AVV}_{\lambda\mu\nu}&=&-{i\over4\pi^2}\epsilon_{\mu\nu\alpha\beta}(k_3-k_1)_\alpha(k_1-k_2)_\beta
\end{eqnarray}

Considering the cross diagram, the last form of anomaly is
\begin{eqnarray}
(k_3-k_2)_\lambda
T^{AVV}_{\lambda\mu\nu}&=&-{i\over2\pi^2}\epsilon_{\mu\nu\alpha\beta}(k_3-k_1)_\alpha(k_1-k_2)_\beta
\end{eqnarray}

However, if the external momentum acting on the amplitude is not
of the axial-vector current but of vector currents, which results
will results will be obtain. To check that, using the relation
\begin{eqnarray}
(k_3-k_1)_\mu\gamma_\mu\gamma_5=(k\hspace{-0.2cm}\slash+k_3\hspace{-0.3cm}\slash)
\gamma_5+\gamma_5(k\hspace{-0.2cm}\slash+k_1\hspace{-0.3cm}\slash)-2\gamma_5k_\perp\hspace{-0.35cm}\slash
\end{eqnarray}
and moving $\gamma_5$ behind $\gamma_\mu$ before analytically
extending the dimension, namely writing the amplitude as
\begin{eqnarray}
T^{(1),AVV}_{\lambda\mu\nu}&=&(-1)\int\frac{d^4k}{(2\pi)^4}{\rm
tr}\{\gamma_\lambda\frac{i}{(k\hspace{-0.2cm}\slash+k\hspace{-0.2cm}\slash_2)}
\gamma_\nu\frac{i}{(k\hspace{-0.2cm}\slash+k\hspace{-0.2cm}\slash_1)}
\gamma_\mu\gamma_5\frac{i}{(k\hspace{-0.2cm}\slash+k\hspace{-0.2cm}\slash_3)}\}
\end{eqnarray}
we then get, along the same calculations as the above, that
\begin{eqnarray}
(k_3-k_1)_\mu
T^{(1),AVV}_{\lambda\mu\nu}&=&-{i\over4\pi^2}\epsilon_{\mu\nu\alpha\beta}(k_3-k_1)_\alpha(k_1-k_2)_\beta
\end{eqnarray}
with the cross diagram, one obtains
\begin{eqnarray}
(k_3-k_1)_\mu
T^{AVV}_{\lambda\mu\nu}&=&-{i\over2\pi^2}\epsilon_{\mu\nu\alpha\beta}(k_3-k_1)_\alpha(k_1-k_2)_\beta
\end{eqnarray}
which shows that in such an evaluation one can not distinguish
with currents which is the axial-vector current and where the
anomaly lives.

\subsubsection{The calculation with Pauli-Villars regularization}

\label{subsubsec ABPV}

In Pauli-Villars regularization\cite{PV}, the ambiguity mentioned
above becomes manifest. After the action of the external momentum
of axial-vector current, we have
\begin{eqnarray}
(k_3-k_2)_\lambda
T^{R,(1),AVV}_{\lambda\mu\nu}&=&i\sum_{i=0}^2\int\frac{d^4k}{(2\pi)^4}{\rm
tr}\{\gamma_5\frac{1}{(k\hspace{-0.2cm}\slash+k\hspace{-0.2cm}\slash_2)-m_i}
\gamma_\nu\frac{1}{(k\hspace{-0.2cm}\slash+k\hspace{-0.2cm}\slash_1)-m_i}\gamma_\mu\}\nonumber\\
&&+i\sum_{i=0}^2\int\frac{d^4k}{(2\pi)^4}{\rm tr}
\{\gamma_5\gamma_\nu\frac{1}{(k\hspace{-0.2cm}\slash+k\hspace{-0.2cm}\slash_1)-m_i}\gamma_\mu
\frac{1}{(k\hspace{-0.2cm}\slash+k\hspace{-0.2cm}\slash_3)-m_i}\}   \\
&&+2i\sum_{i=0}^2m_i\int\frac{d^4k}{(2\pi)^4}\nonumber\\
&&\times{\rm tr}
\{\gamma_5\frac{1}{(k\hspace{-0.2cm}\slash+k\hspace{-0.2cm}\slash_2)-m_i}
\gamma_\nu\frac{1}{(k\hspace{-0.2cm}\slash+k\hspace{-0.2cm}\slash_1)-m_i}\gamma_\mu
\frac{1}{(k\hspace{-0.2cm}\slash+k\hspace{-0.2cm}\slash_3)-m_i}\}
\nonumber
\end{eqnarray}

Again the explicit calculation shows that the first two terms are
zero separately and we obtain the result
\begin{eqnarray}
(k_3-k_2)_\lambda
T^{R,(1),AVV}_{\lambda\mu\nu}&=&2\sum_{i=0}^2m_iT_{\mu\nu}^{(1),PVV}(m_i)\nonumber\\
&=&-{i\over4\pi^2}\epsilon_{\mu\nu\alpha\beta}(k_3-k_1)_\alpha(k_1-k_2)_\beta
\end{eqnarray}
Considering the cross diagram, we have
\begin{eqnarray}
(k_3-k_2)_\lambda
T^{R,AVV}_{\lambda\mu\nu}&=&2\sum_{i=0}^2m_iT_{\mu\nu}^{PVV}(m_i)\nonumber\\
&=&-{i\over2\pi^2}\epsilon_{\mu\nu\alpha\beta}(k_3-k_1)_\alpha(k_1-k_2)_\beta
\end{eqnarray}
which shows that anomaly lives in the axial-vector Ward identity.

However, if the external momentum acting on the amplitude is not
of the axial-vector momentum but the vector momentum
$(k_3-k_1)_\mu$. Then by using the same method, we get
\begin{eqnarray}
(k_3-k_1)_\mu
T^{R,AVV}_{\lambda\mu\nu}&=&2\sum_{i=0}^2m_iT_{\lambda\nu}^{PVV}(m_i)\nonumber\\
&=&-{i\over2\pi^2}\epsilon_{\lambda\nu\alpha\beta}(k_3-k_1)_\alpha(k_1-k_2)_\beta
\end{eqnarray}
which indicates that anomaly lives in the vector Ward identity.

It is demonstrated from the above evaluations that one can not
determine in Pauli-Villars regularization wether anomaly lives in
the axial-vector Ward identity or in the vector Ward identity.


\section{Conclusions and Remarks}

\label{sec CR}

  We have investigated the triangle anomaly by using the
symmetry-preserving loop regularization method developed recently
in refs.\cite{LR}. As the loop regularization is realized in the
initial dimension of theory without modifying the original theory,
it has no $\gamma_5$ problem faced in the dimensional
regularization. Also the loop regularization preserves non-Abelian
gauge symmetry and satisfies a set of consistency conditions, that
clearly distinguishes with the Pauli-Villars scheme. Especially,
the loop regularization allows us to introduce two intrinsic mass
scales without destroying any symmetries of original theory. The
two intrinsic mass scales are corresponding to the characterizing
energy scale $M_c$ and sliding energy scale $\mu_s$ which actually
play the roles of ultra-violet and infrared cut-off energy scales
respectively. It has been shown that when $k^2 \ll \mu_s^2, m^2
\ll M_c^2 \to \infty$ with $k$ the momentum of external
axial-vector state and $m$ the mass of loop fermions, then both
massless and massive spinor quantum gauge theories become anomaly
free. It implies that the loop regularization proposed in
refs.\cite{LR} can be regarded as a fully symmetry-preserving
regularization method by keeping a sufficiently large IR cut-off
scale $\mu_s$. On the other hand, it also becomes clear from loop
regularization that in general the triangle anomaly appears when $
m^2, \mu_s^2 \ll k^2 \ll M_c^2 \to \infty$. The typical case of
triangle anomaly occurs in the massless QED ($m=0$ ) with $\mu_s
=0$ and $M_c\rightarrow\infty$. Namely, the symmetry-preserving
loop regularization can consistently deal with quantum anomaly and
meanwhile allows an anomaly-free treatment.

Comparing loop regularization with Pauli-Villars scheme, it has
been seen that the triangle anomaly in the two regularization
schemes arises from different sources. In loop regularization, the
direct calculation shows that, with the general conclusion of
gamma trace, the triangle anomaly appears only in the axial-vector
Ward identity, that should be the intrinsic property of original
theory as the loop regularization is carried out without modifying
the original theory. When treating the gamma trace with relation
(\ref{twovectorgamma}) in which the Lorentz indices of the two
vector currents were classified in a group, the anomaly appears
only in the two vector currents, while the Ward identity of the
remaining current is still preserved. Such an anomaly can easily
be shifted to the axial-vector Ward identity by a redefinition of
AVV amplitudes. While in the treatment with relation
(\ref{vector-axialvectorgamma}) in which the gamma matrices with
the Lorentz indices of a vector and the axial-vector currents are
grouped to simplify the gamma trace, anomalies live in the
axial-vector Ward identity and the grouped vector Ward identity
while the other vector Ward identity still preserved. This anomaly
can also be shifted to the axial-vector Ward identity and keep
both the vector Ward identities by redefining the physical
amplitude. In Pauli-Villars regularization, in all the three
treatments, the vector Ward identity is preserved since the
anomaly of original theory is cancelled by the heavy regulator
spinors introduced in the modified lagrangian of original theory,
its effect is equivalent to make the redefinition of AVV
amplitudes in loop regularization. Although anomaly only rises in
the axial-vector Ward identity, the cancellation mechanism is
different in different treatments. It is then not difficult to
understand why anomaly in Pauli-Villars scheme appears directly in
the axial-vector Ward identity. But the anomaly is caused by the
regulator spinors, which becomes unclear whether anomaly is the
intrinsic property of the original theory or due to the
introduction of regular fields.

In comparison with dimensional regularization, the explicit
calculation shows that, in all the three treatments, the triangle
anomaly in dimensional regularization receives contributions from
both the $n-4$ dimensions and the original four dimensions.
Consequently, both the vector and axial-vector Ward identities are
violated. Nevertheless, if acting the external momentum of the
axial-vector current on the AVV diagram before evaluating the
integrals, the resulting triangle anomaly only depends on the
extended dimensions and appears in the axial-vector Ward identity.
Namely, in dimensional regularization the triangle anomaly of
vector and axial-vector currents due to quantum loop corrections
depends on the procedures of operation although the total anomaly
when normalizing to the conserved vector current has the same
standard form. Besides this ambiguity, if the acting external
momentum is not operating on the axial-vector current momentum but
on the vector current momentum, the same calculation gives the
anomaly in the vector Ward identity. In this sense, we may state
that the conventional method by acting the external momentum to
reduce the triangle diagram is not a reasonable method for
correctly obtaining the Ward identity anomaly.

In addition, there are also some discussions on the anomaly based
on some models in which the triangle diagram with three
axial-vector current couplings are considered\cite{3A}. Those
results can be obtained from ours by making a redefinition for the
amplitude, this is because we can always contract two $\gamma_5$
and reduce the AAA triangle diagram to the AVV one which has been
considered above. Then one can also make a redefinition for the
amplitude so that all three currents have the same amplitudes.

In conclusion, the symmetry-preserving loop regularization
described in refs.\cite{LR} has been shown to be a useful
regularization method for consistently dealing with anomaly of
theory and providing an anomaly-free treatment of quantum field
theories by keeping a sufficient large infrared cut-off scale
$\mu_s$. A unique solution for the Ward identity anomaly of
axial-vector current is obtained by eliminating the ambiguity
caused by the trace of gamma matrices with $\gamma_5$ through
treating all the three currents symmetrically, which is simply
realized by using the definition of $\gamma_5$ in the trace of
gamma matrices.

\acknowledgments

\label{ACK}

One of authors (YLW) would like to express his thanks to David
Gross for valuable discussions on anomaly-free treatment based on
the loop regularization, and Roman Jackiw for stimulating and
helpful discussions on the treatment of clarifying the
long-standing ambiguities in the perturbative calculations of
triangle anomaly. We would like to thank both Stephen Adler and
Roman Jackiw for reading the manuscript and for their
encouragements. This work was supported in part by the National
Science Foundation of China (NSFC) under the grant 10475105,
10491306, and the Project of Knowledge Innovation Program (PKIP)
of Chinese Academy of Sciences.

\appendix

\label{app.}
\section{Useful Relations of Feynman Parameter Integrals in Loop Regularization}

\label{app.A}

Here we first introduce the definitions
\begin{eqnarray}
Y(p^2,q^2;m^2,\mu_s^2)&=&\int_0^1dz\ln[\frac{p^2z(1-z)-\hat{m}^2}{q^2z(1-z)-\hat{m}^2}]\\
&&+\int_0^1dz\int_0^1\frac{d\sigma}{\sigma}\bigg\{\exp\{\sigma[
\frac{p^2z(1-z)-\hat{m}^2}{-M_c^2}]\}-\exp\{\sigma[\frac{q^2z(1-z)-\hat{m}^2}{-M_c^2}]\}\bigg\}\nonumber\\
I_{-2,(ij)}(m^2,\mu_s^2)&=&{i\over32\pi^2}\int_0^1dx_1\int_0^{x_1}dx_2{x_2^i(x_1-x_2)^j\over
-M^2-\hat{m}^2}[1-y_{-2}({\hat{m}^2+M^2\over M_c^2})]
\end{eqnarray}
with $p_\mu=(k_1-k_2)_\mu,~~q_\nu=(k_3-k_1)_\nu$,
$\hat{m}^2=m^2+\mu_s^2$ and $m$ the mass of loop fermions. It is
easy to see that $I_{-2,(00)}$ is symmetric under the interchange
of $p$ and $q$.

The difference between two logarithemic divergence integrals can
be expressed in terms of the above integrals
\begin{eqnarray}
I_{0,(00)}(m^2,\mu_s^2)&\equiv&{i\over16\pi^2}\{\int^1_0dx_1\int_0^{x_1}dx_2
[\ln(\frac{M_c^2}{M^2+\hat{m}^2})-\gamma_\omega+y_{0}({\hat{m}^2+M^2\over M_c^2})]\nonumber\\
&&\;\;\;\;\;\;\;\;\;\;\;-\int_0^1dxx[\ln(\frac{M_c^2}{M_3^2+\hat{m}^2})
-\gamma_\omega+y_{0}({\hat{m}^2+M_3^2\over M_c^2})]\}\nonumber\\
&&=\frac{i}{16\pi^2}\bigg\{e^{-\hat{m}^2/M_c^2}\int_0^1dx_1\int_0^{x_1}dx_2
e^{-M^2/M_c^2}-{1\over2}Y((p+q)^2,q^2;m^2,\mu_s^2)\bigg\}\nonumber\\
&&\;\;\;\;\;\;\;\;\;\;\;\;\;-q^2I_{-2,(10)}(m^2,\mu_s^2)
-p^2I_{-2,(01)}(m^2,\mu_s^2)+2\hat{m}^2I_{-2,(00)}(m^2,\mu_s^2)\nonumber\\
I_{0,(00)}^\prime(m^2,\mu_s^2)&\equiv&{i\over16\pi^2}\{\int^1_0dx_1\int_0^{x_1}dx_2
[\ln(\frac{M_c^2}{M^2+\hat{m}^2})-\gamma_\omega+y_{0}({\hat{m}^2+M^2\over M_c^2})]\nonumber\\
&&\;\;\;\;\;\;\;\;\;\;\;-\int_0^1dxx[\ln(\frac{M_c^2}{M_1^2+\hat{m}^2})
-\gamma_\omega+y_{0}({\hat{m}^2+M_1^2\over M_c^2})]\}\nonumber\\
&&=\frac{i}{16\pi^2}\bigg\{e^{-\hat{m}^2/M_c^2}\int_0^1dx_1\int_0^{x_1}dx_2
e^{-M^2/M_c^2}-{1\over2}Y((p+q)^2,p^2;m^2,\mu_s^2)\bigg\}\nonumber\\
&&\;\;\;\;\;\;\;\;\;\;\;\;\;-q^2I_{-2,(10)}(m^2,\mu_s^2)-p^2I_{-2,(01)}(m^2,\mu_s^2)
+2\hat{m}^2I_{-2,(00)}(m^2,\mu_s^2)\nonumber
\end{eqnarray}
It should be noticed that although $I_{0,(00)}$ and
$I_{0,(00)}^\prime$ are UV scale dependence, it is actually finite
and independent of $M_c^2$ in the limit $M_c \to \infty$.

Using the tricks of partial integral and changing the integrating
variable, one can easily get the following useful forms
\begin{eqnarray}
I_{-2,(01)}(m^2,\mu_s^2)&=&\frac{-i}{64\pi^2}\frac{(p\cdot
q)q^2}{p^2q^2-(p\cdot q)^2}\nonumber\\
&&\times\{{1\over q^2}Y((p+q)^2,p^2;m^2,\mu_s^2)+{1\over p\cdot
q}Y((p+q)^2,q^2;m^2,\mu_s^2)\}\nonumber\\
&&+\frac{q^2[(p\cdot q)+p^2]}{2[p^2q^2-(p\cdot
q)^2]}I_{-2,(00)}(m^2,\mu_s^2)\nonumber\\
I_{-2,(10)}(m^2,\mu_s^2)&=&I_{-2,(01)}(m^2,\mu_s^2)\bigg|_{p\leftrightarrow q}\nonumber\\
I_{-2,(11)}(m^2,\mu_s^2)&=&\frac{(p\cdot q)q^2}{p^2q^2-(p\cdot
q)^2}\bigg\{-{i\over64\pi^2}{1\over q^2}e^{-\hat{m}^2/M_c^2}
\int_0^1dx_1\int_0^{x_1}dx_2e^{-M^2/M_c^2}\nonumber\\
&&-\frac{i}{128\pi^2}\frac{[2p^4q^2+3p^2q^2p\cdot q+p^2(p\cdot
q)^2]}{2q^2p\cdot q[p^2q^2-(p\cdot
q)^2]}Y((p+q)^2,p^2;m^2,\mu_s^2)\nonumber\\
&&-\frac{i}{128\pi^2}\frac{2p^2q^2+3p^2p\cdot q+(p\cdot
q)^2}{2p\cdot q[p^2q^2-(p\cdot
q)^2]}Y((p+q)^2,q^2;m^2,\mu_s^2)\nonumber\\
&&+\{\frac{p^2[4(p\cdot q)^2+3(p^2+q^2)p\cdot q+2p^2q^2]}{8p\cdot
q[p^2q^2-(p\cdot
q)^2]}-{\hat{m}^2\over2q^2}\}I_{-2,(00)}(m^2,\mu_s^2)\bigg\}\nonumber\\
I_{-2,(20)}(m^2,\mu_s^2)&=&\frac{p^2}{p^2q^2-(p\cdot
q)^2}\bigg\{-{i\over64\pi^2}{1\over p\cdot
q}e^{-\hat{m}^2/M_c^2}\int_0^1dx_1\int_0^{x_1}dx_2e^{-M^2/M_c^2}\nonumber\\
&&-\frac{i}{128\pi^2}\frac{3p^4q^2+3p^4(p\cdot q)+2p^2q^2p\cdot
q-2(p\cdot q)^2}{2p^2[p^2q^2-(p\cdot
q)^2]}Y((p+q)^2,p^2;m^2,\mu_s^2)\nonumber\\
&&-\frac{i}{128\pi^2}\frac{p^2q^2+p\cdot q[3q^2+2p\cdot
q]}{2[p^2q^2-(p\cdot
q)^2]}Y((p+q)^2,q^2;m^2,\mu_s^2)\nonumber\\
&&+\{\frac{p^2[3q^4+q^2p^2+6q^2(p\cdot q)+2(p\cdot
q)^2]}{8[p^2q^2-(p\cdot q)^2]}
-{\hat{m}^2\over2}\}I_{-2,(00)}(m^2,\mu_s^2)\bigg\}\nonumber\\
I_{-2,(02)}(m^2,\mu_s^2)&=&I_{-2,(20)}(m^2,\mu_s^2)\bigg|_{p\leftrightarrow
q}
\end{eqnarray}

 With the above expressions, we can arrive at the following useful relations
\begin{eqnarray}
q^{2}I_{-2,(11)}(m^2,\mu_s^2) -\left( p\cdot q\right)
I_{-2,(02)}(m^2,\mu_s^2)
&=&\frac{-i}{128\pi^2}Y((p+q)^2,p^2;m^2,\mu_s^2)\nonumber\\
&&+\frac{q^{2}}{2}I_{-2,(01)}(m^2,\mu_s^2) \\
q^{2}I_{-2,(20)}(m^2,\mu_s^2) -\left( p\cdot
q\right)I_{-2,(11)}(m^2,\mu_s^2)
&=&\frac{-i}{64\pi^2}e^{-\hat{m}^2/M_c^2}\int_0^1dx_1\int_0^{x_1}dx_2e^{-M^2/M_c^2}\nonumber\\
&&-\frac{\hat{m}^{2}}{2}I_{-2,(00)}(m^2,\mu_s^2)
+\frac{p^{2}}{4}I_{-2,(01)}(m^2,\mu_s^2)\nonumber\\
&&+\frac{3q^{2}}{4}I_{-2,(10)} (m^2,\mu_s^2)\\
p^{2}I_{-2,(02)}(m^2,\mu_s^2) -\left( p\cdot q\right)
I_{-2,(11)}(m^2,\mu_s^2)
&=&\frac{-i}{64\pi^2}e^{-\hat{m}^2/M_c^2}\int_0^1dx_1\int_0^{x_1}dx_2e^{-M^2/M_c^2}\nonumber\\
&&-\frac{\hat{m}^{2}}{2}I_{-2,(00)}(m^2,\mu_s^2)+\frac{q^{2}}{4}I_{-2,(10)}(m^2,\mu_s^2)\nonumber\\
&& +\frac{3p^{2}}{4}I_{-2,(01)}(m^2,\mu_s^2) \\
p^{2}I_{-2,(11)}(m^2,\mu_s^2) -\left( p\cdot
q\right)I_{-2,(20)}(m^2,\mu_s^2)
&=&\frac{-i}{128\pi^2}Y((p+q)^2,q^2;m^2,\mu_s^2)\nonumber\\
&& +\frac{p^{2}}{2}I_{-2,(10)}(m^2,\mu_s^2)\\
q^{2}I_{-2,(10)}(m^2,\mu_s^2) -\left( p\cdot q\right)
I_{-2,(01)}(m^2,\mu_s^2)
&=&\frac{-i}{64\pi^2}Y((p+q)^2,p^2;m^2,\mu_s^2)\nonumber\\
&& +\frac{q^{2}}{2}I_{-2,(00)}(m^2,\mu_s^2) \\
p^{2}I_{-2,(01)}(m^2,\mu_s^2) -\left( p\cdot
q\right)I_{-2,(10)}(m^2,\mu_s^2)
&=&\frac{-i}{64\pi^2}Y((p+q)^2,q^2;m^2,\mu_s^2)\nonumber\\
&&+\frac{p^{2}}{2}I_{-2,(00)}(m^2,\mu_s^2)
\end{eqnarray}

\section{Useful Relations in Pauli-Villars Regularization.}

\label{APP B}

\begin{eqnarray}
\sum_{i=0}^2C_i\int\frac{d^4k}{(2\pi)^4}\frac{m_i^2}{[k^2-M_{1i}^2]^3}
&=&-i\sum_{i=0}^2C_i\int\frac{d^4k_E}{(2\pi)^4}\frac{m_i^2}{[k_E^2+M_{i}^2]^3}
=\frac{-i}{32\pi^2}\sum_{i=0}^2C_i\frac{m_i^2}{M_{i}^2}\\
\sum_{i=0}^2C_i\int\frac{d^4k}{(2\pi)^4}\frac{k^2}{[k^2-M_{1i}^2]^3}
&=&\frac{-i}{8\pi^2}\int_0^1dz\{{(M_{12}^2-M_{10}^2)\over
M_{12}^2z+M_{10}^2(1-z)}+{(M_{12}^2-M_{11}^2)\over
M_{12}^2z+M_{11}^2(1-z)}\}\nonumber\\
&=&\frac{-i}{8\pi^2}\{\ln(\frac{M_{12}^2}{M_{10}^2})+\ln(\frac{M_{12}^2}{M_{11}^2})\}
=\frac{i}{8\pi^2}\sum_{i=0}^2\ln(\frac{M_{1i}^2}{\mu^{\prime2}})\\
\sum_{i=0}^2C_i\int\frac{d^4k}{(2\pi)^4}\frac{1}{[k^2-M_{2i}^2]^2}
&=&{i\over16\pi^2}\int_0^1dz\{{(M_{22}^2-M_{20}^2)\over
M_{22}^2z+M_{20}^2(1-z)}+{(M_{22}^2-M_{21}^2)\over
M_{22}^2z+M_{21}^2(1-z)}\}\nonumber\\
&=&{i\over16\pi^2}\{\ln(\frac{M_{22}^2}{M_{20}^2})+\ln(\frac{M_{22}^2}{M_{21}^2})\}
=\frac{-i}{16\pi^2}\sum_{i=0}^2\ln(\frac{M_{2i}^2}{\mu^{\prime2}})\\
\sum_{i=0}^2C_i\int\frac{d^4k}{(2\pi)^4}\frac{1}{[k^2-M_{3i}^2]^2}
&=&{i\over16\pi^2}\int_0^1dz\{{(M_{32}^2-M_{30}^2)\over
M_{32}^2z+M_{30}^2(1-z)}+{(M_{32}^2-M_{32}^2)\over
M_{32}^2z+M_{31}^2(1-z)}\}\nonumber\\
&=&{i\over16\pi^2}\{\ln(\frac{M_{32}^2}{M_{30}^2})+\ln(\frac{M_{32}^2}{M_{31}^2})\}
=\frac{-i}{16\pi^2}\sum_{i=0}^2\ln(\frac{M_{3i}^2}{\mu^{\prime2}})
\end{eqnarray}

\section{momentum integral in dimensional regualarization}

\label{app C}
\begin{eqnarray}
\int\frac{d^nk}{(2\pi)^n}\frac{1}{(k^2-\Delta)^3}
&=&-i\int\frac{d^nk_E}{(2\pi)^n}\frac{1}{(k_E^2+\Delta)^3}
\nonumber \\
& = &
\frac{-i}{(4\pi)^{n/2}}\frac{\Gamma(3-{n\over2})}{\Gamma(3)}({1\over\Delta})^{3-{n\over2}}
=\frac{-i}{32\pi^2}{1\over\Delta} \\
\int\frac{d^nk}{(2\pi)^n}\frac{1}{(k^2-\Delta)^2}
&=&i\int\frac{d^nk_E}{(2\pi)^n}\frac{1}{(k_E^2+\Delta)^2}
\nonumber \\
& = & \frac{i}{(4\pi)^{n/2}}\frac{\Gamma(2-{n\over2})}{\Gamma(2)}
({1\over\Delta})^{2-{n\over2}} ={i\over16\pi^2}[-\ln\Delta+C] \\
\int\frac{d^nk}{(2\pi)^n}\frac{k^2}{(k^2-\Delta)^3}
&=&\int\frac{d^nk}{(2\pi)^n}[\frac{1}{(k^2-\Delta)^2}+\frac{\Delta}{(k^2-\Delta)^3}]
\nonumber \\
& = & {i\over16\pi^2}[-\ln\Delta+C]+{-i\over32\pi^2} \\
\int\frac{d^nk}{(2\pi)^n}\frac{(k_\perp)^2}{[k^2-M_1^2]^3}&=&{-i\over32\pi^2}
\end{eqnarray}


\end {document}